\documentclass[12pt]{iopart}
\usepackage{graphicx}
\usepackage{stmaryrd}
\usepackage{amssymb}
\usepackage[number=none]{glossary}

\def \d{\mathrm{d}}
\makeglossary
\begin{document}

\title{The apparent surface roughness of moving sand transported by wind}

\author{T P\"ahtz$^1$, J F Kok$^2$, H J Herrmann$^1$}

\address{$^1$ Computational Physics, IfB, ETH Z\"urich, Schafmattstrasse 6, 8093 Z\"urich, Switzerland}
\eads{\mailto{tpaehtz@ethz.ch}, \mailto{hans@ifb.baug.ethz.ch}}
\address{$^2$ Department of Earth and Atmospheric Sciences, Cornell University, 1104B Bradfield Hall, Ithaca, New York 14850}
\ead{jk2457@cornell.edu}
\begin{abstract}
We present a comprehensive analytical model of aeolian sand transport in saltation. It quantifies the momentum transfer from the wind to the transported sand by providing expressions for the thickness of the saltation layer and the apparent surface roughness. These expressions are for the first time entirely derived from basic physical principles. The model further predicts the sand transport rate (mass flux) and the impact threshold shear velocity. We show that the model predictions are in very good agreement with experiments and numerical state of the art simulations of aeolian saltation.
\end{abstract}
\submitto{\NJP}

\section{Introduction}
Saltation is the dominant mechanism of aeolian sand transport on Earth's deserts under turbulent wind flow. Unidirectional wind accelerates sand grains, which perform hops of typical shapes. During their hops the wind continuously transfers momentum to the grains. Therefore the wind momentum decreases, resulting in reduced wind velocities not only within, but also above the saltation layer. Above the saltation layer the average horizontal wind velocity profile $u(z)$ follows the well known law \cite{Bagnold41},
\begin{eqnarray}
 u(z)=\frac{u_*}{\kappa}\ln\frac{z}{z_o^*}, \label{velprofile}
\end{eqnarray}
where $u_*$ is the wind shear velocity, $\kappa=0.4$ is the von K\'arm\'an constant, and $z_o^*$ is the apparent roughness of the moving saltation layer. In the absence of sand transport $z_o^*$ becomes $z_o$, which is the surface roughness of a quiescent sand bed. In the presence of sand transport the magnitude of $z_o^*$ depends on how much momentum is absorbed by the saltation layer. It is crucial for the development of sand transport models, but also for landscape modelers and coastal managers to know $z_o^*$ as function of $u_*$ and $z_o$. For instance in aeolian dune models $z_o^*$ is a key quantity in the computation of the wind field over a non-flat topography, in which the shear velocity $u_*$ varies with the spatial position \cite{Kroyetal02a,Kroyetal02b,ParteliHerrmann07,Partelietal07}. The main purpose of this paper is to derive a novel prediction of $z_o^*$, which is entirely based on physical principles.

Deriving a scaling law for $z_o^*$ was also approached by previous studies. First Owen \cite{Owen64} suggested
\begin{eqnarray}
 z_o^*\propto\frac{u_*^2}{g}, \label{Oweneq}
\end{eqnarray}
where $g$ is the gravity constant. (\ref{Oweneq}) is also known as the Charnock relation, since Charnock \cite{Charnock55} derived (\ref{Oweneq}) for the roughness of a wind-blown water surface. Owen \cite{Owen64} based his formula on the assumptions that the average lift-off velocity $v_l$, with which a grain leaves the bed, is proportional to $u_*$, that opposing drag forces can be neglected, and that $z_o^*$ is proportional to the average hop height of grains $h$, also called saltation height. However the author's assumptions fail to agree with measurements. Experimental studies \cite{Namikas03,RasmussenSorensen08,Creysselsetal09,Hoetal11} found $v_l$ and consequently $h$ to be almost independent of $u_*$ within the measured range. Furthermore $z_0^*$ cannot be proportional to $h$, since $z_o^*$ was found to strongly vary with $u_*$ in experiments \cite{Rasmussenetal96,Dongetal03,ShermanFarrell08} in contrast to $h$. In addition Sherman \cite{Sherman92} found that (\ref{Oweneq}) leads to strong discrepancies with experiments close to the impact threshold $u_t$, which is the threshold shear velocity at which saltation can be sustained through the splash process. On Earth $u_t$ is below the fluid entrainment shear velocity, needed to entrain sand grains from the soil by fluid lift \cite{Bagnold41,Kok10a}. Sherman \cite{Sherman92} therefore extended (\ref{Oweneq}) to the so-called modified Charnock relation,
\begin{eqnarray}
 z_o^*-z_o\propto\frac{(u_*-u_t)^2}{g}, \label{modCharnock}
\end{eqnarray}
which ensures that at the threshold $u_*=u_t$, where no particles are moving, the roughness is unchanged, $z_o^*=z_o$. Although (\ref{modCharnock}) was successfully validated with the data set of Sherman and Farrell \cite{ShermanFarrell08} for $z_o^*$, it shares the same lack of physical justification as (\ref{Oweneq}).

A much more physical approach was presented by Raupach \cite{Raupach91}. From the mixing length approximation \cite{Prandtl25}, the author derived
\begin{eqnarray}
 \ln\frac{z_o^*}{z_o}=\left(1-\frac{u_t}{u_*}\right)\ln\frac{z_s}{1.78z_o}, \label{Raupacheq}
\end{eqnarray}
where $z_s$ is the decay height of the grain shear stress profile $\tau_g(z)$, which the author assumed to be exponentially decreasing,
\begin{eqnarray}
 \tau_g(z)=\tau_{go}e^{-z/z_s}, \label{zsexp}
\end{eqnarray}
where $\tau_{go}=\tau_g(0)$. $\tau_g(z)$ describes how much momentum is transferred, at each height $z$, from the fluid to the grains per unit soil area and time. The difficulty in the usage of (\ref{Raupacheq}) is the undetermined quantity $z_s$. Raupach \cite{Raupach91} therefore assumed $z_s$ to be proportional to the saltation height $h$, which he in turn assumed to be proportional to $u_*^2/g$ like before Owen \cite{Owen64}. However, as we discussed before, Owen's assumption is in disagreement with experiments. Relations very similar to (\ref{Raupacheq}) were also obtained in two further studies \cite{Andreotti04,DuranHerrmann06}. These studies achieved agreement with the experimental data of Rasmussen et al. \cite{Rasmussenetal96}, by introducing an ad-hoc fit relation for $z_s$. Andreotti \cite{Andreotti04} found that the data set can be well fitted if $z_s$ scales with $\sqrt{d}$, where $d$ is the mean particle diameter, and he therefore suggested
\begin{eqnarray}
 z_s\propto\sqrt{sgd}\left(\frac{\mu}{\rho_wg^2}\right)^{1/3}, \label{zsAndreotti}
\end{eqnarray}
where $s=\rho_s/\rho_w$ is the ratio of sand density $\rho_s$ and fluid density $\rho_w$, and $\mu$ is the kinematic viscosity. A scaling law very similar to (\ref{zsAndreotti}) was also used by Duran and Herrmann \cite{DuranHerrmann06}. However (\ref{zsAndreotti}) is very weakly founded on physics. Its only justification is the resulting agreement of (\ref{Raupacheq}) with the data set of Rasmussen et al. \cite{Rasmussenetal96}. Therefore there is a great necessity to either validate (\ref{zsAndreotti}) or to derive a new expression for $z_s$ from physical principles.

Within this study we do the latter. We present a comprehensive analytical model of aeolian saltation, which aims to significantly improve previous analytical models \cite{Bagnold41,Owen64,Bagnold73,Sorensen91,Sauermannetal01}. It for the first time provides expressions for $z_s$ and $z_o^*$ entirely derived from physical principles. Our analysis will reveal that $z_s$ is a measure for the thickness of the saltation layer and not proportional to the average hop height $h$. The model furthermore incorporates expressions for other important sand transport quantities, such as the mass flux $Q$ and the impact threshold $u_t$. The model is based on the concept of mean motion, meaning that average quantities are used for its description. In our model we separately consider the horizontal and vertical transport of grains. For each we analytically balance the average force and work rate per unit soil area applied by the wind on a grain during a trajectory versus the respective amounts applied by the soil on the grains during an impact. This results in a model parameter $\alpha$, describing the ratio between the average vertical and horizontal force per unit soil area, and another model parameter $\beta$, describing the ratio between the average work rate per unit soil area in the vertical motion and the horizontal motion. With theoretical arguments it is shown that $\alpha$ and $\beta$ are nearly independent of $u_*$ and the atmospheric conditions and only slightly varying with the buoyancy-reduced gravity $\tilde g$ and the particle diameter $d$. The model further contains a third parameter $\gamma$, defined as the ratio between $z_s$ and the effective height of the mean motion $z_m$. The final relations for $z_s$, $z_o^*$, and $Q$ are functions of $\alpha$, $\beta$, $\gamma$, and $u_t$. We afterwards extend our model in such a way that also $u_t$ can be computed. Thereby a fourth parameter $\eta$ comes into play, describing the ratio between the average particle velocity, reduced by the particle slip velocity, and the average wind velocity. In our model we use the assumption that the grain shear stress profile is exponentially decaying, equivalent to what was assumed in previous studies \cite{Raupach91,DuranHerrmann06} (see (\ref{zsexp})).

We validate our model with the apparent roughness data of Rasmussen et al. \cite{Rasmussenetal96} for five different grain sizes, with combined impact threshold data of several studies \cite{Bagnold37,Chepil45,Rasmussenetal94}, and with mass flux data of Creyssels et al. \cite{Creysselsetal09}. Furthermore, through simulations with the numerical state of the art model of Kok and Renno \cite{KokRenno09} we support our derived expressions and our statement that the model parameters are nearly independent of $u_*$, the atmospheric conditions, as well as $\tilde g$ and $d$.

The manuscript is structured in the following way. It starts with a comprehensive model description in Section \ref{secmodeldes}, which is followed by the model validation in Section \ref{secmodelval} and a discussion of the results in Section \ref{discussion}. The appendices incorporate long calculations and side information. There is also a glossary at the end of the manuscript, which helps to keep track of the mathematical symbols.
\section{Model description} \label{secmodeldes}
It is the main purpose of our paper to derive a novel expression for the apparent roughness $z_o^*$ during aeolian sand transport in steady state. The main focus of our model lies therefore in the analytical description of the momentum and energy transfer from the wind to the grains. Momentum and energy transfer are the main causes for the increase of the surface roughness $z_o$ of a quiescent sand bed to the apparent roughness $z_o^*$ of a moving saltation layer. In detail we use Newton's law to obtain equations, which balance the average force and work rate per unit soil area applied during a grains trajectory with the respective amounts applied during an impact. The ratio of the average force (work rate) per unit soil area for the vertical motion and force (work rate) per unit soil area for the horizontal motion is the definition of our model parameter $\alpha$ ($\beta$). After applying the balance laws we show that the decay height $z_s$ of the grain shear stress profile $\tau_g(z)$ and subsequently $z_o^*$ as well as the mass flux $Q$ can be calculated from the impact threshold $u_t$ and our model parameters. Afterwards the model is extended in such a way that also $u_t$ can be calculated as function of the model parameters. As previous studies \cite{Raupach91,DuranHerrmann06} we assume an exponentially decreasing $\tau_g(z)$ (see (\ref{zsexp})) and extensively use it in our calculations. This assumption is therefore discussed in a separate paragraph.

For the balance laws, we only consider wind drag and gravity as driving forces, but neglect turbulent lift forces, the Magnus force, electrostatic forces, and momentum as well as energy changes by mid-air collisions between grains for simplicity reasons and because gravity and drag dominate the sand transport \cite{KokRenno09}. Furthermore we simplify the description, by only considering average quantities, which implies that we neglect turbulent fluctuations of the wind velocities. Further simplifications are the use of monodisperse, spherical sand grains, being transported above a horizontal sand bed. The probably most crucial of all these simplifications is the negligence of mid-air collisions between saltating grains. The effect of such collisions has only rarely been subject of scientific studies \cite{SorensenMcEwan96,Dongetal05,Huangetal07,RenHuang10}, because for a comparison between saltation with and without mid-air collisions, one has to turn off mid-air collisions, what is possible (and common) in numerical simulations, but impossible in experiments. According to the most recent numerical study (Figure 5 in Ren and Huang \cite{RenHuang10}), the change of the mass flux due to mid-air collisions is less than $10\%$ for $u_*\approx3.5u_t$. This is below the typical measurement error of mass flux measurements ($>10\%$). According to this study the neglegance of mid-air collisions and therefore our model simplifications are acceptable up to at least $u_*\approx3.5u_t$.

This section is separated in several subsections. It starts with the presentation of notations and definitions, which are used for the description of our model, in Section \ref{notations}. In Section \ref{modelassumption} follows a short discussion of our main model assumption, the exponentially decreasing grain shear stress profile. After that the balance laws are applied, first for the force per unit soil area in Section \ref{momentum} and then for the work rate per unit soil area in Section \ref{energy}. Subsequently we discuss the invariance of the model parameters $\alpha$ and $\beta$ in Section \ref{invariance}. Afterwards we obtain a novel relation for $z_s$ in Section \ref{zsrelation}, which is further discussed in Section \ref{zsmeaning}. Then in Section \ref{zorelation} relations for $z_o^*$ and $Q$ as a function of $u_t$ and the model parameters are obtained. In Section \ref{utrelation} the model is extended, in order to allow for the computation of $u_t$ as well.

\subsection{Notations and definitions} \label{notations}
For the coming analytical calculations, we henceforth use the following notations: An index $x$ refers to the horizontal component of a given quantity, which coincides with the direction of the wind, an index $z$ to the vertical direction, whereby $z$ is also the height above the sand bed. Furthermore we differentiate between the upward and downward part of a grain's trajectory by indices $\uparrow$ and $\downarrow$, respectively. Quantities evaluated at the sand bed $z=0$ incorporate an additional index $o$. In particular quantities, which refer to a grain's impact, consist of the indices $o$ and $\downarrow$, if the quantity is evaluated before the impact, and the indices $o$ and $\uparrow$, if the quantity is evaluated after the impact.

In order to keep the manuscript simple, it is advantageous to predefine quantities, which are used in the following calculations. One quantity is the average particle mass per unit volume $\rho(z)$, transported at height $z$. $\rho(z)$ integrated over the whole saltation layer describes the mass $M$ of transported sand per unit soil area.
\begin{eqnarray}
 M=\int\limits_0^\infty\rho(z)\d z. \label{M}
\end{eqnarray}
Since we differentiate between upward and downward movement $\rho(z)$ can be divided in the mass of upward and downward moving particles per unit volume
\begin{eqnarray}
 \rho(z)=\rho_{\uparrow}(z)+\rho_{\downarrow}(z). \label{rho}
\end{eqnarray}
Other important quantities are the average vectorial wind velocity profile, $\mathbf{u}(z)$, whose $z$-component is zero $u(z):=u_x(z)=|\mathbf{u}(z)|$, and the average vectorial particle velocity profile for the upward (downward) part of the trajectory $\mathbf{v}_{\uparrow(\downarrow)}(z)$. The difference between both velocities is denoted as
\begin{eqnarray}
 \mathbf{v_{r\uparrow(\downarrow)}}(z)=\mathbf{u}(z)-\mathbf{v}_{\uparrow(\downarrow)}(z).
\end{eqnarray}
Based on these definitions, we further define the following velocity differences by
\begin{eqnarray}
 \Delta v_x(z)=v_{x\downarrow}(z)-v_{x\uparrow}(z), \label{deltavx}\\
 \Delta v_z(z)=v_{z\downarrow}(z)-v_{z\uparrow}(z),
\end{eqnarray}
and
\begin{eqnarray}
 \Delta v_x^2(z)=v_{x\downarrow}^2(z)-v_{x\uparrow}^2(z), \\
 \Delta v_z^2(z)=v_{z\downarrow}^2(z)-v_{z\uparrow}^2(z), \label{deltavz2}
\end{eqnarray}
as well as the local vertical mass flux $\phi(z)$ by
\begin{eqnarray}
 \phi(z)=\rho_\uparrow(z)v_{z\uparrow}(z)=-\rho_\downarrow(z)v_{z\downarrow}(z), \label{phi}
\end{eqnarray}
where we used that the vertical upward flux must exactly compensate the downward flux in steady state. Note that $v_{z\downarrow}(z)$ and thus $\Delta v_z(z)$ are negative. Using (\ref{rho}), (\ref{phi}) can be rewritten as
\begin{eqnarray}
 \phi(z)=\rho(z)\frac{v_{z\downarrow}(z)v_{z\uparrow}(z)}{\Delta v_z(z)}. \label{phi2}
\end{eqnarray}
With these definitions the average gain of horizontal and vertical momentum of a transported grain per unit soil area and time between the two times it crosses height $z$ can be written as
\begin{eqnarray}
 \tau_g(z)=\phi(z)\Delta v_x(z) \label{taugspez}
\end{eqnarray}
for the horizontal and
\begin{eqnarray}
 p_g(z)=\phi(z)\Delta v_z(z) \label{pgspez}
\end{eqnarray}
for the vertical momentum gain per unit soil area and time. $\tau_g(z)$ is also known as the grain shear stress profile \cite{Raupach91,DuranHerrmann06,Sauermannetal01} and $p_g(z)$ can be seen as a grain normal stress (grain pressure) profile. Note that, by inserting (\ref{phi2}) in (\ref{pgspez}), one obtains
\begin{eqnarray}
 p_g(z)=\rho(z)v_{z\downarrow}(z)v_{z\uparrow}(z), \label{pressure}
\end{eqnarray}
which is identical to the definition of the granular pressure in previous studies \cite{Creysselsetal09,Jenkinsetal10}, if vertical drag is neglected.
\subsection{Grain shear stress profile} \label{modelassumption}
\begin{figure}
 \begin{center}
  \includegraphics[scale=0.29]{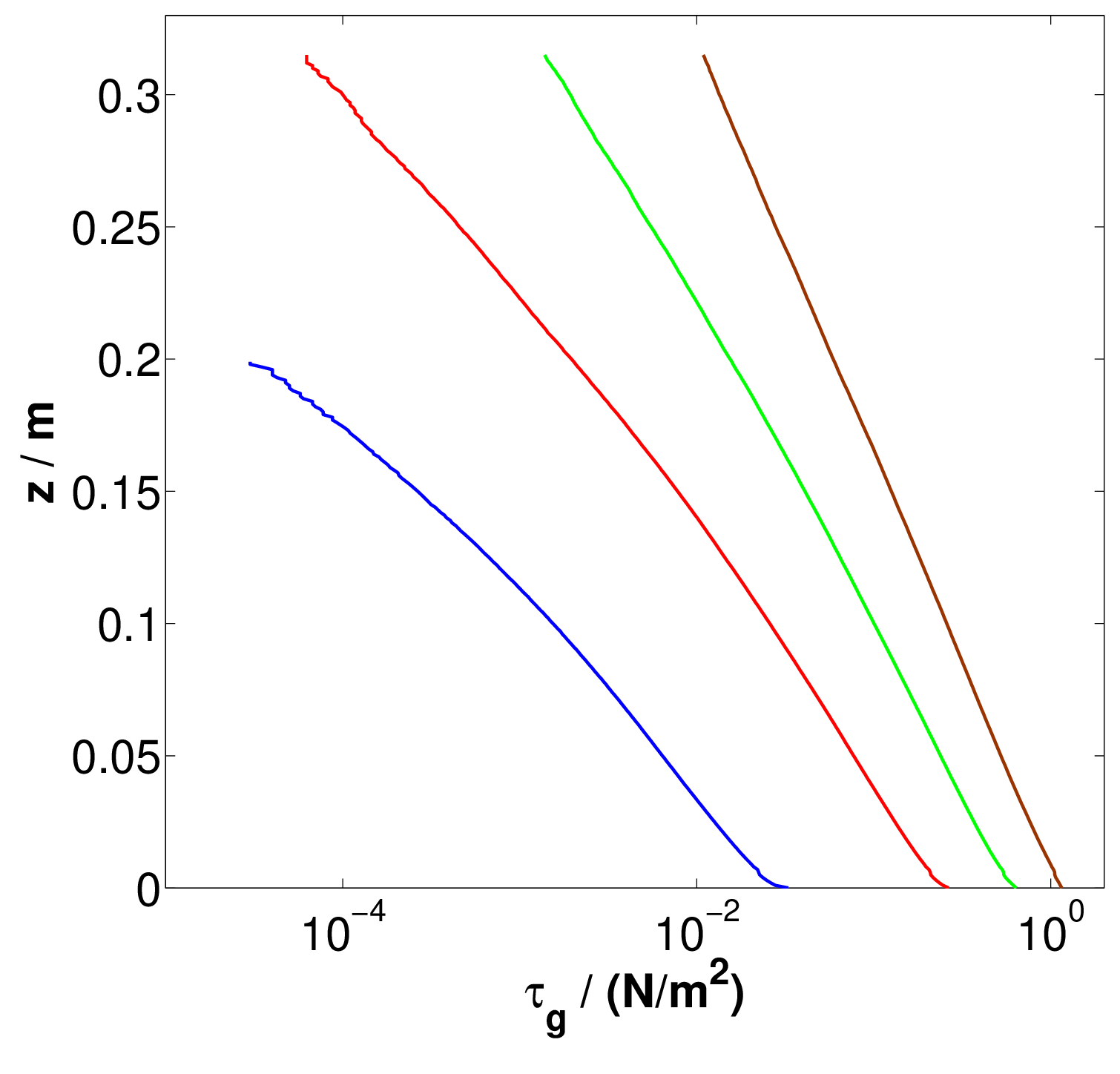}
 \end{center}
 \caption{Plot of the grain shear stress profile $\tau_g(z)$ obtained from numerical simulations with the model of Kok and Renno \cite{KokRenno09} for $u_*=0.25m/s$ (blue), $u_*=0.5m/s$ (red), $u_*=0.75m/s$ (green), and $u_*=1m/s$ (brown). The simulations are performed under Earth conditions with a mean diameter $d=250\mu m$. Over a large part $\tau_g(z)$ decays exponentially for all shear velocities.}
 \label{taugfig}
\end{figure}

In this section we discuss and motivate our main model assumption of an exponentially decreasing grain shear stress profile (see also (\ref{zsexp}))
\begin{eqnarray}
 \tau_g(z)=\tau_{go}e^{-z/z_s}. \label{zsexp2}
\end{eqnarray}
(\ref{zsexp2}) is justified in the following manner. First, an approximately exponentially decreasing mass density profile $\rho(z)$ was measured in wind tunnels \cite{Creysselsetal09,Hoetal11}. Although not necessarily identical, the profiles $\tau_g(z)$ and $\rho(z)$ should at least behave in a similar manner. Therefore it is very reasonable that also $\tau_g(z)$ decreases approximately exponentially. Second, $\tau_g(z)$ has been obtained from numerical simulations \cite{KokRenno09,AnderssonHaff91}, which indeed showed an approximately exponential decrease. This is shown in Figure \ref{taugfig} for simulation results with the numerical model of Kok and Renno \cite{KokRenno09}.
It should be noted that the mass density profile $\rho(z)$ strongly deviates from the exponential shape at very small heights in the simulations. Such a deviation is also present for the grain shear stress profile $\tau_g(z)$ (see Figure \ref{taugfig} at heights very close to zero), however to a much lesser extent.
\subsection{Force balance} \label{momentum}
As already pointed out, the description of the momentum transfer from the wind to the grains is a key ingredient towards a description of the feedback effect of sand transport on the wind profile and thus a first step towards a prediction of the apparent roughness $z_o^*$. Newton's second law for grains moving in a particular trajectory, indicated by a lower index '1', can be written as ($\d z=v_{1z\uparrow(\downarrow)}\d t$)
\begin{eqnarray}
 \rho_{1\uparrow(\downarrow)}v_{1z\uparrow(\downarrow)}\frac{\d v_{1x\uparrow(\downarrow)}}{\d z}=f_{1x\uparrow(\downarrow)}, \label{mom1x} \\
 \rho_{1\uparrow(\downarrow)}v_{1z\uparrow(\downarrow)}\frac{\d v_{1z\uparrow(\downarrow)}}{\d z}=f_{1z\uparrow(\downarrow)}, \label{mom1z}
\end{eqnarray}
for the upward (downward) part of this trajectory, where $f_{1x\uparrow(\downarrow)}$ and $f_{1z\uparrow(\downarrow)}$ are the horizontal and vertical components of the total average force $\mathbf{f_{1\uparrow(\downarrow)}}$ per unit volume acting on the grain in the upward (downward) part of this trajectory. For the single-trajectory case $\rho_{1\uparrow}v_{1z\uparrow}=-\rho_{1\downarrow}v_{1z\downarrow}=\phi(0)$ is constant with height $z$ \cite{UngarHaff87}. Summing the upward and downward part of (\ref{mom1x}) and (\ref{mom1z}), respectively, followed by averaging over all trajectories and integration over height therefore yields
\begin{eqnarray}
 \tau_g(z)=\phi(z)\Delta v_x(z)=\int\limits_z^\infty f_x(z')\d z', \label{taugo} \\
 p_g(z)=\phi(z)\Delta v_z(z)=\int\limits_z^\infty f_z(z')\d z', \label{pgo}
\end{eqnarray}
where we approximated the trajectory average of products as the product of trajectory averages in $\phi\Delta v_x$ and $\phi\Delta v_z$. Here $\Delta v_{xo}=\Delta v_x(0)$ and $\Delta v_{zo}=\Delta v_z(0)$, and $f_x(z)$ and $f_z(z)$ are the averages of $f_{1x}=f_{1x\uparrow}+f_{1x\downarrow}$ and $f_{1z}=f_{1z\uparrow}+f_{1z\downarrow}$ over all trajectories, respectively. The terms on the left hand side of (\ref{taugo}) and (\ref{pgo}) describe the average horizontal and vertical force per unit soil area applied by the soil on the grains during an impact and the right hand side the average horizontal and vertical force per unit soil area applied by the wind on the grains during a trajectory, respectively. In the next steps we evaluate the integrals in (\ref{taugo}) and (\ref{pgo}). Therefore we first need an expression for the total trajectory-averaged force $\mathbf{f}$. Since we neglect the Magnus force, turbulent lift forces, and momentum transfer through collisions, $\mathbf{f}$ is only composed of the drag and the gravity force. Further approximating the trajectory average of products as the product of trajectory averages, $\mathbf{f}$ can be written as
\begin{eqnarray}
 \mathbf{f}=\frac{3}{4sd}\left(\rho_{\uparrow}C_d(v_{r\uparrow})v_{r\uparrow}\mathbf{v_{r\uparrow}}+\rho_{\downarrow}C_d(v_{r\downarrow})v_{r\downarrow}\mathbf{v_{r\downarrow}}\right)-\rho\tilde g\mathbf{e_z}, \label{drag}
\end{eqnarray}
where $\mathbf{e_z}$ is the unit vector in $z$-direction, $\tilde g=\frac{s-1}{s}g$ with $s=\rho_s/\rho_w$ is the buoyancy-reduced gravity (for most atmospheres $\tilde g\approxeq g$), and $v_{r\uparrow(\downarrow)}=|\mathbf{v_{r\uparrow(\downarrow)}}|$. $C_d$ is the drag coefficient, which is a function of the particle Reynolds number and therefore of $v_{r\uparrow(\downarrow)}$. For many drag laws in the literature e.g. \cite{Rubey33,Julien95} the dependency of $C_d$ on a velocity difference $V$ can be described by a law of the type
\begin{eqnarray}
 C_d(V)=\frac{C_o\mu}{V\rho_wd}+C_\infty, \label{Cd}
\end{eqnarray}
where $C_o$ and $C_\infty=C_d(\infty)$ are dimensionless parameters. Such a drag law strongly simplifies $f_z$, which becomes
\begin{eqnarray}
 f_z=-\rho\tilde g-\frac{3C_\infty\tau_g}{4sd}, \label{fz}
\end{eqnarray}
where (\ref{phi}), (\ref{taugspez}), and $v_{rz\uparrow(\downarrow)}=-v_{z\uparrow(\downarrow)}$ were used. If a drag law of another type than (\ref{Cd}) was used e.g. \cite{Cheng97}, (\ref{fz}) would still be valid in very good approximation. On the other hand we rewrite $f_x$ as
\begin{eqnarray}
 f_x=\frac{3\rho}{4sd}\langle C_d(v_r)v_rv_{rx}\rangle, \label{fx}
\end{eqnarray}
where $\langle \rangle$ denotes a weighted average of a quantity $f$ between the upward and downward movement, $\langle f\rangle=(\rho_{\uparrow}f_{\uparrow}+\rho_{\downarrow} f_{\downarrow})/\rho$. Now we can evaluate the integral in (\ref{taugo}) using the expression we derived for $f_x$. We obtain
\begin{eqnarray}
 \tau_{go}=\tau_g(0)=\int\limits_0^\infty\frac{3\rho}{4sd}\langle C_d(v_r)v_rv_{rx}\rangle\d z\approxeq\frac{3C_d(\overline V_r)\overline V_r^2M}{4sd}, \label{taug1}
\end{eqnarray}
where the overbar and the capital letters denote the average of a quantity $f$ over height, $\overline F=\int_0^\infty\rho f\d z/\int_0^\infty\rho\d z$. We further used (\ref{M}), the approximation $\overline V_r\approxeq\overline V_{rx}$, which is reasonable since in aeolian saltation the horizontal motion dominates the vertical one, and we approximated the height average of the products by the products of the height averages $\overline{\langle C_d(v_r)v_rv_{rx}\rangle}\approxeq C_d(\overline V_r)\overline V_r^2$. On the other hand $p_{go}$ can now be calculated from our expression for $f_z$. It becomes
\begin{eqnarray}
 p_{go}=p_g(0)=-\tilde gM-\frac{3C_\infty z_s\tau_{go}}{4sd}, \label{pg1}
 \end{eqnarray}
where we used our assumption (\ref{zsexp2}). With the evaluation of the integrals, we can now define $\alpha'$ and our first model parameter $\alpha$ as
\begin{eqnarray}
 \alpha'=-\frac{p_{go}}{\tau_{go}}=\frac{-\Delta v_{zo}}{\Delta v_{xo}}, \label{alphadef} \\
 \alpha=\alpha'-\frac{3C_\infty z_s}{4sd}=\frac{4s\tilde gd}{3C_d(\overline V_r)\overline V_r^2}, \label{alphadef2}
\end{eqnarray}
where we used (\ref{taugspez}) and (\ref{pgspez}) as well as the notations $\Delta v_{xo}=\Delta v_x(0)$ and $\Delta v_{zo}=\Delta v_z(0)$. The advantage of this definition of $\alpha$ lies in the fact that $\alpha'$ is almost independent of $u_*$, atmospheric conditions, $\tilde g$, and $d$ as we show later in a separate chapter. For many conditions, we can approximate
\begin{eqnarray}
 \alpha&\approxeq&\alpha', \label{alphaapprox}
\end{eqnarray}
since $\alpha'$ is typically much larger than $C_\infty z_s/(sd)$, as will be verified later. It mainly means that the gravity force is large in comparison to the vertical drag force, $f_z\approx-\tilde g\rho$. We can thus formulate relevant sand transport quantities as a function of a constant $\alpha$. For instance from (\ref{alphadef2}) we obtain a direct relation between the average velocity difference $\overline V_r$ and $\alpha$, writing
\begin{eqnarray}
 C_d(\overline V_r)\overline V_r^2=\frac{4s\tilde gd}{3\alpha}, \label{vr}
\end{eqnarray}
and further, using (\ref{pg1}), (\ref{alphadef}), and (\ref{alphaapprox}), a direct relation between the grain shear stress $\tau_{go}$ at the bed and the mass of transported grains per unit soil area $M$, writing
\begin{eqnarray}
 \tau_{go}=\alpha^{-1}\tilde gM. \label{taugM}
\end{eqnarray}
\subsection{Work rate balance} \label{energy}
The second important ingredient towards a description of the feedback effect of the grain motion on the wind profile and towards a prediction of $z_o^*$ is the description of the energy transfer from the fluid to the grains. Since we discuss a purely Newtonian problem, we separate the horizontal and vertical motion. The work rate balance with respect to the horizontal (vertical) motion can be obtained by multiplying (\ref{mom1x}) ((\ref{mom1z})) with $v_{1x\uparrow(\downarrow)}$ ($v_{1z\uparrow(\downarrow)}$), summing the upward and downward part, integrating over height, and averaging over all trajectories. It yields
\begin{eqnarray}
 \frac{1}{2}\phi(0)\Delta v_{xo}^2=\int\limits_0^\infty(f_{x\uparrow}v_{x\uparrow}+f_{x\downarrow}v_{x\downarrow})\d z, \label{Egx} \\
 \frac{1}{2}\phi(0)\Delta v_{zo}^2=\int\limits_0^\infty(f_{z\uparrow}v_{z\uparrow}+f_{z\downarrow}v_{z\downarrow})\d z, \label{Egz}
\end{eqnarray}
where $\Delta v_{xo}^2=\Delta v_x^2(0)$, $\Delta v_{zo}^2=\Delta v_z^2(0)$, and we approximated the trajectory average of products as the product of trajectory averages in $\phi(0)\Delta v_{xo}^2$ and $\phi(0)\Delta v_{zo}^2$. The terms on the left hand side of (\ref{Egx}) and (\ref{Egz}) describe the average work rate during an impact and the right hand side the average work rate during a trajectory for the horizontal and vertical motion, respectively. Analogous to (\ref{drag}) and (\ref{fx}) we can now write
\begin{eqnarray}
 f_{x\uparrow}v_{x\uparrow}+f_{x\downarrow}v_{x\downarrow}=\frac{3\rho}{4sd}\langle C_d(v_r)v_rv_{rx}v_x\rangle, \\
  f_{z\uparrow}v_{z\uparrow}+f_{z\downarrow}v_{z\downarrow}=-\frac{3\rho}{4sd}\langle C_d(v_r)v_rv_z^2\rangle, \label{Pz}
\end{eqnarray}
where we used (\ref{phi}) in (\ref{Pz}). Analogous to (\ref{taug1}), integration now approximately yields
\begin{eqnarray}
 \frac{1}{2}\phi(0)\Delta v_{xo}^2&\approxeq&\frac{3M}{4sd}C_d(\overline V_r)\overline V_r^2\overline V, \label{Egx2} \\
 \frac{1}{2}\phi(0)\Delta v_{zo}^2&\approxeq&-\frac{3M}{4sd}C_d(\overline V_r)\overline V_r\overline {V_z^2}, \label{Egz2}
\end{eqnarray}
where $\overline V=\overline V_x$ describes the average particle velocity. Note that (\ref{Egz2}) would write $0=0$, if vertical drag is neglected ($\Delta v_{zo}^2=0$ and $\langle C_d(v_r)v_rv_z^2\rangle=0$). Evaluating the integrals, we now define $\beta'$ as
\begin{eqnarray}
 \beta'=\sqrt{\frac{-\frac{1}{2}\phi(0)\Delta v_{zo}^2}{\frac{1}{2}\phi(0)\Delta v_{xo}^2}}=\sqrt{\frac{-\Delta v_{zo}^2}{\Delta v_{xo}^2}}=\sqrt{\frac{\overline {V_z^2}}{\overline V_r\overline V}}. \label{betadef}
\end{eqnarray}
(\ref{betadef}) means that the average granular temperature $\overline{V_z^2}$ is proportional to $\overline V_r\overline V$. This is different from Creyssels et al. \cite{Creysselsetal09} who found $\overline{V_z^2}$ to be approximately equal to $\langle{v_z^2}\rangle(0)$. The main reason for this difference is that the authors neglected vertical drag, whereas our description considers it (see (\ref{Egz2})). As before for $\alpha'$, the advantage of (\ref{betadef}) lies in the fact that $\beta'$ is almost independent of $u_*$, atmospheric conditions, $\tilde g$, and $d$ as we will show in the following.
\subsection{Invariance of $\alpha'$ and $\beta'$} \label{invariance}
Since our model relations, including the final relation for $z_o^*$, will be expressed as functions of $\alpha'$ and $\beta'$, it is important to discuss, how these parameters change with varying conditions. As can be seen from (\ref{alphadef}) and (\ref{betadef}), both parameters are ratios of certain velocity differences evaluated at the soil $z=0$ and therefore related to the splash-entrainment process. The splash-entrainment process dominates the entrainment of bed grains in aeolian steady state saltation, because the entrainment by wind is small due to a strong reduction of the wind velocity close to the sand bed, which even leads to decreasing near-surface velocities with increasing $u_*$ \cite{DuranHerrmann06}. Since fluid-entrainment is not relevant, each impacting grain must exactly lead to one grain leaving the surface (rebound or ejection of new grains) on average for steady state sand transport. The average number grains leaving the surface per impacting grain can however only depend on the average impact velocity $v_i$, angle $\theta_i$, and the relevant bed properties $\tilde g$ and $d$. Furthermore, for given values of $\tilde g$ and $d$, the average velocity $v_l$ and angle $\theta_l$ of a grain leaving the surface after an impact, called lift-off velocity and angle, can only depend on $v_i$ and $\theta_i$. This means, for given values of $\tilde g$, $d$, and $\theta_i$, there are unique values $v_i$, $v_l$, and $\theta_l$, which fulfill that one impacting grain makes one grain leave the surface on average.

Another necessary condition, which must hold, is that $v_{zo\downarrow}=v_i\sin\theta_i$ must be smaller than $v_{zo\uparrow}=v_l\sin\theta_l$ due to friction with the air, it would be equal in the absence of drag. The validity of this condition was observed in saltation experiments \cite{Riceetal95}. Further, from collision experiments Oger et al. \cite{Ogeretal08} found that this condition is only fulfilled for small impact angles $\theta_i\lessapprox15^{\circ}$. The authors also found that the number of ejected particles significantly decreases with decreasing $\theta_i$, meaning that the range of $\theta_i$-values, in which saltation can be sustained, should be rather narrow. In fact, it has been measured in experiments \cite{Namikas03} that the average impact angle is approximately constant, $\theta_i\approx11^{\circ}$, with increasing $u_*$, and thus the other splash quantities as well stay approximately constant with varying $u_*$. Here and henceforth we refer to situations above the impact threshold $u_*\geq u_t$ and within our model limits (not too large $u_*$), when mentioning dependencies on $u_*$. This means that in particular the parameters $\alpha'$ and $\beta'$ are approximately independent of $u_*$ and atmospheric properties, and thus only functions of $\tilde g$ and $d$.

\begin{figure}
 \begin{center}
 \begin{tabular}{ll}
  (a)&(b)\\
  \includegraphics[scale=0.29]{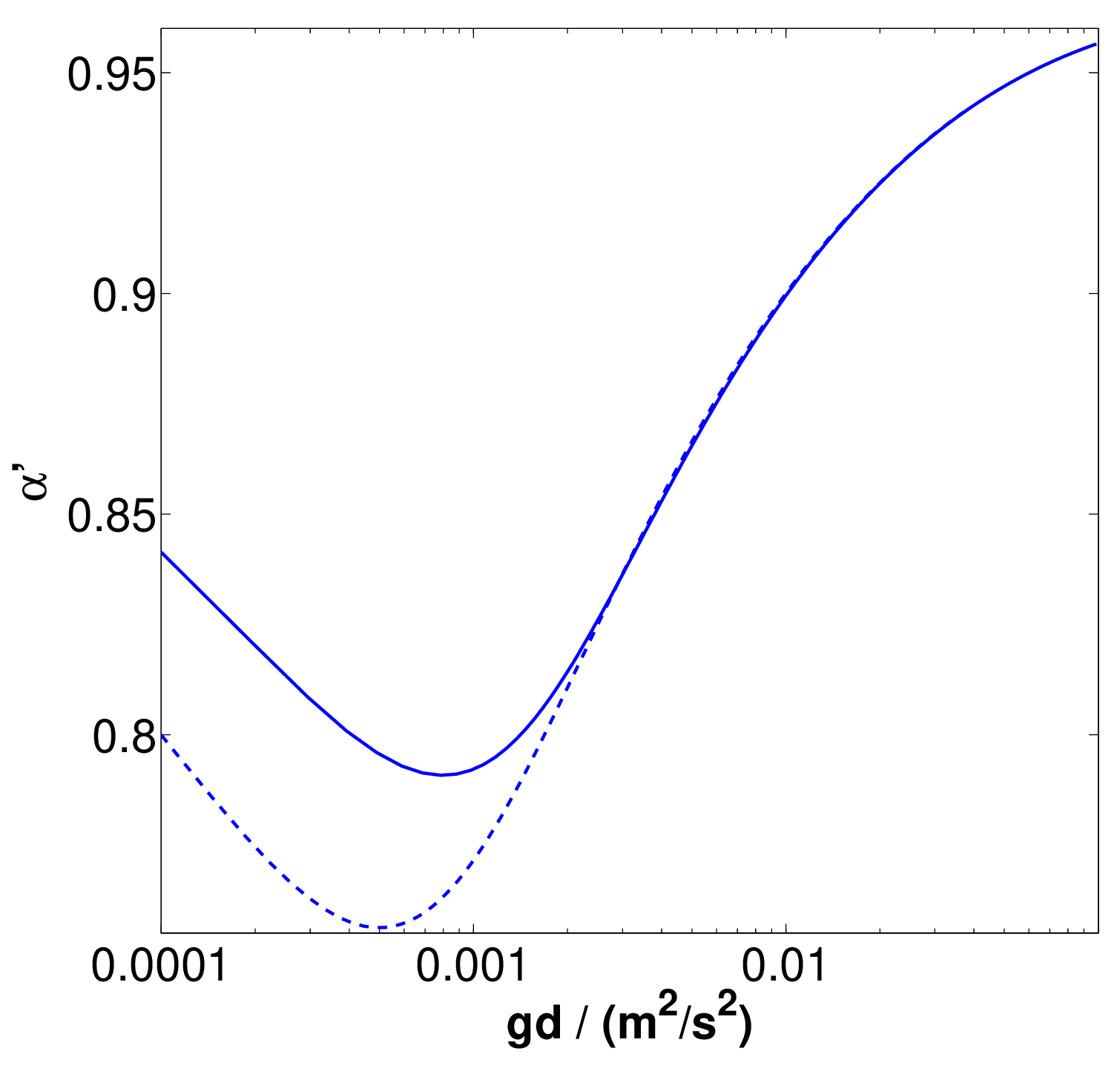} &\includegraphics[scale=0.29]{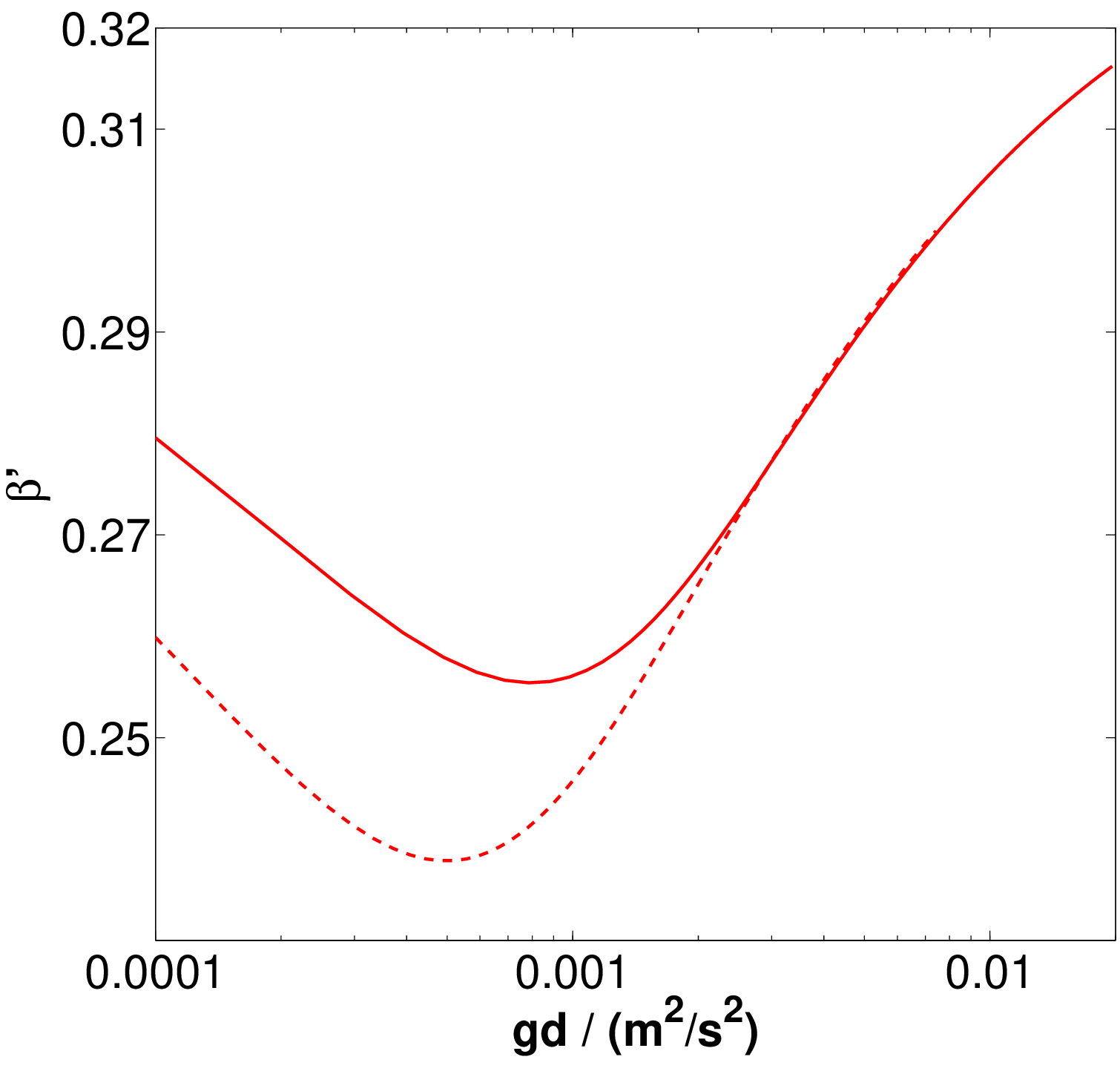}
  \end{tabular}
 \end{center}
 \caption{$\alpha'$ (blue) and $\beta'$ (red) computed with the model of Kok \cite{Kok10b} (see \ref{appkok}) plotted versus $\tilde gd$. Here the gravity was fixed to either the Earth (solid lines) or the Mars value (dashed lines) and $d$ varied.}
 \label{alphabetafig}
\end{figure}

In order to get an idea of how $\alpha'$ and $\beta'$ behave as functions of $\tilde g$ and $d$, we use the model of Kok \cite{Kok10b}, with which $v_{zo\downarrow}$ and $v_{zo\uparrow}$ and thus $\alpha'$ and $\beta'$ can be computed. The model is briefly described in \ref{appkok}. The model results in functions $\alpha'(\tilde gd)$ and $\beta'(\tilde gd)$ plotted in Figure \ref{alphabetafig}.
For the plots the gravities of Earth, $\tilde g=9.81m/s^2$, and Mars, $\tilde g=3.71m/s^2$, were fixed and $d$ varied. As can be seen, even as functions of $\tilde g$ and $d$ the variance of $\alpha'$ and $\beta'$ is small according to this model. Furthermore $\alpha'$ is of the order of unity, what justifies the approximation (\ref{alphaapprox}) for many conditions. For instance for Earth conditions the neglected term can be estimated as being of the order of $1/25$, since $z_s$ is the same order of magnitude as the saltation height $h$, which has been measured as being equal to about $40d$ \cite{Hoetal11}, $C_\infty\approx1$, and $s\approx1000$ on Earth.
\subsection{A novel relation for $z_s$} \label{zsrelation}
\begin{figure}
 \begin{center}
  \includegraphics[scale=0.29]{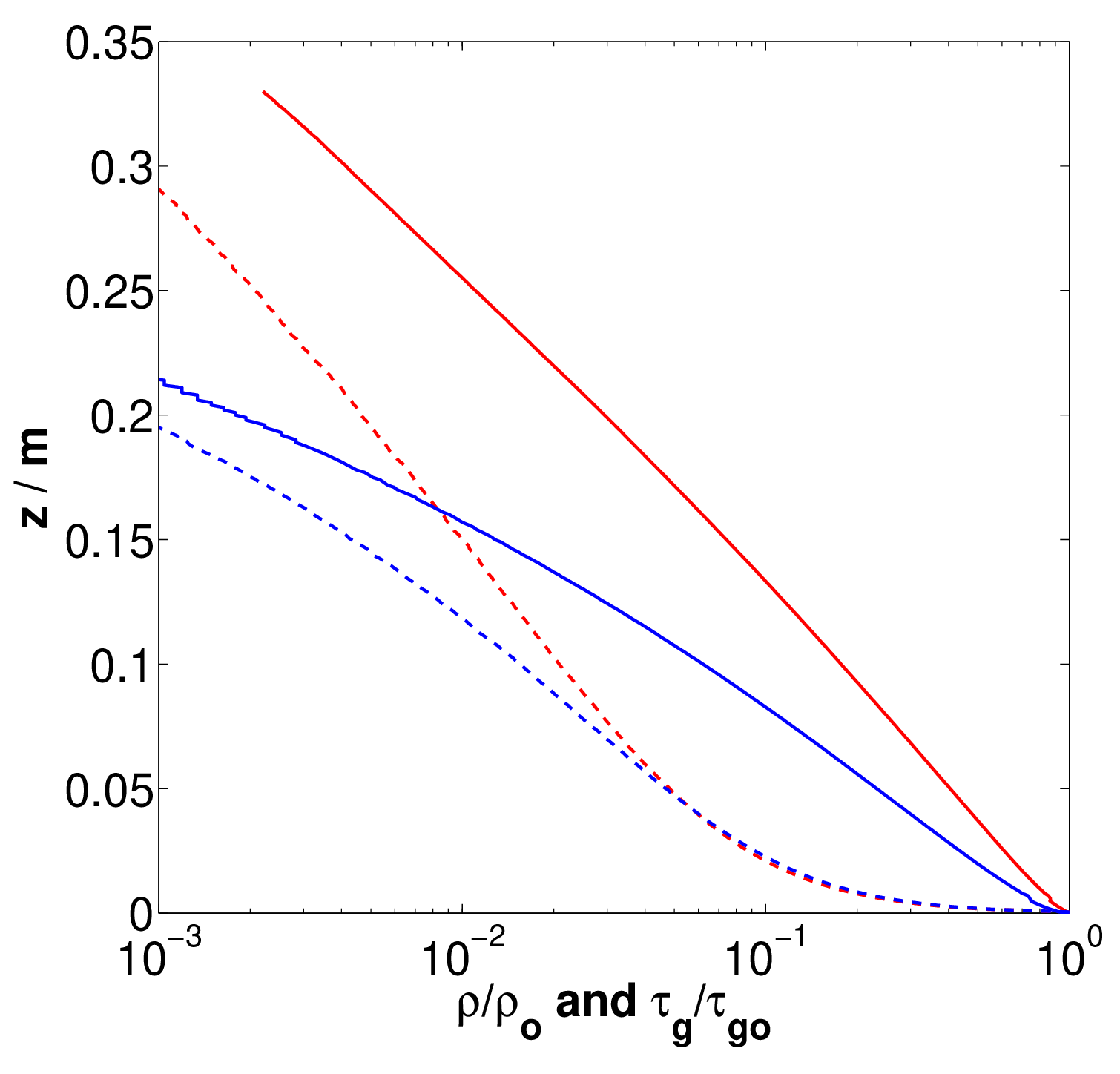}
 \end{center}
 \caption{$\tau_g/\tau_{go}$ (solid lines) and $\rho/\rho_o$ (dashed lines), where $\rho_o=\rho(0)$, plotted versus height for Earth conditions with $d=250\mu m$ and two different shear velocities, $u_*=0.3m/s$ (blue) and $u_*=0.8m/s$ (red).}
 \label{rhovstaug}
\end{figure}

The definitions of the parameters $\alpha$ and $\beta'$ obtained from the momentum and energy balances can now be used to express the decay height $z_s$ of the grain shear stress profile $\tau_g(z)$ as a function of $\alpha$ and $\beta'$. As already pointed out $z_s$ is the key quantity towards a prediction of the apparent roughness $z_o^*$. For the calculation of $z_s$ we use
\begin{eqnarray}
 \langle v_z^2\rangle=\frac{\rho_\uparrow v_{z\uparrow}+\rho_\downarrow v_{z\downarrow}}{\rho_\uparrow+\rho_\downarrow}=-v_{z\uparrow}v_{z\downarrow}, \label{vz2}
\end{eqnarray}
where we inserted (\ref{phi}). We further approximate the arithmetic average of $|v_{z\uparrow}|$ and $|v_{z\downarrow}|$ by their geometric average
\begin{eqnarray}
 \frac{-\Delta v_z}{2}=\frac{|v_{z\uparrow}|+|v_{z\downarrow}|}{2}\approxeq\sqrt{|v_{z\uparrow}||v_{z\downarrow}|}=\sqrt{-v_{z\uparrow}v_{z\downarrow}}, \label{aritgeom}
\end{eqnarray}
what is reasonable since $|v_{z\uparrow}|/|v_{z\downarrow}|<|v_{zo\uparrow}|/|v_{zo\downarrow}|$ and $|v_{zo\uparrow}|/|v_{zo\downarrow}|$ is about $1.5$ as measurements indicate \cite{Ogeretal08}, which means that the error of this approximation is less than $3\%$. Using (\ref{phi}), (\ref{taugspez}), (\ref{vz2}) and (\ref{aritgeom}), we express $\tau_g$ as
\begin{eqnarray}
 \tau_g=\frac{1}{2}\rho\sqrt{\langle v_z^2\rangle}\Delta v_x, \label{taugnew}
\end{eqnarray}
This allows us to rewrite (\ref{taugo}) as
\begin{eqnarray}
 -\frac{\d\tau_g}{\d z}=\frac{\tau_g}{z_s}=\frac{\rho\sqrt{\langle v_z^2\rangle}\Delta v_x}{2z_s}=f_x=\frac{3\rho}{4sd}\langle C_d(v_r)v_rv_{rx}\rangle, \label{difftau}
\end{eqnarray}
where we used (\ref{zsexp2}). Since (\ref{difftau}) is valid for all heights $z$, it must be particularly valid for the average over height. We therefore approximately calculate $z_s$ as
\begin{eqnarray}
 z_s=\frac{2sd\overline{\sqrt{\langle v_z^2\rangle}\Delta v_x}}{3\overline{\langle C_d(v_r)v_rv_{rx}\rangle}}\approxeq\frac{2sd\overline{\Delta V_x}\sqrt{\overline{V_z^2}}}{3C_d(\overline V_r)\overline V_r^2}=\alpha\beta\frac{\overline V_r^{\frac{1}{2}}\overline V^{\frac{3}{2}}}{\tilde g}, \label{zs2}
\end{eqnarray}
where $\beta=\beta'\overline{\Delta V_x}/(2\overline V)$ and we used (\ref{alphadef2}) and (\ref{betadef}). $\beta$ is our second model parameter and like $\beta'$ approximately constant, since we expect that $\overline{\Delta v_x}$ is in leading order proportional to $\overline V$. (\ref{zs2}) is the main contribution of our paper, since it is, to our knowledge, the first physically based prediction of $z_s$ and therefore the most important part towards a novel physically based prediction of $z_o^*$. If the values of the model parameters $\alpha$ and $\beta$ are known, $\overline V$ remains the only undetermined quantity in (\ref{zs2}), since $\overline V_r$ can be calculated by (\ref{vr}) as a function of $\alpha$. There is evidence that $z_s$ does not only describe the decay of $\tau_g(z)$, but also the decay of $\rho(z)$ for large $z$. This can be seen from (\ref{difftau}), which says that $\tau_g(z)$ decays in the same way as $\rho\langle C_d(v_r)v_rv_{rx}\rangle$. Since $\langle C_d(v_r)v_rv_{rx}\rangle$ is only slightly decaying with height for large $z$, the decaying behaviors of $\rho(z)$ and $\tau_g(z)$ are very similar to each other. This is shown in Figure \ref{rhovstaug} for simulations with the numerical model of Kok and Renno \cite{KokRenno09}.
Since the profile $\rho(z)/M$ describes the hop height distribution of saltating grains, $z_s$ is related to the saltation height $h$, which is the average hop height of the grains. This is discussed in detail in the following section.
\subsection{Physical meaning of $z_s$} \label{zsmeaning}
It was assumed in previous studies that $z_s$ and the saltation height $h$ are proportional to each other \cite{Raupach91,Andreotti04,DuranHerrmann06}. This very natural assumption is however not valid for the saltation simulated with the numerical model of Kok and Renno \cite{KokRenno09}, since the normalized profiles $\rho(z)/\rho_o$ for $u_*=0.3m/s$ and $u_*=0.8m/s$ in Figure \ref{rhovstaug} almost coincide with each other at small heights. This means, although $z_s$ is larger for  $u_*=0.8m/s$, the hop height of grains transported close to the surface is almost the same and hence $h$ increases weaker with $u_*$ than $z_s$. It is therefore reasonable to interpret $z_s$ as the height of high-energy saltons and the height $z_r$ up to which the profiles $\rho(z)/\rho_o$ coincide as the height of low-energy saltons \cite{Andreotti04}, which remains unchanged with $u_*$. In the following we explain the reason for this behavior of $\rho(z)$.

In our model $\rho(z)$ decays approximately exponentially, if and only if $\langle v_z^2\rangle$ does not vary much with height $z$. This can be seen from (\ref{pgo}), (\ref{fz}), and (\ref{alphaapprox}), which allow us to write the differential equation, using $p_g=-\rho\langle v_z^2\rangle$ (analogous to (\ref{taugnew})),
\begin{eqnarray}
 \frac{\d(\rho\langle v_z^2\rangle)}{\d z}=-\frac{\d p_g}{\d z}=f_z\approxeq-\rho\tilde g, \label{diffrho}
\end{eqnarray}
whose solution is an exponential decrease, if and only if $\langle v_z^2\rangle$ is constant with $z$. However, there can be a huge difference between the value $\langle v_z^2\rangle(0)$ at the soil, which is fixed by the splash-entrainment process, and the value of $\langle v_z^2\rangle$ at larger heights, which is proportional to $z_s\tilde g$. Figure \ref{rhovstaug} shows indeed a strong deviation of $\rho(z)$ from the exponential shape at very small heights within the low energy layer $z\ll z_r$. In the light of our analysis, this deviation corresponds to a strong increase of $\langle v_z^2\rangle(z)$ from $\langle v_z^2\rangle(0)$ towards about $z_s\tilde g$ at larger heights. Note that wind tunnel studies \cite{Creysselsetal09,Hoetal11} did not notice such a deviation from the exponential shape in their measurements of $\rho(z)$. A possible cause is that their lowest measurement points, $z\approx20d$ and $z\approx40d$, respectively, were already too high, and they measured only in a region, where $\langle v_z^2\rangle(z)$ was not varying much anymore. Note further that a very similar reason, namely that $-\langle v_xv_z\rangle(0)$ is fixed by the splash-entrainment process, is the probable cause for the very slight deviation of $\tau_g(z)$ from the exponential shape at very small heights (see Figure \ref{taugfig}). Finally note that mass flux profiles $\rho(z)\langle v_x\rangle(z)$, which are often measured in experiments \cite{Namikas03,RasmussenSorensen08,Greeleyetal96,Dongetal04,DongQian07}, decay slightly weaker than $\rho(z)$, since the particle velocity $\langle v_x\rangle(z)$ increases weakly with height.

After explaining $\rho(z)$, we now calculate $h$ and $z_r$. First, using (\ref{betadef}) and (\ref{diffrho}) as well as partial integration, $h$ can be calculated as
\begin{eqnarray}
 h=\overline z=\frac{1}{M}\int\limits_0^\infty z\rho(z)\d z=\frac{1}{M\tilde g}\int\limits_0^\infty\rho\langle v_z^2\rangle\d z=\frac{\overline{V_z^2}}{\tilde g}=\beta'^2\frac{\overline V_r\overline V}{\tilde g}.
\end{eqnarray}
Since $z_r$ does not depend on $u_*$ and since it must be of the same structure as $z_s$ and $h$, namely proportional to $\langle v_z^2\rangle(z)/\tilde g$ at a typical height $z$, and since $z=0$ is the only height where $\langle v_z^2\rangle(z)/\tilde g$ does not depend on $u_*$ due to the splash-entrainment process, $z_r$ must write
\begin{eqnarray}
 z_r&\propto&\frac{\langle v_z^2\rangle(0)}{\tilde g}=-\frac{v_{zo\uparrow}v_{zo\downarrow}}{\tilde g},
\end{eqnarray}
where we used (\ref{phi}).

Our analysis confirms the picture of Andreotti \cite{Andreotti04}, who hypothesized that one can essentially distinguish two species in aeolian saltation, low-energy saltons (reptons) slowly moving in small hops, and high-energy saltons moving fast in huge hops. The author hypothesized that $z_r$ is a measure for the height of the focal-region, a region in which steady state wind profiles for different shear velocities $u_*$ intersect.

In this section we showed that the saltation layer can be characterized by three heights. The height of low-energy saltatons $z_r$, the saltation height $h$, and the height of high-energy saltons $z_s$, which is also a measure for the thickness of the saltation layer. In contrast to the first height, the latter two change with $\overline V$ and thus with $u_*$. The prediction of $\overline V$ is therefore subject of the following section.
\subsection{Calculation of $z_o^*$} \label{zorelation}
The last step towards the calculation of $z_o^*$ is to derive an expression for $\overline V$, the last undetermined quantity in (\ref{zs2}), our relation for $z_s$. $z_s$ is the main parameter in our final relation for $z_o^*$, which will be of a similar structure as (\ref{Raupacheq}), the relation of Raupach \cite{Raupach91}. For deriving $\overline V$ we use the following strategy. We first approximately calculate $\overline U$ as the wind velocity at an height $z_m$, which denotes the height of the mean motion, $\overline U=u(z_m)$, with the mixing length approximation \cite{Prandtl25}. Then we compute $\overline V$ by
\begin{eqnarray}
 \overline V=\overline U-\overline V_r, \label{vm1}
\end{eqnarray}
where we use that $\overline V_r$ does not change with $u_*$ (see (\ref{vr})).

Following the outlined strategy, we calculate $\overline U$ from the mixing length approximation \cite{Raupach91,Prandtl25,DuranHerrmann06,Sauermannetal01,AnderssonHaff91} as
\begin{eqnarray}
 \frac{\d u(z)}{dz}=\frac{u_*}{\kappa z}\sqrt{1-\frac{\tau_g(z)}{\rho_wu_*^2}}, \label{mixinglength}
\end{eqnarray}
with
\begin{eqnarray}
 u(z_o)=0.
\end{eqnarray}
In the absence of sand transport, $\tau_g(z)=0$, the mixing length approximation yields the undisturbed logarithmic velocity profile ((\ref{velprofile}) with $z_o^*=z_o$). In the presence of sand transport, $\tau_g(z)\neq0$, the velocity profile deviates from the logarithmic shape.  In \ref{appwind} $u(z)$ as well as $z_o^*$ are calculated based on our model assumption of an exponentially decreasing grain shear stress profile, (\ref{zsexp2}), and the calculation approximately yields
\begin{eqnarray}
 \ln\frac{z_o^*}{z_o}=\left(1-\frac{u_b}{u_*}\right)\ln\frac{z_s}{1.78z_o}-G\left(\frac{u_b}{u_*}\right) \label{lnzo*}
\end{eqnarray}
and for $z>0.1z_s$
\begin{eqnarray}
 u(z)=\frac{u_*}{\kappa}\ln\frac{z}{z_o^*}+\frac{u_*^2-u_b^2}{2\kappa u_*}\mathrm{E_1}\left(\frac{z}{z_s}\right), \label{Um}
\end{eqnarray}
where $u_b$ is the reduced wind shear velocity at the bed, defined by
\begin{eqnarray}
 u_b=u_a(0), \label{ub} \\
 u_a(z)=u_*\sqrt{1-\frac{\tau_g(z)}{\rho_wu_*^2}}, \label{ua}
\end{eqnarray}
and the exponential integral $\mathrm{E_1}(x)$ as well as $G(x)$ are defined by
\begin{eqnarray}
 \mathrm{E_1}(x)=\int\limits_x^\infty\frac{e^{-x'}}{x'}\d x'=-0.577-\ln x+\sum_{l=1}^\infty\frac{(-1)^{l+1}x^l}{ll!}, \label{expint} \\
 G(x)=1.154(1+x\ln x)(1-x)^{2.56}. \label{G}
\end{eqnarray}
Note that (\ref{Um}) is only an approximation of $u(z)$ for $z>0.1z_s$. In \ref{appwind} one can also find an approximation for $z<z_s$, which however has a much more complicated structure. For the coming calculations, we are only interested in the value $\overline U=u(z_m)$, for whose calculation both approximations perform similarly well, since $z_m$ is between $0.1z_s$ and $0.4z_s$ as we show later.

\begin{figure}
 \begin{center}
 \begin{tabular}{ll}
  (a)&(b)\\
  \includegraphics[scale=0.29]{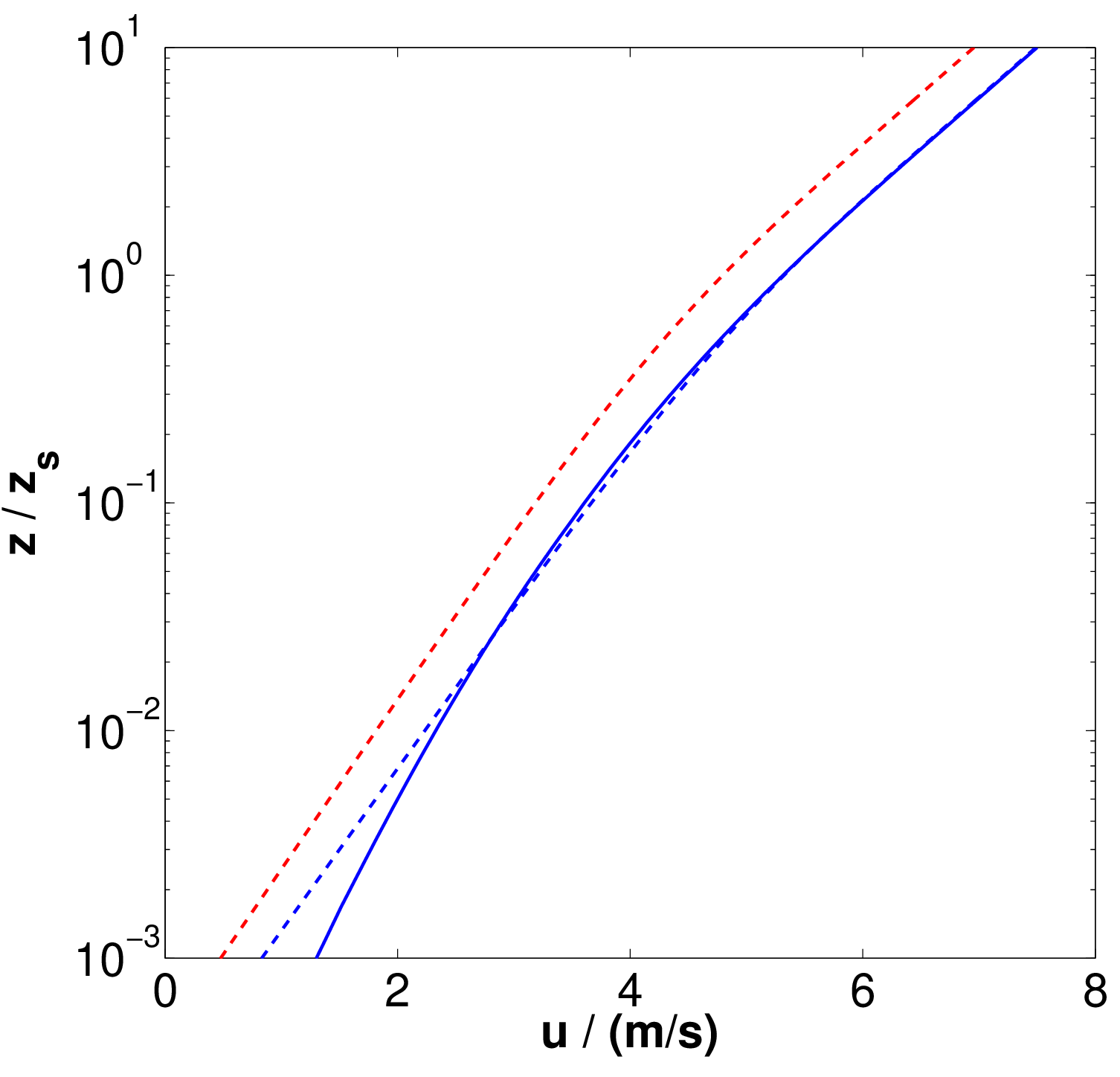} &\includegraphics[scale=0.29]{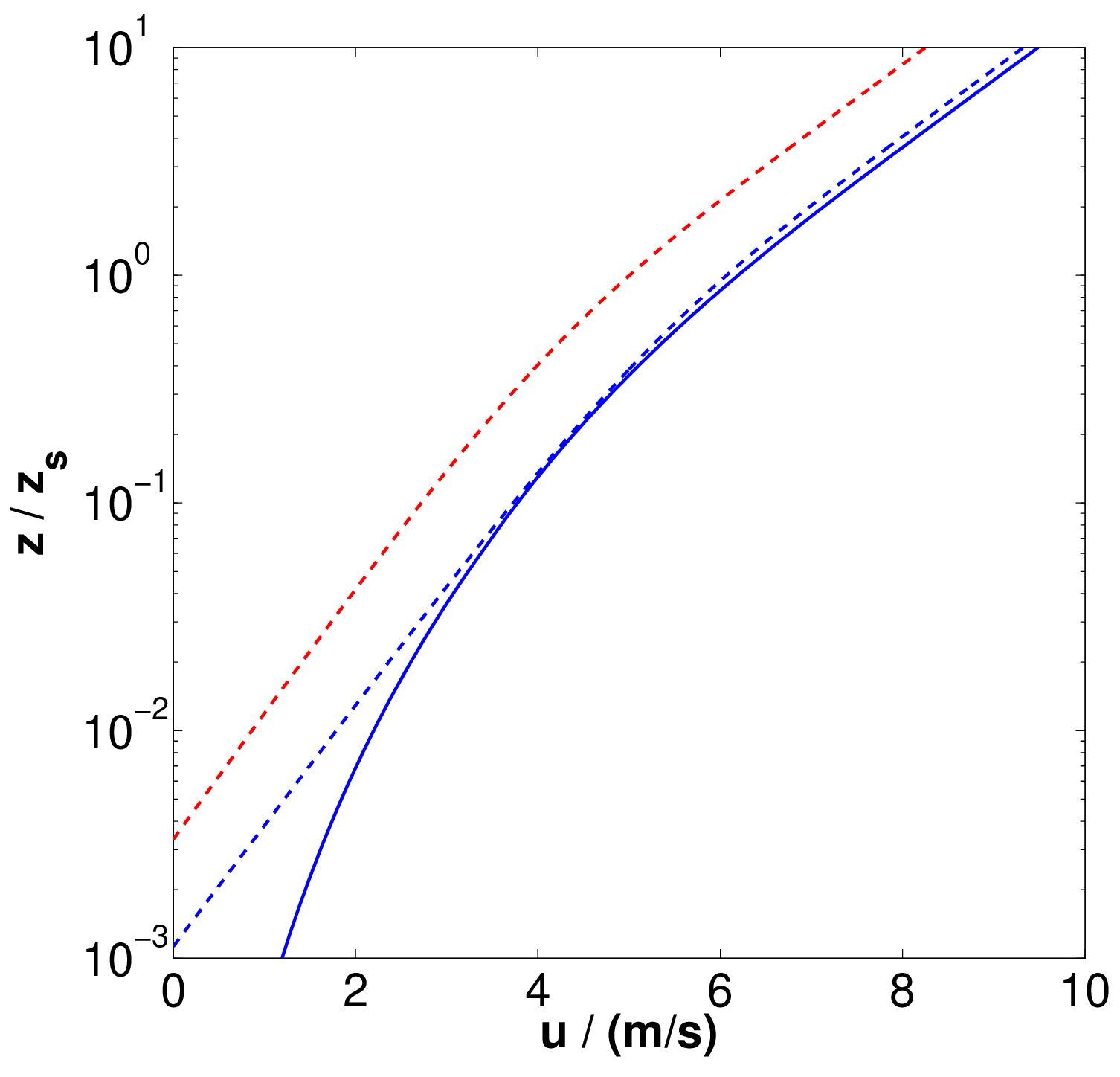}
  \end{tabular}
 \end{center}
 \caption{$u$ plotted versus $z/z_s$ for Mars conditions with $d=250\mu m$ and two different shear velocities, $u_*=0.39m/s$ (a) and $u_*=0.59m/s$ (b) . The solid lines shows the simulated wind profiles \cite{KokRenno09}, the dashed lines show the wind profiles computed by (\ref{lnzo*}) and (\ref{Um}) with $u_b=u_t$ (blue) and the simulated values of $u_b$ (red), respectively.}
 \label{windutub}
\end{figure}

In (\ref{lnzo*}) and (\ref{Um}) the shear velocity at the bed $u_b$ remains undetermined. It was shown by Duran and Herrmann \cite{DuranHerrmann06} that $u_b$ must decrease with $u_*$ starting from $u_b(u_t)=u_t$, however with a small slope. The reason is that one observes a focal region in aeolian steady state saltation, called Bagnold-focus, below which the wind velocities decrease with increasing $u_*$ \cite{Bagnold41,Bagnold38}. Andreotti \cite{Andreotti04} argued that this strong decrease of wind velocities, the presence of the Bagnold-focus, and consequently the value of $u_b$ are mainly caused by grains transported in the low-energy layer. As we explained in Section \ref{zsmeaning}, at very small heights within the low-energy layer $z\ll z_r$ the mass density profile $\rho(z)$ and also the grain shear stress profile $\tau_g(z)$ deviate from the exponential decrease. We however used this exponential decrease of $\tau_g(z)$ (see (\ref{zsexp2})), our main model assumption, for the computation of the apparent roughness and the wind profile in (\ref{lnzo*}) and (\ref{Um}). This does not mean that (\ref{lnzo*}) and (\ref{Um}) are wrong, because the deviation of $\tau_g(z)$ from the exponential shape is very small (almost invisible in Figure \ref{taugfig}). But it means that we cannot use the real value of $u_b$, which is influenced by the low-energy layer. Instead we must use a value of $u_b$, which corresponds to the extrapolation of the exponential shape of $\tau_g(z)$ above $z=z_r$ to the height $z=0$. Such a value of $u_b$ would be larger than the real value, because $\rho(z)$ and thus $\tau_g(z)$ deviate from the exponential shape towards higher values within the low-energy layer. We therefore propose that this value is close to $u_t$,
\begin{eqnarray}
 u_b&\approxeq&u_t. \label{ubut}
\end{eqnarray}
This is supported by simulations using the numerical model of Kok and Renno \cite{KokRenno09}. Figure \ref{windutub} shows that wind profiles calculated by (\ref{Um}) with the hypothesis (\ref{ubut}) are much closer to the simulated profiles than those in which the simulated values of $u_b$ were used, and this although this hypothesis eliminates the Bagnold-focus. Especially in the region which we are interested in, $z>0.1z_s$, (\ref{lnzo*}) and (\ref{Um}) provide an excellent approximation of the simulated wind profile, if using (\ref{ubut}).
Note that (\ref{ubut}) is also known as Owen's second hypothesis, and it has been used in many previous models e.g. \cite{Owen64,Raupach91,Bagnold73,Sorensen91,Sauermannetal01}. From (\ref{lnzo*}), (\ref{Um}), and (\ref{ubut}) we now obtain
\begin{eqnarray}
 \ln\frac{z_o^*}{z_o}=\left(1-\frac{u_t}{u_*}\right)\ln\frac{z_s}{1.78z_o}-G\left(\frac{u_t}{u_*}\right) \label{lnzo*2}
\end{eqnarray}
and
\begin{eqnarray}
 \overline U=\frac{u_*}{\kappa}\ln\frac{z_m}{z_o^*}+\frac{u_*^2-u_t^2}{2\kappa u_*}\mathrm{E_1}(\gamma), \label{Um2}
\end{eqnarray}
where $\gamma$ is the third model parameter and defined by
\begin{eqnarray}
 \gamma=\frac{z_m}{z_s}. \label{A}
\end{eqnarray}
The first terms on the right hand side of (\ref{lnzo*}) and (\ref{lnzo*2}) are identical to the relations of previous studies \cite{Raupach91,DuranHerrmann06} (see also (\ref{Raupacheq})). The second term appears, because we did not approximate the right hand side of (\ref{mixinglength}) before the integration as done in these studies. Our solution is therefore more precise. Note that the height of the mean motion $z_m$ should be in leading order proportional to the decay height $z_s$ of the grain shear stress profile, because the mean motion is dominated by the motion of high-energy saltons \cite{Andreotti04}, and $z_s$ is a measure for the height of high-energy saltons (see Section \ref{zsmeaning}). Consequently $\gamma$ is in leading order constant.

Having obtained a relation for $\overline U$, we can calculate $\overline V$ with (\ref{vm1}), insert $\overline V$ in (\ref{zs2}) to obtain $z_s$, and insert $z_s$ in (\ref{lnzo*2}) to obtain $z_o^*$. The calculation of $z_o^*$ can therefore be summarized as
\numparts
\begin{eqnarray}
 \ln\frac{z_o^*}{z_o}=\left(1-\frac{u_t}{u_*}\right)\ln\frac{z_m}{1.78\gamma z_o}-G\left(\frac{u_t}{u_*}\right), \label{lnzo*3} \\
 z_m=\frac{\alpha\beta\gamma \overline V_r^\frac{1}{2}\left(\overline U-\overline V_r\right)^\frac{3}{2}}{\tilde g}, \label{zm} \\ 
 \overline U=\frac{u_*}{\kappa}\ln\frac{z_m}{z_o^*}+\frac{u_*^2-u_t^2}{2\kappa u_*}\mathrm{E_1}(\gamma), \label{Um3} \\
 C_d(\overline V_r)\overline V_r^2=\frac{4s\tilde gd}{3\alpha}, \label{vr2}
\end{eqnarray}
\endnumparts
where we used (\ref{lnzo*2}) and (\ref{A}) for (\ref{lnzo*3}), (\ref{zs2}), (\ref{vm1}), and (\ref{A}) for (\ref{zm}), and (\ref{vr2}) and (\ref{Um3}) are the same as (\ref{vr}) and (\ref{Um2}), respectively. (\ref{zm}) and (\ref{Um3}) can be solved iteratively for $\overline U$ and $z_m$, and (\ref{vr2}) can be solved iteratively given a certain drag law $C_d(V)$. This is our novel prediction of $z_o^*$ in the most general version. If the impact threshold $u_t$ is known, $z_o^*$ can be calculated using (\ref{lnzo*3}-\ref{vr2}) as function of the model parameters $\alpha$, $\beta$, and $\gamma$. It is furthermore possible to compute the mass flux $Q$ as function of the same parameters. For this purpose we first compute the mass of transported sand per unit soil area $M$ from (\ref{taugM}) and (\ref{ubut}) as
\begin{eqnarray}
 M=\frac{\alpha\rho_w}{\tilde g}(u_*^2-u_t^2). \label{M2}
\end{eqnarray}
Then the mass flux $Q$ becomes
\begin{eqnarray}
 Q=\int\limits_0^\infty\rho(z)v_x(z)\d z=M\overline V, \\
 Q=\frac{\alpha\rho_w}{\tilde g}(u_*^2-u_t^2)(\overline U-\overline V_r), \label{Q}
\end{eqnarray}
where we inserted (\ref{M2}) for $M$ and (\ref{vm1}) for $\overline V$. $\overline U$ and $\overline V_r$ can be computed by (\ref{zm}-\ref{vr2}). At the moment all model equations are functions of the model parameters and $u_t$. In order to close the model, $u_t$ must be calculated as function of the model parameters as well. This is done in the following with the help of a closing assumption.
\subsection{Closing the model - a relation for $u_t$} \label{utrelation}
(\ref{lnzo*3}-\ref{vr2}) are already a novel expression for $z_o^*$. Like previous expressions in the literature (see (\ref{Oweneq})-(\ref{Raupacheq}) and (\ref{zsAndreotti})) it needs the impact threshold $u_t$ as an input parameter. We therefore close the model by deriving a relation for $u_t$ in this section. The strategy is as follows. We first motivate a simple expression for the particle velocity at the threshold, $\overline V_t=\overline V(u_t)$. We then compute $\overline V_r=\overline U_t-\overline V_t$, where $\overline U_t=\overline U(u_t)$, and combine the result with our previous expression for $\overline V_r$, (\ref{vr2}). The resulting equation can be rearranged to compute $u_t$.

\begin{figure}
 \begin{center}
  \includegraphics[scale=0.29]{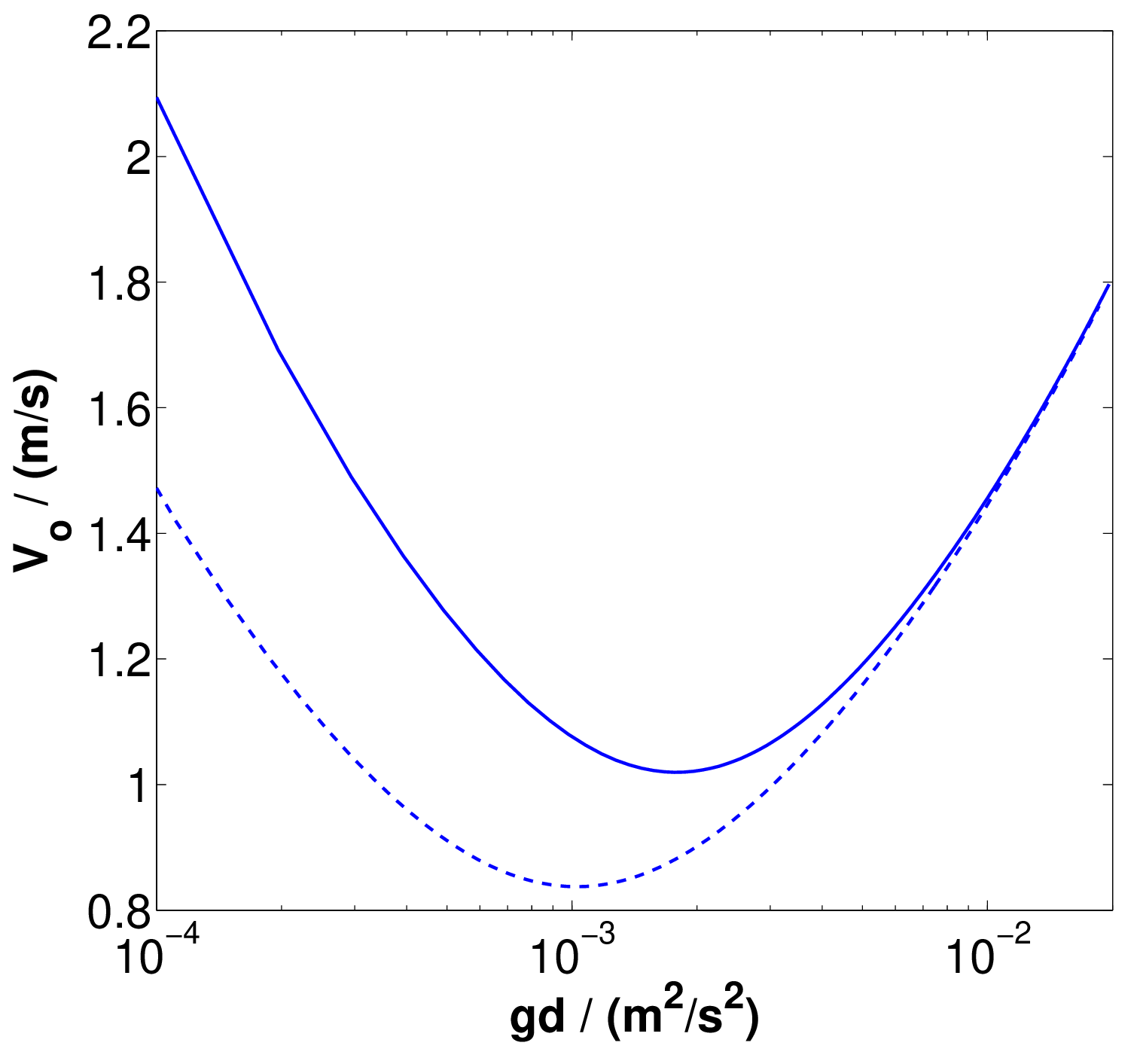}
 \end{center}
 \caption{$V_o$ computed with the model of Kok \cite{Kok10b} (see \ref{appkok}) plotted versus $\tilde gd$. Here $\tilde g=9.81m/s^2$ (solid line) and $\tilde g=3.71m/s^2$ (dashed line) are fixed, respectively, whereas $d$ is varied.}
 \label{vofig}
\end{figure}

As outlined before, we motivate the following closing expression,
\begin{eqnarray}
 \overline V_t=\eta\overline U_t+V_o=\eta\frac{u_t}{\kappa}\ln\frac{z_{mt}}{z_o}+V_o, \label{vm}
\end{eqnarray}
where $z_{mt}=z_m(u_t)$, $V_o=(\rho_{o\uparrow}v_{xo\uparrow}+\rho_{o\downarrow}v_{xo\downarrow})/\rho_o$ is the average particle slip velocity (i.e., the particle speed at the surface), where $v_{xo\uparrow(\downarrow)}=v_{x\uparrow(\downarrow)}(0)$ and $\rho_{o\uparrow(\downarrow)}=\rho_{\uparrow(\downarrow)}(0)$, and $\eta$ is the fourth model parameter. (\ref{vm}) means that the difference between average particle and slip velocity under threshold conditions is proportional to the average wind velocity. This is justified in the following manner. The particle slip velocity $V_o$ is a quantity that like the model parameters $\alpha$ and $\beta$ only depends on the impact-entrainment process. In particular $V_o$ is independent of the average wind velocity $\overline U_t$. From theoretical and experimental studies it is known that the profile of the average horizontal particle velocity $v_x(z)$ starts with $V_o$ at $z=0$ and increases with height $z$ \cite{Creysselsetal09,Hoetal11,KokRenno09}. The average increase with $z$ mainly depends on the average wind velocity $\overline U_t$. Consequently, the average particle velocity $\overline V_t$ is a function of the average wind velocity $\overline U_t$ plus an offset $V_o$. The simplest possible relation with such a behavior is given by (\ref{vm}). Thereby $\eta$ describes how efficiently the wind accelerates transported grains under threshold conditions. Rearranging (\ref{vm}), $\eta$ can be written as
\begin{eqnarray}
 \eta=\frac{\overline V_t-V_o}{\overline U_t}. \label{gammadef}
\end{eqnarray}
We calculate the particle slip velocity $V_o$ with the model of Kok \cite{Kok10b}, explained in \ref{appkok}. The result, $V_o$ as function of $\tilde gd$, is plotted in Figure \ref{vofig}.

Now we can use (\ref{vm}) to calculate $\overline V_r$. According to (\ref{vr}), $\overline V_r$ does not depend on $u_*$. We can therefore write
\begin{eqnarray}
 \overline V_r=\overline V_r(u_t)=\overline U_t-\overline V_t=(1-\eta)\frac{u_t}{\kappa}\ln\frac{z_{mt}}{z_o}-V_o, \label{vr3}
\end{eqnarray}
Rearranging and using (\ref{zs2}), (\ref{A}), and (\ref{vm}) finally yields an expression for $u_t$, which can be summarized as
\numparts
\begin{eqnarray}
 u_t=\frac{\kappa(\overline V_r+V_o)}{(1-\eta)\ln\frac{z_{mt}}{z_o}}, \label{ut} \\
 z_{mt}=\frac{\alpha\beta\gamma\overline V_r^\frac{1}{2}(V_o+\eta\overline V_r)^\frac{3}{2}}{(1-\eta)^\frac{3}{2}\tilde g}, \label{zm2} \\
 C_d(\overline V_r)\overline V_r^2=\frac{4s\tilde gd}{3\alpha}. \label{vr4}
\end{eqnarray}
\endnumparts
\section{Model Validation} \label{secmodelval}
The model is validated by simulation results with the numerical state of the art model of Kok and Renno \cite{KokRenno09} and by several experiments. The simulation results validate (\ref{zs2}), (\ref{vm1}), (\ref{lnzo*3}), (\ref{vr2}), (\ref{M2}), and (\ref{vr3}), and confirm our statements that the model parameters $\alpha$ and $\beta$ are approximately independent of the shear velocity and atmospheric conditions, and further indicate that $\gamma$ and $\eta$ are not varying much as well. In detail we show that all model parameters always adopt approximately the same values for whatever conditions are simulated. The experiments confirm our expressions (\ref{lnzo*3}-\ref{vr2}), (\ref{Q}), and (\ref{ut}-\ref{vr4}) by showing that the same set of model parameters can explain the following experiments: the mass flux data of Creyssels et al. \cite{Creysselsetal09}, the apparent roughness data of Rasmussen et al. \cite{Rasmussenetal96}, and the combined impact threshold data of different studies \cite{Bagnold37,Chepil45,Rasmussenetal94}.
\subsection{Independence of the model constants of atmosphere and grain properties}
For the verification of the independence of the model parameters we use simulation results of the numerical model of Kok and Renno \cite{KokRenno09} for conditions, which are summarized in Table (\ref{conditions}).
\begin{table*}
	\centering
		\begin{tabular}{|c|c|c|c|c|c|c|c|c|c|c|c|}
             $\frac{\rho_s}{kg/m^3}$&$\frac{\rho_w}{kg/m^3}$&$\frac{10^5\mu}{kg/(ms)}$&$\frac{\tilde g}{m/s^2}$&$\frac{d}{\mu m}$&$\frac{d}{z_o}$&$\frac{u_t}{m/s}$&$\frac{V_o}{m/s}$&$\alpha$&$\beta$&$\gamma$&$\eta$ \\
             \hline
             $2650$&$1.174$&$1.87$&9.81&100&30&0.126&0.88&0.72&0.12&0.27&0.26 \\
             \hline
             $2650$&$1.174$&$1.87$&9.81&200&30&0.176&1.14&0.88&0.13&0.32&0.19 \\
             \hline
             $2650$&$1.174$&$1.87$&9.81&250&30&0.196&1.23&0.94&0.125&0.33&0.21 \\
             \hline
             $2650$&$1.174$&$1.87$&9.81&300&30&0.22&1.3&0.97&0.125&0.33&0.22 \\
             \hline
             $2650$&$1.174$&$1.87$&9.81&500&30&0.288&1.57&1.1&0.125&0.32&0.18 \\
             \hline
             $3000$&$0.0145$&$1.49$&3.71&200&30&0.158&0.95&0.91&0.135&0.26&0.28 \\
             \hline
             $3000$&$0.0145$&$1.49$&3.71&250&30&0.194&0.95&0.96&0.135&0.27&0.2 \\
             \hline
             $3000$&$0.0145$&$1.49$&3.71&300&30&0.233&1.01&1&0.135&0.26&0.25 \\
             \hline
             $3000$&$0.0145$&$1.49$&3.71&500&30&0.42&1.56&1.17&0.135&0.23&0.17 \\
             \hline
             $2650$&$1.174$&$1.87$&3.71&250&30&0.128&0.88&0.84&0.135&0.33&0.27 \\
             \hline
             $2650$&$1.174$&$1.87$&3.71&500&30&0.182&1.15&1&0.135&0.34&0.19 \\
             \hline
             $3000$&$2.3353$&$1.49$&3.71&250&30&0.123&0.87&0.84&0.13&0.35&0.28 \\
             \hline
             $2650$&$0.0579$&$1.87$&9.81&250&30&0.274&1.31&1&0.13&0.3&0.23 \\
             \hline
             $2650$&$0.1159$&$1.87$&9.81&250&30&0.258&1.3&1&0.13&0.31&0.23 \\
             \hline
             $2650$&$0.5793$&$1.87$&9.81&250&30&0.216&1.25&0.94&0.13&0.35&0.21 \\
             \hline
             $5000$&$1.174$&$1.87$&9.81&250&30&0.249&1.08&0.99&0.12&0.34&0.29 \\
             \hline
             $2650$&$1.174$&$1.87$&9.81&250&10&0.236&1.25&0.95&0.125&0.33&0.23 \\
             \hline
             $2650$&$1.174$&$1.87$&9.81&250&90&0.169&1.23&0.91&0.125&0.36&0.18 \\
             \hline
		\end{tabular}
	\caption{Conditions simulated with the numerical program of \cite{KokRenno09}. The first five conditions describe an Earth atmosphere with five different particle diameters $d$. The next four conditions describe a Mars atmosphere with four different values of $d$. The remaining nine simulated conditions are imaginary conditions, where one or more of the atmospheric parameters were varied.}
	\label{conditions}
\end{table*}
Fluid densities $\rho_w$ and viscosities $\mu$ as well as particle densities $\rho_s$ are varied between Earth and Mars values and particle diameters $d$ are varied between $100\mu m$ and $500\mu m$. The surface roughnesses $z_o$ in the absence of saltation is chosen to be $z_o=d/30$, except in two cases ($z_o=d/10$ and $z_o=d/90$) which allow us to check, whether the predictive performance of our model equations is sensitive to the value of $z_o$. The model of Kok and Renno \cite{KokRenno09} uses the drag law of Cheng \cite{Cheng97}, namely
\begin{eqnarray}
 C_d(V)=\left(\left(\frac{32\mu}{V\rho_wd}\right)^{2/3}+1\right)^{3/2}, \label{dragcheng}
\end{eqnarray}
which we use to compute $\overline V_r$ in (\ref{vr2}) and (\ref{vr4}).

\begin{figure}
 \begin{center}
  \includegraphics[scale=0.29]{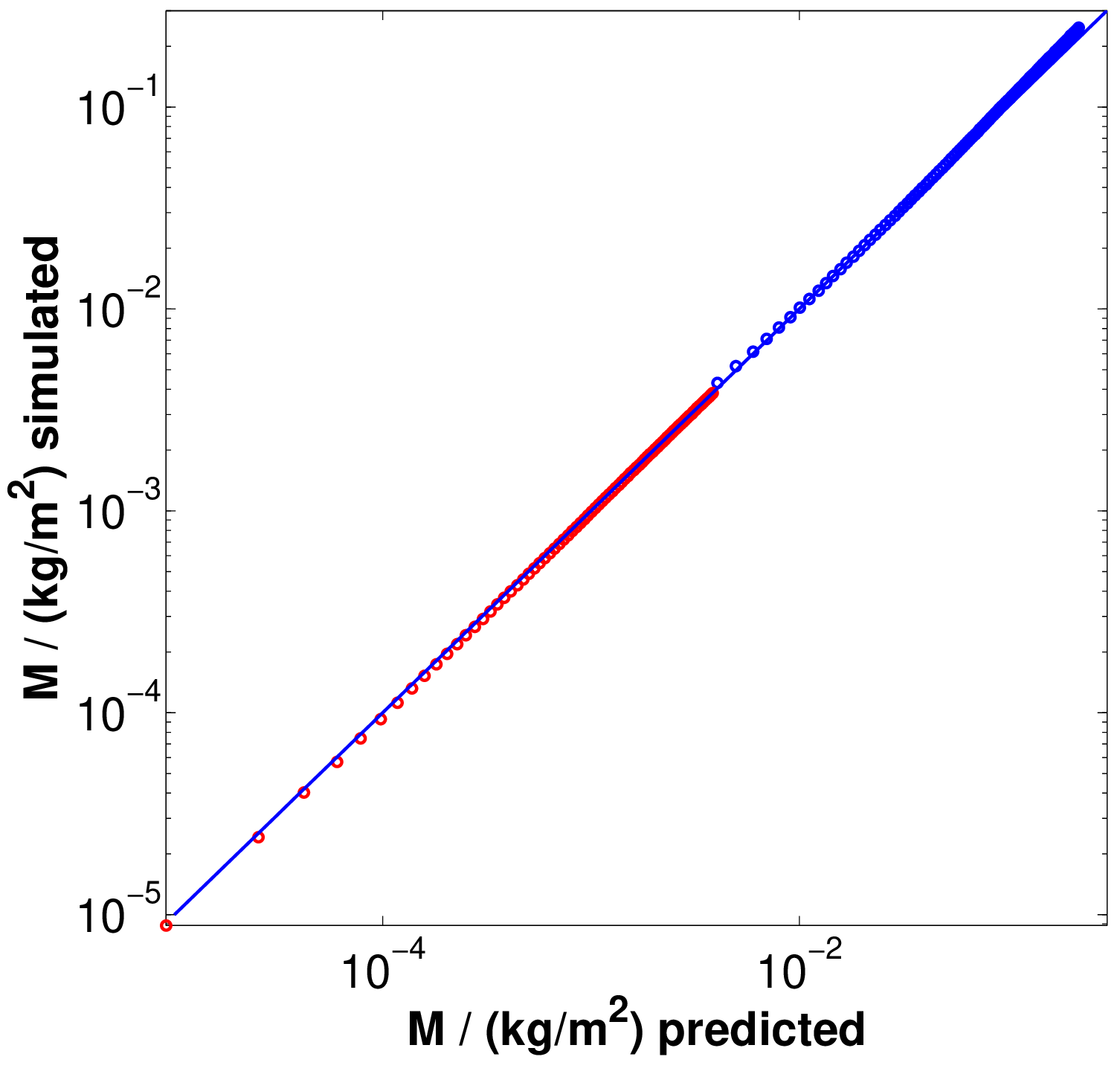}
 \end{center}
 \caption{$M$ calculated using (\ref{M2}) with $\alpha$ as given in Table \ref{conditions} versus the values of $M$, obtained directly from the simulation for Earth conditions with $d=500\mu m$ (blue) and Mars conditions with $d=250\mu m$ (red). Each circle belongs to a different shear velocity $u_*$. The solid line indicates perfect agreement. The parameter $u_t$, which appears in (\ref{M2}), was not calculated by our model, but directly obtained from the simulations (see Table \ref{conditions}).}
 \label{alpha}
\end{figure}

\begin{figure}
 \begin{center}
 \begin{tabular}{ll}
  (a)&(b)\\
  \includegraphics[scale=0.29]{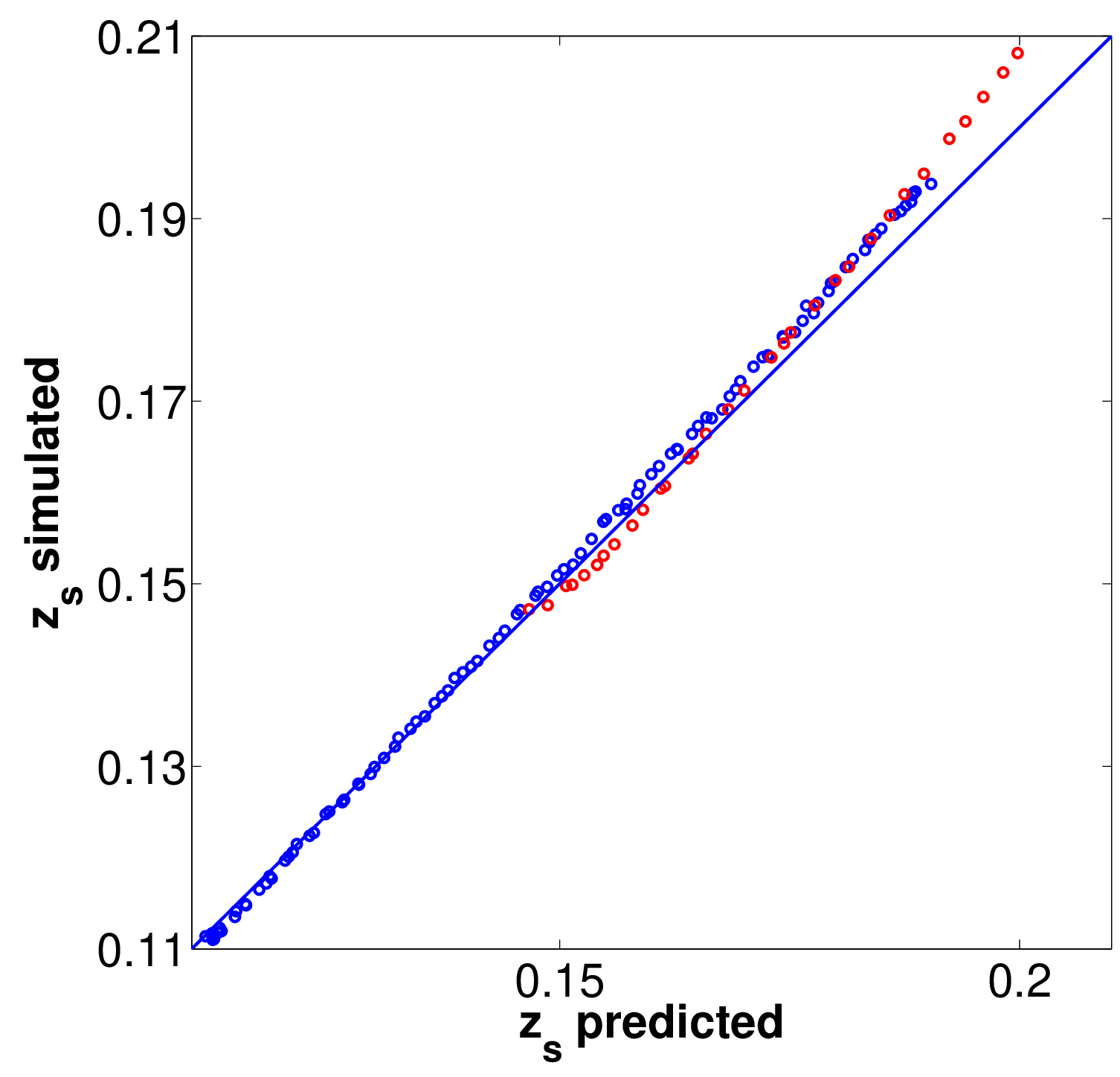} &\includegraphics[scale=0.29]{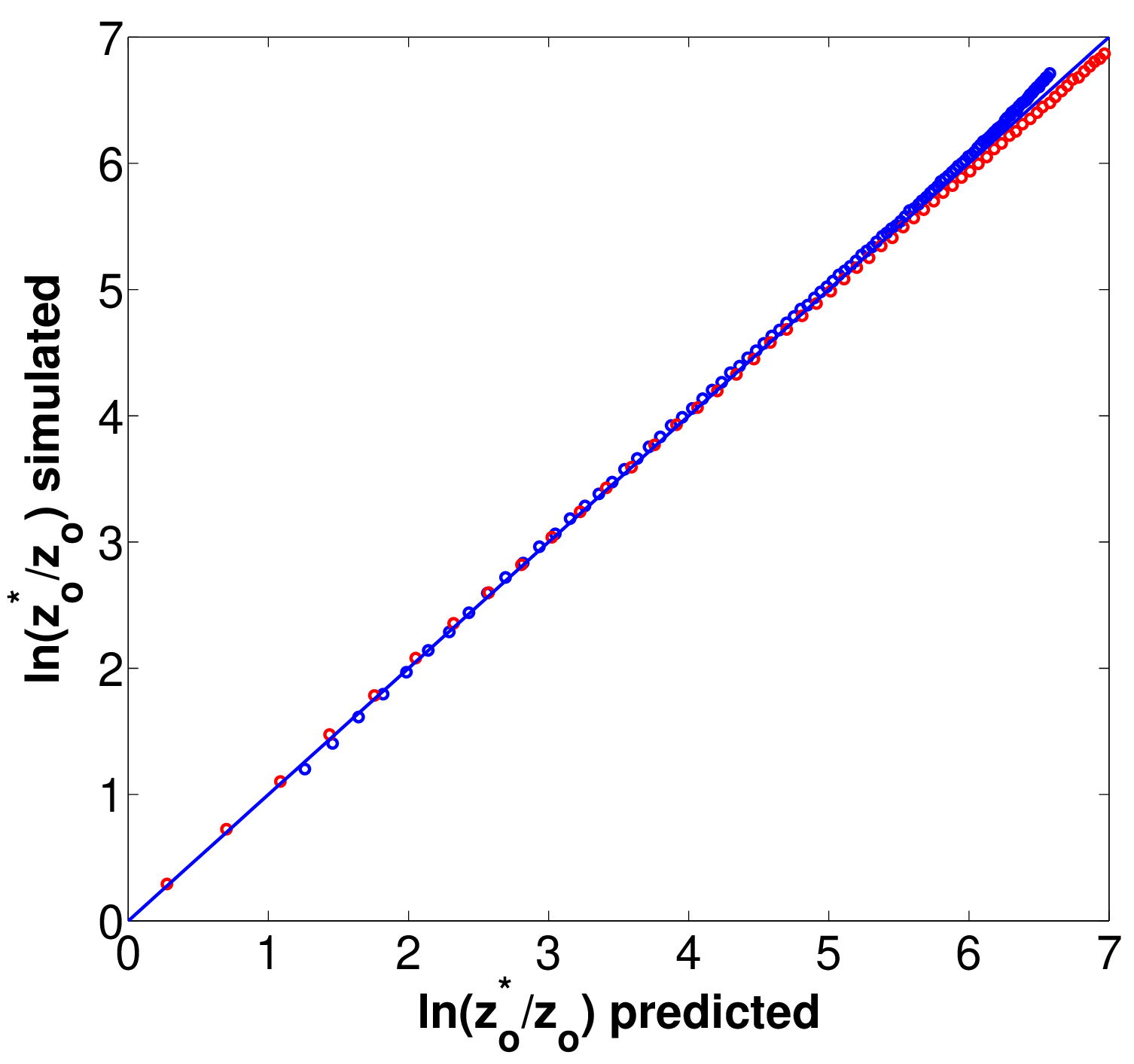}
  \end{tabular}
 \end{center}
 \caption{$z_s$ (a) and $\ln(z_o^*/z_o)$ (b) calculated using (\ref{zs2}), (\ref{lnzo*3}), and (\ref{vr2}) with $\beta$ as given in Table \ref{conditions} versus the values of $z_s$ and $\ln(z_o^*/z_o)$, obtained directly from the simulations for Earth conditions with $d=500\mu m$ (blue) and Mars conditions with $d=250\mu m$ (red). Each circle belongs to a different shear velocity $u_*$. The solid line indicates perfect agreement. The parameters $u_t$ and $\alpha$ are taken from Table (\ref{conditions}) and the simulated values of $\overline V$ are used in (\ref{zs2}), instead of calculating them with the parameter $\gamma$.}
 \label{zozo}
\end{figure}

\begin{figure}
 \begin{center}
  \includegraphics[scale=0.29]{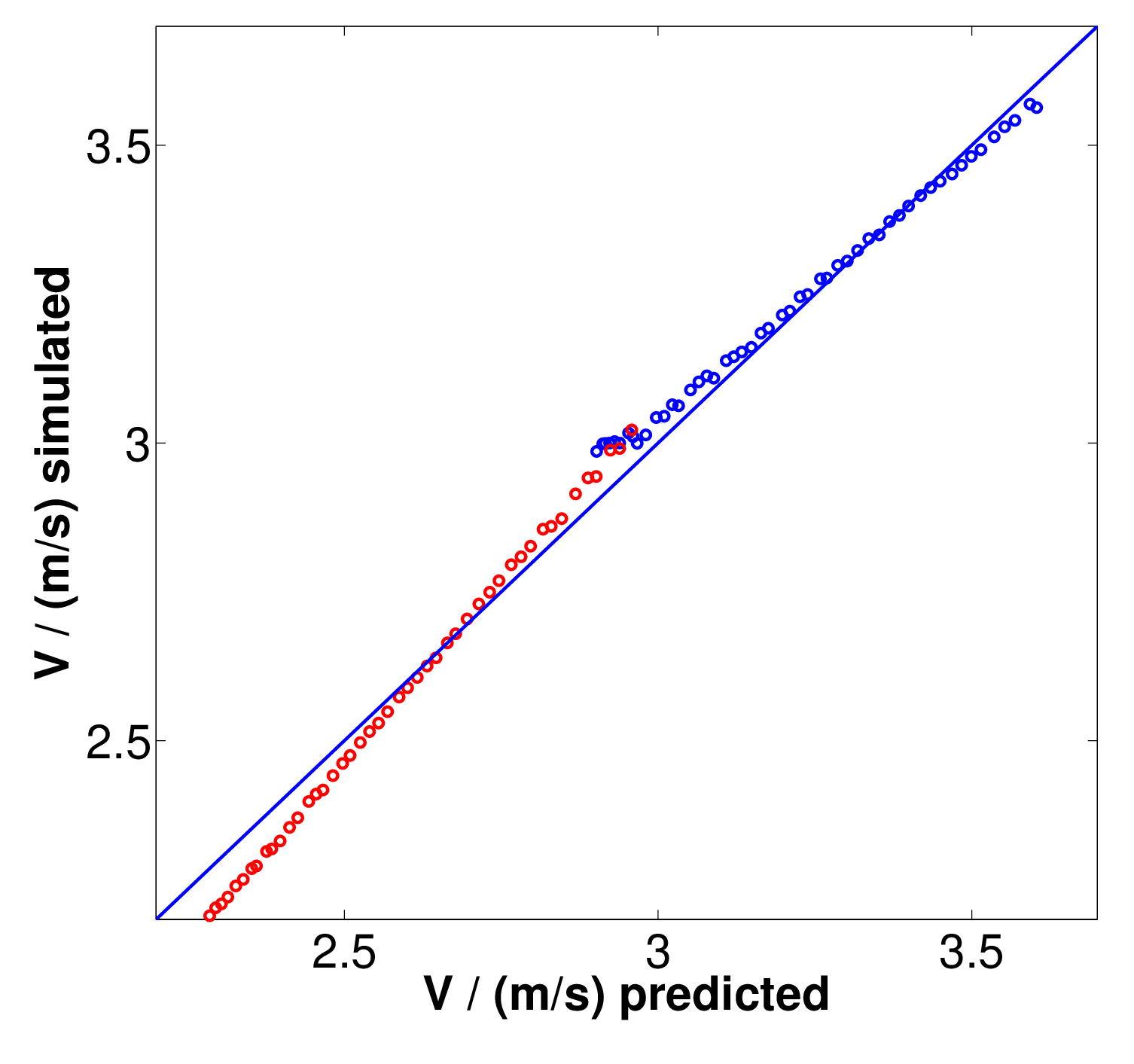}
 \end{center}
 \caption{$\overline V$ calculated using (\ref{vm1}), (\ref{Um3}), and (\ref{vr2}) with $\gamma$ as given in Table \ref{conditions} versus the values of $\overline V$, obtained directly from the simulations for Earth conditions with $d=500\mu m$ (blue) and Mars conditions with $d=250\mu m$ (red). Each circle belongs to a different shear velocity $u_*$. The solid line indicates perfect agreement. The parameters $u_t$ and $\alpha$ are taken from Table (\ref{conditions}). The parameter $\beta$ is not used, instead the simulated values of $z_s$ are taken for the computation of $\overline V$.}
 \label{vmfig}
\end{figure}

We evaluate (\ref{zs2}), (\ref{vm1}), (\ref{lnzo*3}), (\ref{vr2}), and (\ref{M2}) in order to show the approximate independence of the model parameters $\alpha$, $\beta$, and $\gamma$, of atmospheric conditions and grain properties. Exemplary for Earth conditions with $d=500\mu m$ and Mars conditions with $d=250\mu m$, Figure \ref{alpha} shows $M$ calculated using (\ref{M2}) with $\alpha$ as given in Table \ref{conditions} versus the values of $M$, obtained directly from the simulations.
Furthermore, for the same conditions, Figure \ref{zozo} shows $z_s$ and $\ln(z_o^*/z_o)$ calculated using (\ref{zs2}), (\ref{lnzo*3}), and (\ref{vr2}) with $\beta$ as given in Table \ref{conditions} versus the values of $z_s$ and $\ln(z_o^*/z_o)$, obtained directly from the simulations.
And finally, for the same conditions, Figure \ref{vmfig} shows $\overline V$ calculated using (\ref{vm1}), (\ref{Um3}), and (\ref{vr2}) versus the values of $\overline V$, obtained directly from the simulations.
For all plots, each circle belongs to a different shear velocity $u_*$ and the solid line indicates perfect agreement. Conditions different from those plotted in Figs. (\ref{alpha}-\ref{vmfig}) show in most cases the same good agreement. The values of $\alpha$, $\beta$, and $\gamma$ for all conditions are given in Table (\ref{conditions}). It shows that $\alpha$ is between $0.9$ and $1$ for most of the tested conditions. A variance of $\alpha$ and thus $M$ of about $10\%$ is however small compared to the degree of uncertainty one usually faces in saltation mass(flux) measurements. Furthermore, $\beta$ is between $0.12$ and $0.135$ for all tested conditions, the variance of $\beta$ is therefore even less than the variance of $\alpha$. Furthermore, Table (\ref{conditions}) shows that $\gamma$ is between $0.23$ and $0.34$ for all of the tested conditions. Therefore we can confirm that $\alpha$, $\beta$, and to a lesser extend $\gamma$ can indeed be used as approximately constant parameters for saltation simulated by the model of Kok and Renno \cite{KokRenno09} at least within the range of conditions displayed in Table (\ref{conditions}). We want to emphasize that in particular (\ref{zs2}), which is the main contribution of our paper, well describes the behavior of the simulated decay heights $z_s$ (see Figure \ref{zozo}). Note that the slight disagreement of the Mars simulations from the perfect agreement in Figure \ref{vmfig} is probably due to turbulent fluctuations of the wind velocities, which are considered by the numerical model of Kok and Renno \cite{KokRenno09}, but not by our analytical model. These fluctuations are much more important for small $\tilde gd$ like Mars conditions with $d=250\mu m$ than for large $\tilde gd$ like Earth conditions with $d=500\mu m$.

Finally we check (\ref{vr3}) by plotting $\overline V_r+V_o$ over $\overline U_t=\kappa^{-1}u_t\ln(z_{mt}/z_o)$ for all simulated conditions. Thereby $\overline V_r$ is calculated by (\ref{vr2}) with the values of $\alpha$ in Table (\ref{conditions}) and the values $V_o$ are also given in Table (\ref{conditions}). Note that in the simulations $V_o$ does not change significantly with $u_*$ as we also stated before from a theoretical point of view.
\begin{figure}
 \begin{center}
  \includegraphics[scale=0.29]{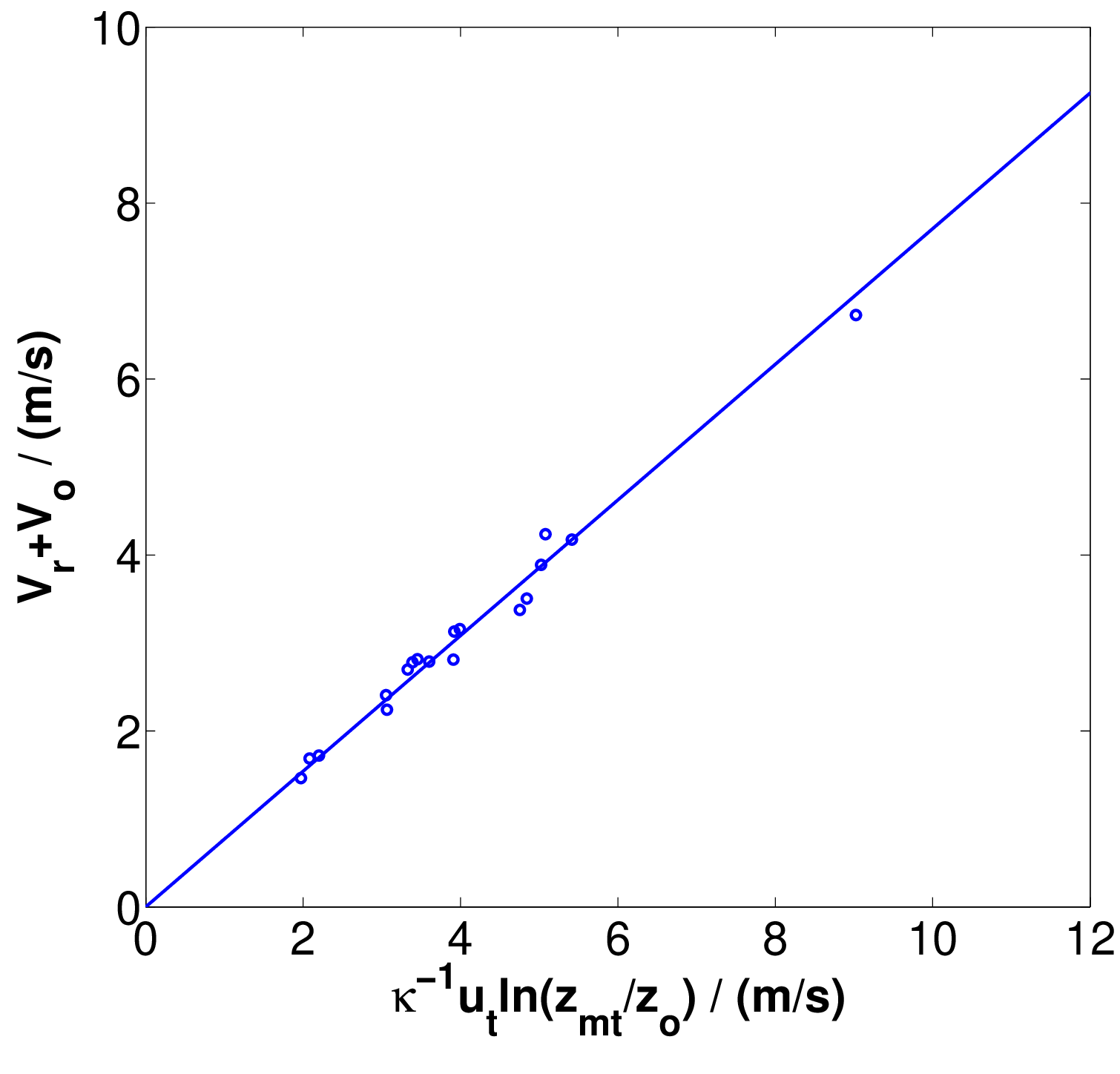}
 \end{center}
 \caption{Plot of $\overline U_t=\kappa^{-1}u_t\ln(z_{mt}/z_o)$ versus $\overline V_r+V_o$. According to (\ref{vr3}) $\overline V_r+V_o$ and $\overline U_t$ are proportional to each other with a proportionality constant $1-\eta$. Each circle corresponds to one of the conditions in Table (\ref{conditions}).}
 \label{gamma}
\end{figure}
Figure \ref{gamma} shows that the plotted circles, each of them corresponding to one of the conditions in Table (\ref{conditions}), approximately lie on a straight line through the origin. This indicates an approximately universal behavior of $\eta$, because according to (\ref{vr3}) $\overline V_r+V_o$ and $\kappa^{-1}u_t\ln(z_{mt}/z_o)$ are proportional to each other with a proportionality constant $1-\eta$. The values of $\eta$ are also given in Tab. (\ref{conditions}).

In conclusion, the simulation results of the numerical state of the art model of \cite{KokRenno09} can be very well described by our analytical model indicating only slight variances of all four model parameters with changing conditions. Note that the simulated values of $\alpha$, $\beta$, and $V_o$ do not follow the predictions of the model of Kok \cite{Kok10b}, which were plotted in Figs. (\ref{alphabetafig}) and (\ref{vofig}).

We also want to emphasize that we entirely failed to fit the numerical data for the apparent roughness $z_o^*$, when using (\ref{zsAndreotti}), the scaling for $z_s$ proposed by Andreotti \cite{Andreotti04}. Earth and Mars conditions could not be fitted simultaneously by (\ref{Raupacheq}) and (\ref{zsAndreotti}), or alternatively (\ref{lnzo*3}) and (\ref{zsAndreotti}) with a single proportionality constant in (\ref{zsAndreotti}). Fitting to good agreement with the numerical Earth data, led to a disagreement with the Mars data by almost two orders of magnitude, even after trying several modifications like for instance replacing $\sqrt{sgd}$ in (\ref{zsAndreotti}) by $u_t$. This strongly indicates that (\ref{zsAndreotti}) does not describe the physics sufficiently well. Andreotti \cite{Andreotti04} mentioned himself that (\ref{zsAndreotti}) is nothing but a guess: "We have not found any simple explanation of this scaling with $\nu=\mu/\rho_w$. The only indication that we have identified the good parameter is the prefactor of order unity."
\subsubsection{Explicit relation for mass flux $Q$ of monodisperse sand, $d=250\mu m$, on Earth}
Many mass flux relations in the literature are given for free field Earth conditions with $d=250\mu m$ e.g. \cite{Sauermannetal01}. In this section, we provide a further explicit prediction of $Q$ for these conditions, based on the parameter values in the third row in Table (\ref{conditions}). In contrast to the parameter values which we obtained from wind tunnel experiments discussed in the following section, the values given in Table (\ref{conditions}) correspond to free field conditions, because the numerical model of Kok and Renno \cite{KokRenno09} was adjusted to such conditions. From $\alpha=0.94$ we obtain $\overline V_r=1.55m/s$ using (\ref{vr2}) and (\ref{dragcheng}). From $u_t=0.196m/s$, $\beta=0.125$, and $\gamma=0.33$ we further obtain $z_{mt}=0.0153m$ using (\ref{zm}) evaluated at $u_*=u_t$. Inserting all values in (\ref{Q}) then yields
\begin{eqnarray}
 \fl \frac{Q\tilde g}{\rho_wu_*^3}=0.94\left(1-\frac{u_t^2}{u_*^2}\right)\left[\frac{u_t}{\kappa u_*}\ln\frac{z_m}{z_o}+2.5G\left(\frac{u_t}{u_*}\right)-1.33\left(1-\frac{u_t}{u_*}\right)\right. \nonumber\\ 
 +\left.1.05\left(1-\frac{u_t^2}{u_*^2}\right)-\frac{\overline V_r}{u_*}\right], \label{Q250}
\end{eqnarray}
where we used (\ref{lnzo*3}-\ref{Um3}). This is however not an explicit relation for $Q$, since $z_m$ increases with $u_*$ as well. In order to obtain an explicit relation, we approximate $z_m\approxeq z_{mt}$. This is reasonable, since the increase of $Q$ with $z_m$ is only logarithmic. Further inserting $z_o=d/30$ (see Table \ref{conditions}) and using that the first non-vanishing term $0.0125\ln 2(1-u_t^2/u_*^2)^2$ of $G(u_t/u_*)$ (see \ref{appwind}) is very small compared to the other terms close to $u_t$, $G\approx0$, we finally obtain
\begin{eqnarray}
 \fl \frac{Q\tilde g}{\rho_wu_*^3}=\left(1-\frac{u_t^2}{u_*^2}\right)\left[17.67\frac{u_t}{u_*}-1.25\left(1-\frac{u_t}{u_*}\right)+0.98\left(1-\frac{u_t^2}{u_*^2}\right)-\frac{\overline V_r}{u_*}\right]. \label{Q250-2}
\end{eqnarray}
\subsection{Experimental validation}
For the validation with experiments we use the drag law (\ref{Cd}) of Cheng \cite{Cheng97}, which we also used before. Furthermore, we need to account for the fact that the surface roughness $z_o$ of a quiescent sand bed is a function of the roughness Reynolds number. This is in particular important when comparing with experimental impact thresholds, because some experiments were made with very small particle diameters $d$ in the aerodynamically smooth regime. The whole context is explained in \ref{appzo} including equations, which are used to compute $z_o$.

In this section we validate our apparent roughness prediction, (\ref{lnzo*3}-\ref{vr2}), with the experiments of Rasmussen et al. \cite{Rasmussenetal96} for five different particle diameters $d$. The chosen data set has the advantage that the scatter in the data is small in comparison to other data sets \cite{Dongetal03,ShermanFarrell08}. Furthermore we use a combination of several data sets \cite{Bagnold37,Chepil45,Rasmussenetal94}, in order to evaluate our impact threshold prediction, (\ref{ut}-\ref{vr4}) and the data set of Creyssels et al. \cite{Creysselsetal09} for our mass flux prediction (\ref{Q}). The latter choice is motivated by the fact that Creyssels et al. \cite{Creysselsetal09} used particle tracking methods, which are more accurate than measurements of the mass flux $Q$ with sand traps, which underestimate $Q$ by up to $50\%$ \cite{Greeleyetal96,RasmussenMikkelsen98}. Furthermore the experiments were performed in the same wind tunnel as the experiments of Rasmussen et al. \cite{Rasmussenetal96} and mass flux measurements typically vary from wind tunnel to wind tunnel. For instance the two recent measurements of the mass flux with particle tracking methods \cite{Creysselsetal09,Hoetal11} show the same qualitative behavior, an approximate scaling of $Q$ with $u_*^2-u_t^2$, but quite different magnitudes of $Q$, although they used the same sand in their experiments. We want to strongly emphasize that we use only one single set of parameters, $\alpha$, $\beta$, $\gamma$, and $\eta$ to fit all data sets at the same time and that the values of $u_t$, obtained from our prediction (\ref{ut}-\ref{vr4}), are used in (\ref{lnzo*3}-\ref{vr2}) and (\ref{Q}).

\begin{figure}
 \begin{center}
   \begin{tabular}{ll}
  	(a)&(b)\\
	  \includegraphics[scale=0.25]{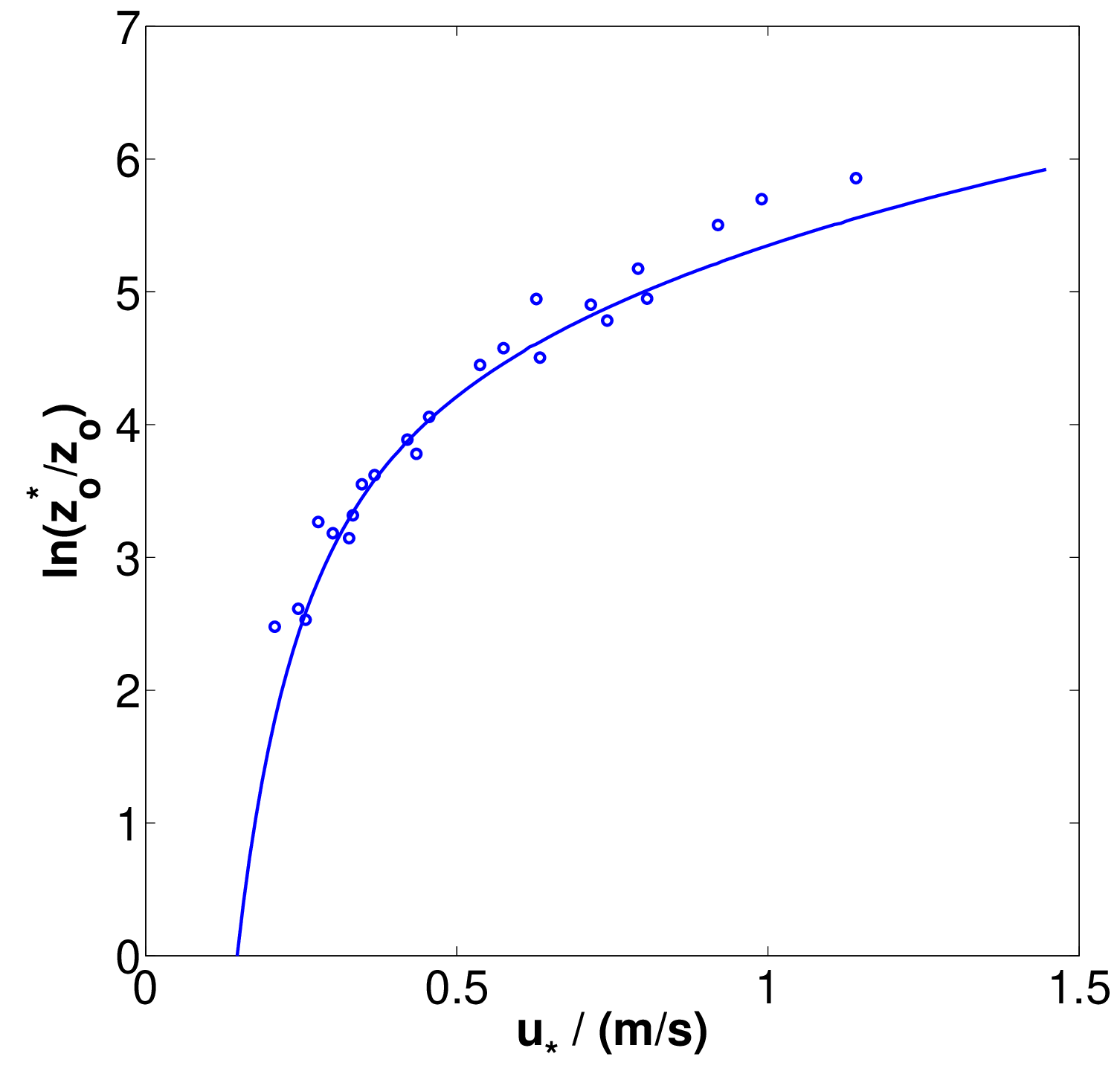} &\includegraphics[scale=0.25]{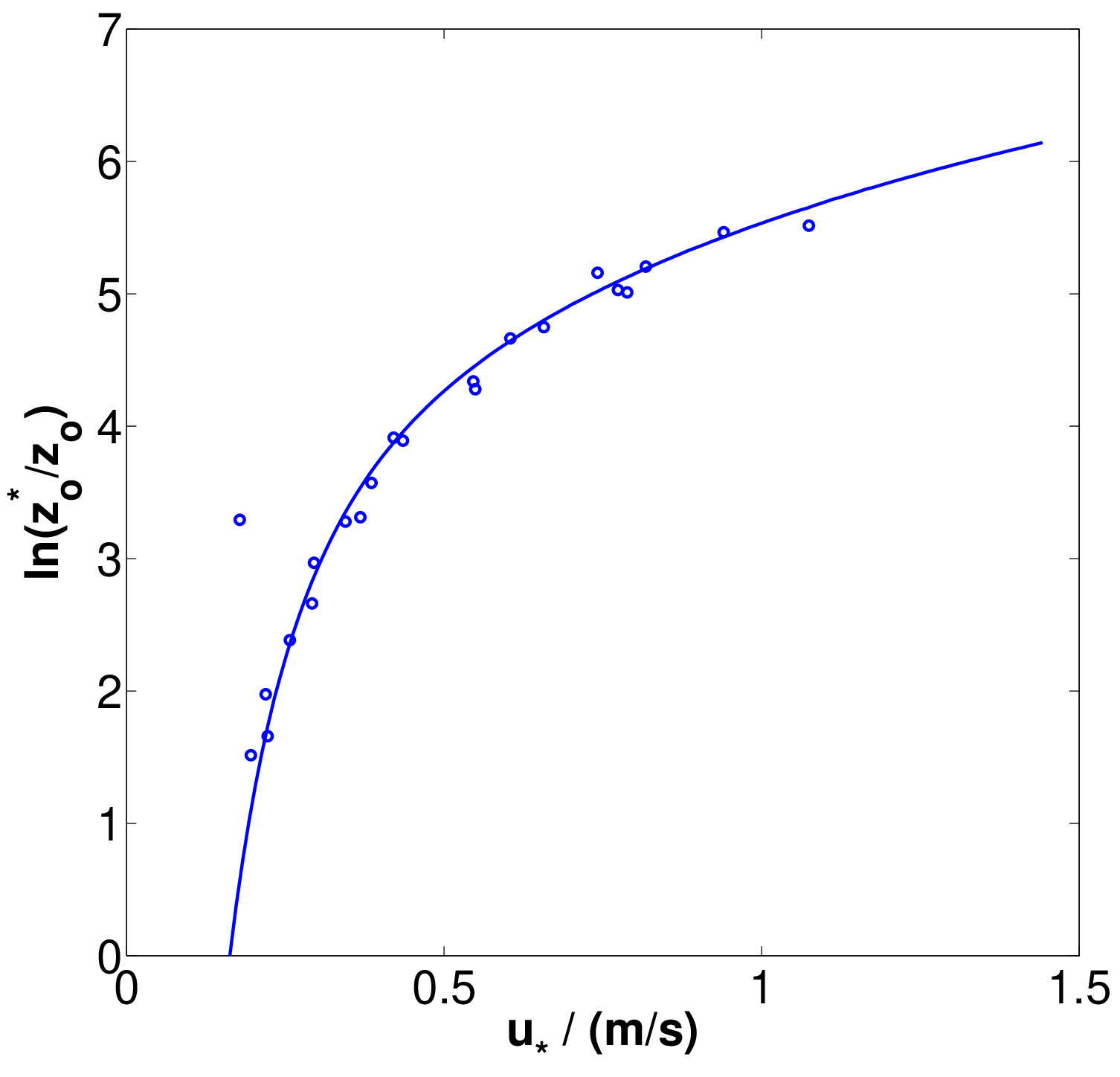} \\
	  (c)&(d)\\
	  \includegraphics[scale=0.25]{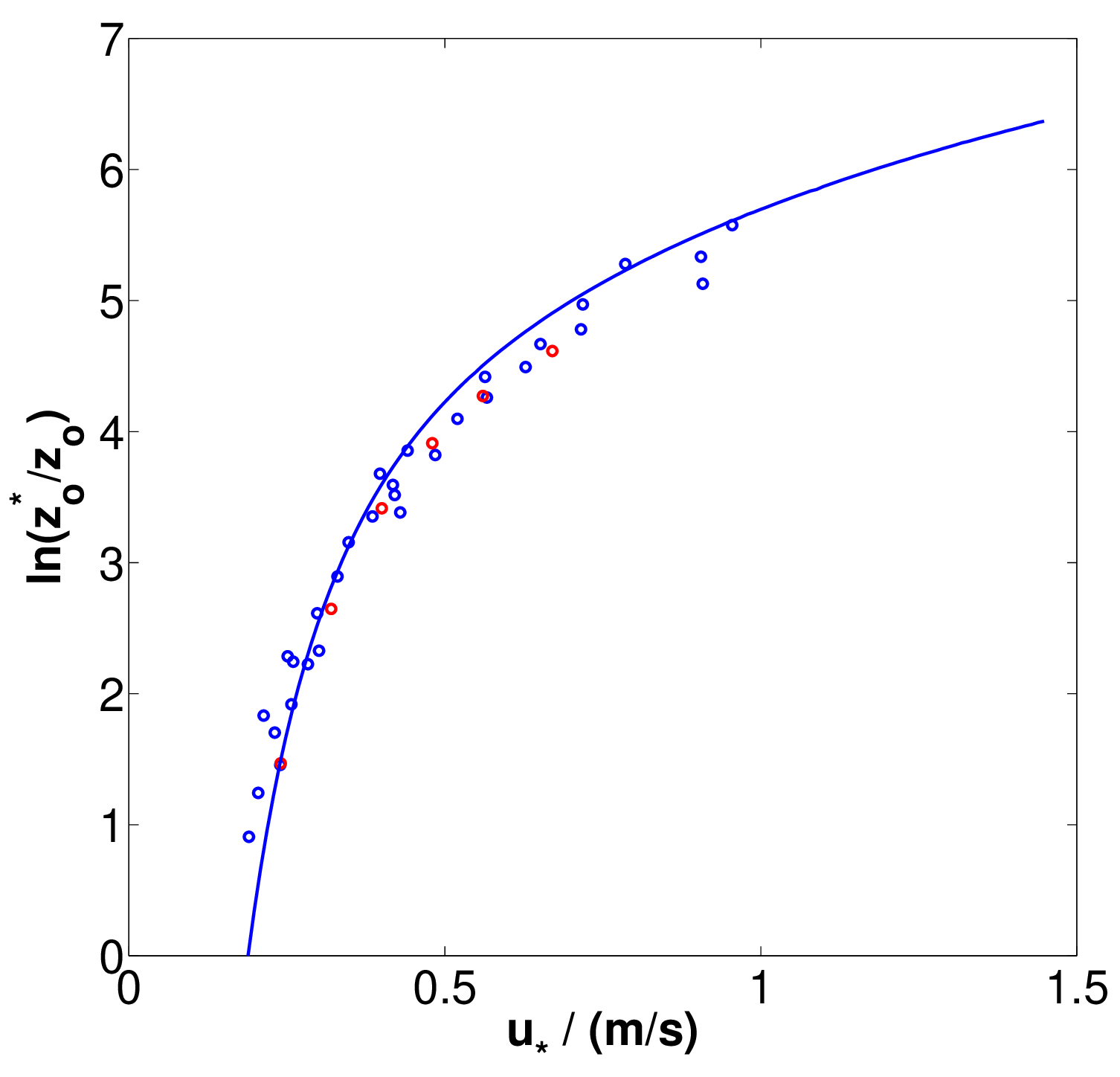} &\includegraphics[scale=0.25]{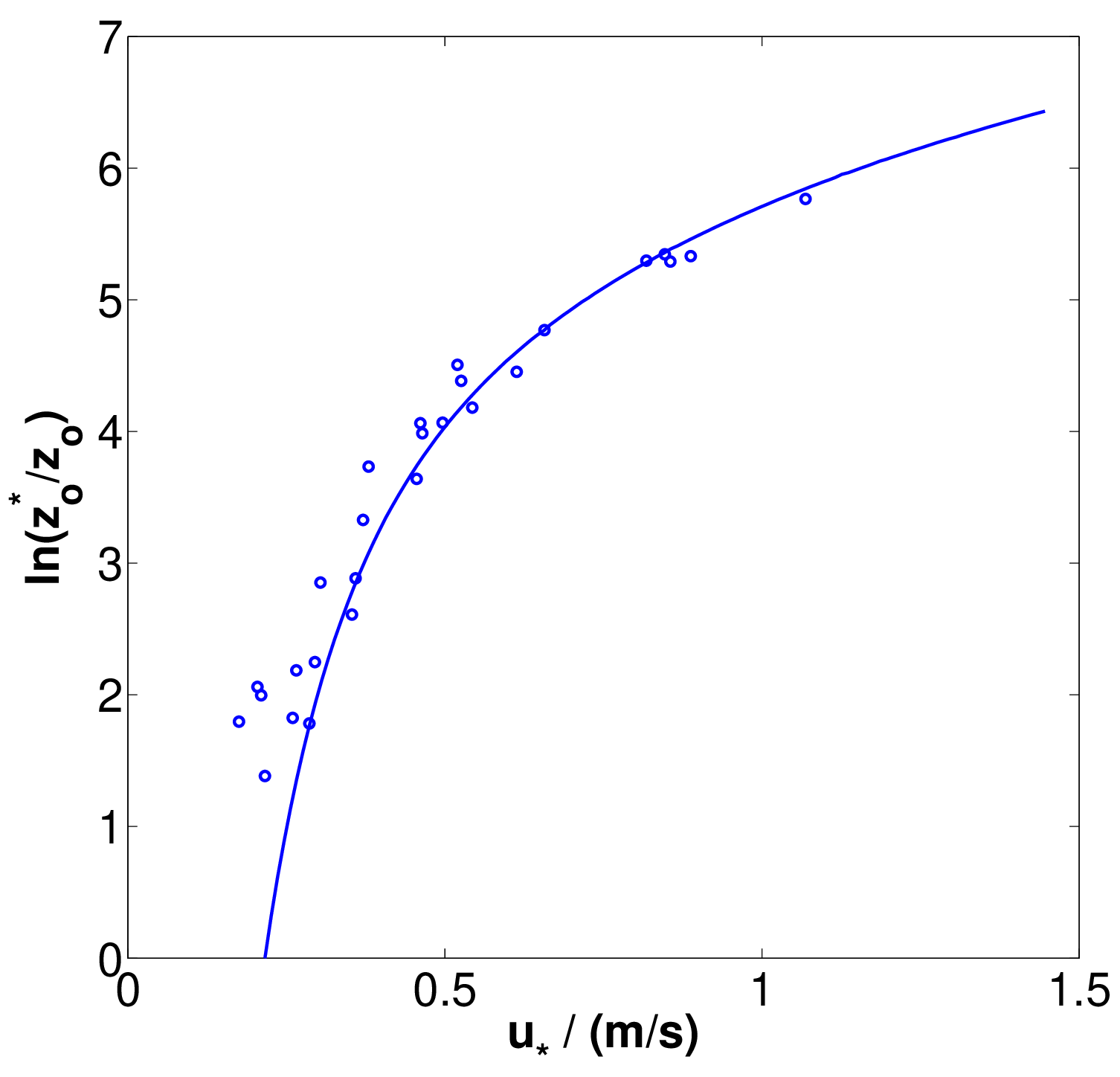}
   \end{tabular}  
    \\(e)\\
   \includegraphics[scale=0.25]{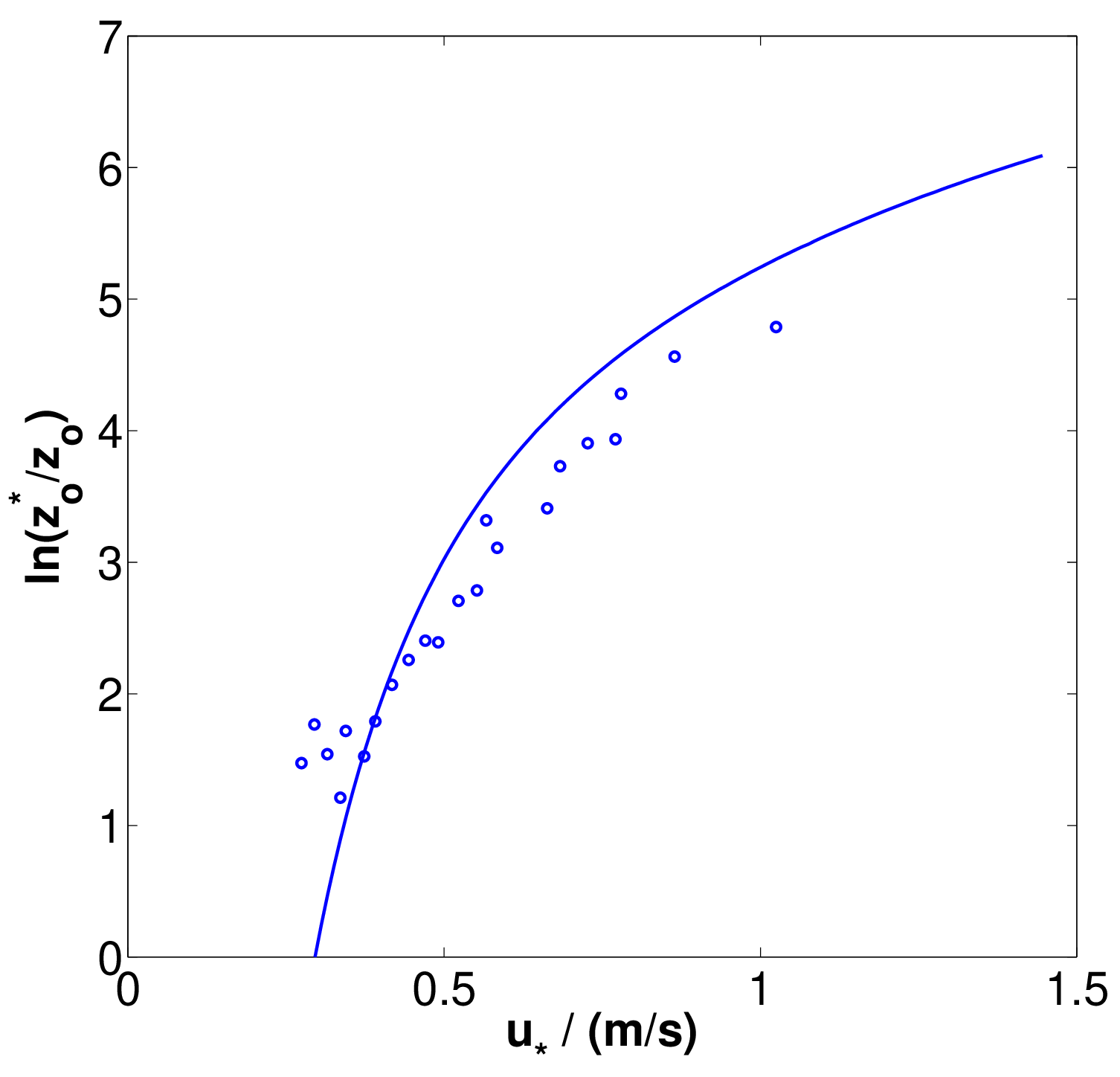}
 \end{center}
 \caption{Comparison of our model (\ref{lnzo*3}-\ref{vr2}) (solid lines) with experimental data of Rasmussen et al. \cite{Rasmussenetal96} (circles) for five different particle sizes, (a) $d=125\mu m$, (b) $d=170\mu m$, (c) $d=242\mu m$, (d) $d=320\mu m$, and (e) $d=544\mu m$. (c) includes six $z_o^*$ data points (red circles) measured by Creyssels et al. \cite{Creysselsetal09}.}
 \label{rasges}
\end{figure}

\begin{figure}
 \begin{center}
  \includegraphics[scale=0.29]{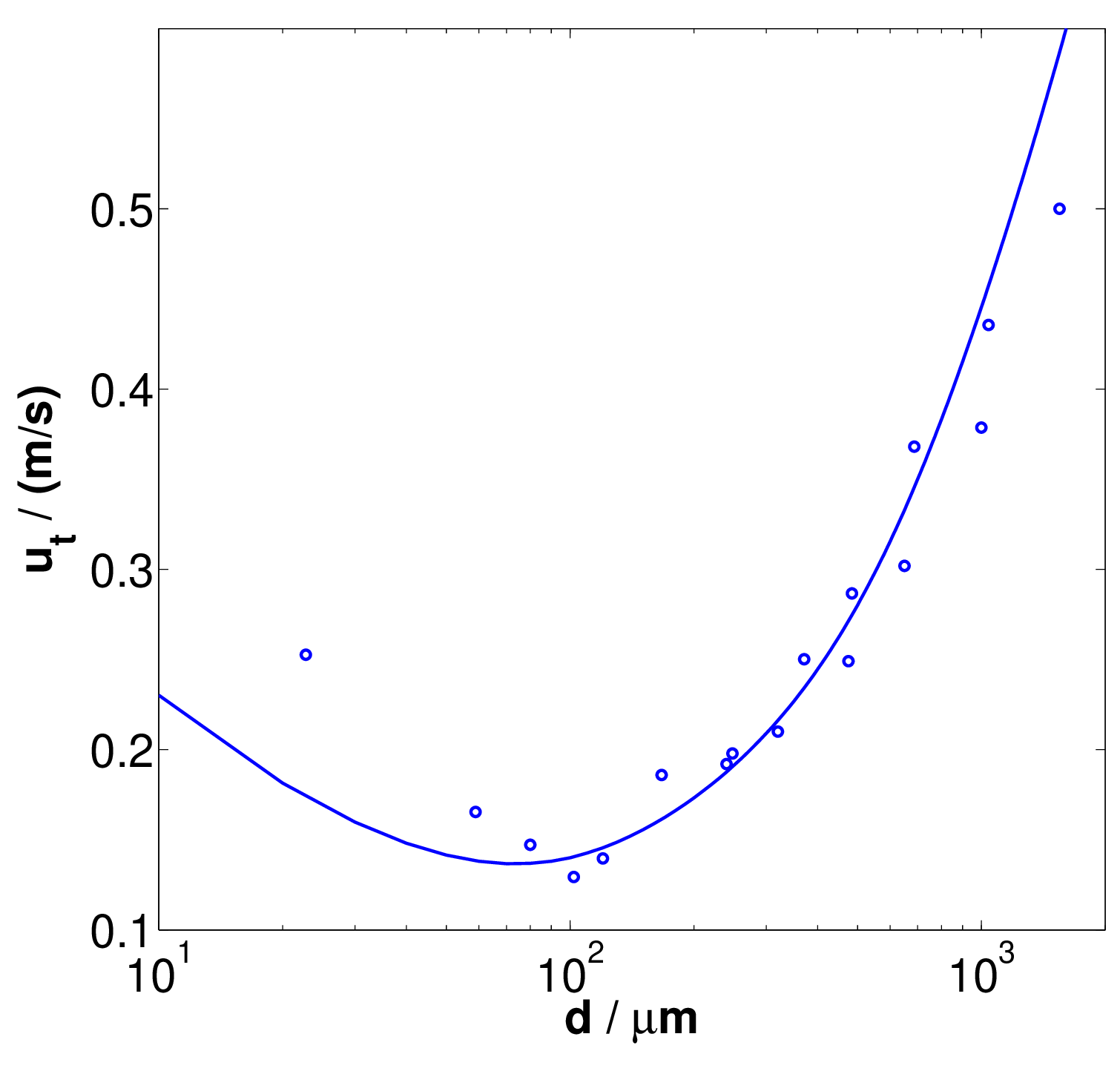}
 \end{center}
 \caption{Comparison of our model (\ref{ut}-\ref{vr4}) (solid line) with experimental data \cite{Bagnold37,Chepil45,Rasmussenetal94} (circles).}
 \label{utcomp}
\end{figure}

\begin{figure}
 \begin{center}
  \includegraphics[scale=0.3]{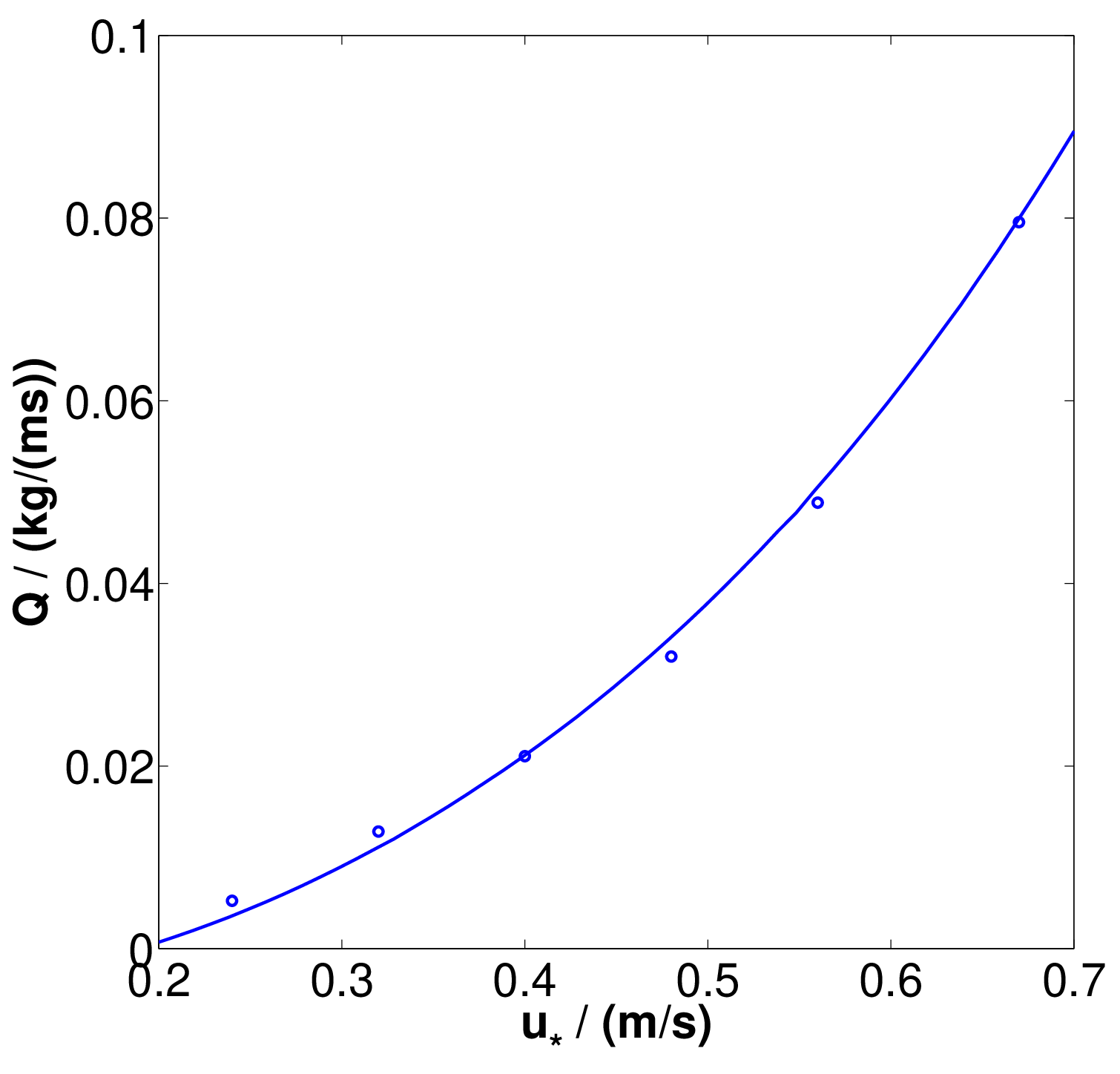}
 \end{center}
 \caption{Comparison of our model (\ref{Q}) (solid line) with experimental data of Creyssels et al. \cite{Creysselsetal09} (circles).}
 \label{crey}
\end{figure}

By fitting the model parameters to $\alpha=1.02$, $\beta=0.095$, $\gamma=0.17$, and $\eta=0.1$ we obtain good to excellent agreement with all data sets. This is shown in Figs.(\ref{rasges}-\ref{crey}), which present the comparison of (\ref{lnzo*3}-\ref{vr2}) with the data of Rasmussen et al. \cite{Rasmussenetal96}, the comparison of (\ref{ut}-\ref{vr4}) with the impact threshold data sets \cite{Bagnold37,Chepil45,Rasmussenetal94}, and the comparison of (\ref{Q}) with the mass flux data of Creyssels et al. \cite{Creysselsetal09}, respectively.
It is important to note that the fitted values of the model parameters significantly differ from those in Table (\ref{conditions}), which were adjusted to match the numerical data obtained from simulation with the model of Kok and Renno \cite{KokRenno09}. The very likely cause is that the numerical model of Kok and Renno \cite{KokRenno09} was adjusted to field data, whereas our model is adjusted to wind tunnel data. According to Sherman and Farrell \cite{ShermanFarrell08}, field and wind tunnel data of the apparent roughness $z_o^*$ can differ by up to one order of magnitude. The reason for this is not fully understood. It was speculated by Raupach \cite{Raupach91} that differences between wind tunnel and field data are attributed to not fully equilibrated saltation in wind tunnels. It is however likely that this is not the only reason, since even in large wind tunnels like the one used by Creyssels et al. \cite{Creysselsetal09} and Rasmussen et al. \cite{Rasmussenetal96}, for which we adjusted our model, the differences still exist. We propose that differences in the grain size distribution could be another cause. Sand in field experiments has typically a much broader grain size distribution than sand with the same mean diameter $d$ typically used in wind tunnels. For instance the five different sands used by Rasmussen et al. \cite{Rasmussenetal96} are the same as the ones used by Iversen and Rasmussen \cite{IversenRasmussen99}, which the latter described as "closely sized sand samples" and referred to as "uniform samples". Bagnold \cite{Bagnold73} argued that the average particle velocity $\overline V$ in aeolian saltation is found to be about two times higher for broadly in comparison to narrowly distributed grain sizes with the same $d$, because "elastic rebound of smaller grains off more massive ones is superimposed on the splash process", leading to higher saltation heights. According to (\ref{zs2}), two times higher $\overline V$ in field experiments than in wind tunnel experiments would lead to about three times higher $z_s$ and therefore to a much higher apparent roughness $z_o^*$.
\section{Discussion} \label{discussion}
We have presented a comprehensive analytical saltation model, which provides a new level of understanding of the change of the average wind profile during aeolian saltation and of aeolian saltation in general. The model incorporates (\ref{lnzo*3}-\ref{vr2}), a novel expression for the apparent roughness $z_o^*$, and (\ref{zs2}), a new expression for the thickness of the saltation layer $z_s$. (\ref{zs2}) can be seen as the main contribution of our study and it is to our knowledge the first relation for $z_s$ entirely derived from physical principles. It is an improvement of the relation (\ref{zsAndreotti}), proposed by Andreotti \cite{Andreotti04}, which had no physical foundations. Besides, our saltation model also provides (\ref{Q}), a novel expression for the mass flux $Q$, and (\ref{ut}-\ref{vr4}), an expression for the impact threshold $u_t$. All relations are extensively evaluated. They are in very good agreement with the recent numerical model of Kok and Renno \cite{KokRenno09} and with many experiments \cite{Creysselsetal09,Rasmussenetal96,Bagnold37,Chepil45,Rasmussenetal94}.

In particular our expression for the mass flux $Q$, (\ref{Q}), is in excellent agreement with the data set of Creyssels et al. \cite{Creysselsetal09} (see Figure \ref{crey}). The authors found a scaling of $Q$ with $u_*^2-u_t^2$, and it can be seen that this is also the dominant term in (\ref{Q}). The same scaling was also found by theoretical studies \cite{Andreotti04,UngarHaff87,Almeidaetal07,Almeidaetal08}, and recently measured by Ho et al. \cite{Hoetal11}, who showed that this scaling relation is a consequence of the bed being erodible instead of fixed. The splash-entrainment mechanism on erodible beds causes the particle slip velocity $V_o$ to be constant with increasing $u_*$, what in turn hinders the average particle velocity $\overline V$ to increase fast with $u_*$. The main increase of the mass flux $Q=M\overline V$ with $u_*$ comes instead from the average transported mass per unit soil area $M$, which scales with $u_*^2-u_t^2$, as we and other studies e.g. \cite{Sauermannetal01} showed.

In conclusion, the facts that our model stays on sound physical foundations and that it is in very good agreement with experiments and state of the art simulations make us confident that it for the first time reliably quantifies the feedback of the sand transport on the wind momentum, even on planets different from Earth.
\section*{Acknowledgement}
We acknowledge the support of ETH Grant ETH-10 09-2. We further acknowledge fruitful discussions with Dirk Kadau, Mathias Fuhr, and Beat L\"uthi.
\appendix
\section{Computing impact and lift-off velocities with the model of Kok \cite{Kok10b}} \label{appkok}
Kok \cite{Kok10b} has recently derived analytic expressions for the average impact and lift-off velocities, from which we obtain the following expressions for the constants $\alpha'$ and $\beta'$ and for the particle slip velocity $V_o$. They write
\begin{eqnarray}
 \alpha'=\frac{-\Delta v_{zo}}{\Delta v_{xo}}=\frac{\sin\theta_iv_i+\sin\theta_lv_l}{\cos\theta_iv_i-\cos\theta_lv_l}, \\
 \beta'=\frac{-\Delta v_{zo}^2}{\Delta v_{xo}^2}=\frac{(\sin\theta_iv_i)^2-(\sin\theta_lv_l)^2}{(\cos\theta_iv_i)^2-(\cos\theta_lv_l)^2}, \\
 V_o=\frac{\rho_{o\uparrow}v_{xo\uparrow}+\rho_{o\downarrow}v_{xo\downarrow}}{\rho_{o\uparrow}+\rho_{o\downarrow}}=\frac{\sin\theta_iv_i\cos\theta_lv_l+\sin\theta_lv_l\cos\theta_iv_i}{\sin\theta_iv_i+\sin\theta_lv_l}, \label{deltav}
\end{eqnarray}
where $\theta_i\approx11^\circ$, $\theta_l\approx40^\circ$, and $\rho_{o\uparrow}v_{zo\uparrow}+\rho_{o\downarrow}v_{zo\downarrow}=0$ were used (see (\ref{phi})). $v_i$ and $v_l$ are furthermore given by
\begin{eqnarray}
 v_i=\frac{1-F}{2r}\sqrt{g_fd}-\frac{1}{2\epsilon}+\sqrt{\frac{1}{4\epsilon^2}+\left(\frac{1-F}{2r}\right)^2g_fd+\frac{1+F}{2r\epsilon}\sqrt{g_fd}}, \label{vimp}
\end{eqnarray}
and
\begin{eqnarray}
 v_l=F\alpha_Rv_{\mathrm{imp}}\left(1-\frac{1}{(1+\beta v_i)^2}\right)+\alpha_{ej}v_i\left(1-\exp\left(-\frac{v_i}{40\sqrt{g_fd}}\right)\right), \label{lift}
\end{eqnarray}
where the parameters are given by $r=0.02$, $F=0.96$, $\epsilon\approx1s/m$, $\alpha_R=0.55$, $\alpha_{ej}=0.15\sqrt{\tilde g/g_f}$ and $g_f$ is an effective gravity, which incorporates the effect of cohesion at small particle diameters $d$,
\begin{eqnarray}
 g_f=\tilde g+\frac{6\zeta}{\pi\rho_sd^2}. \label{cohesion1}
\end{eqnarray}
Here $\zeta$ is the dimensional cohesion parameter. (\ref{cohesion1}) expresses that cohesive forces scale with $d^1$ in contrast to the gravity force, which scales with $d^3$, which means that at small $d$ cohesive forces become dominant. Shao and Lu \cite{ShaoLu00} estimated the magnitude of $\zeta$ to be between about $1\mathrm{x}10^{-4}N/m$ and $5\mathrm{x}10^{-4}N/m$, we use $\zeta=5\mathrm{x}10^{-4}N/m$.
\section{Calculation of the average wind velocity profile and the apparent roughness} \label{appwind}
\begin{figure}
 \begin{center}
  \includegraphics[scale=0.29]{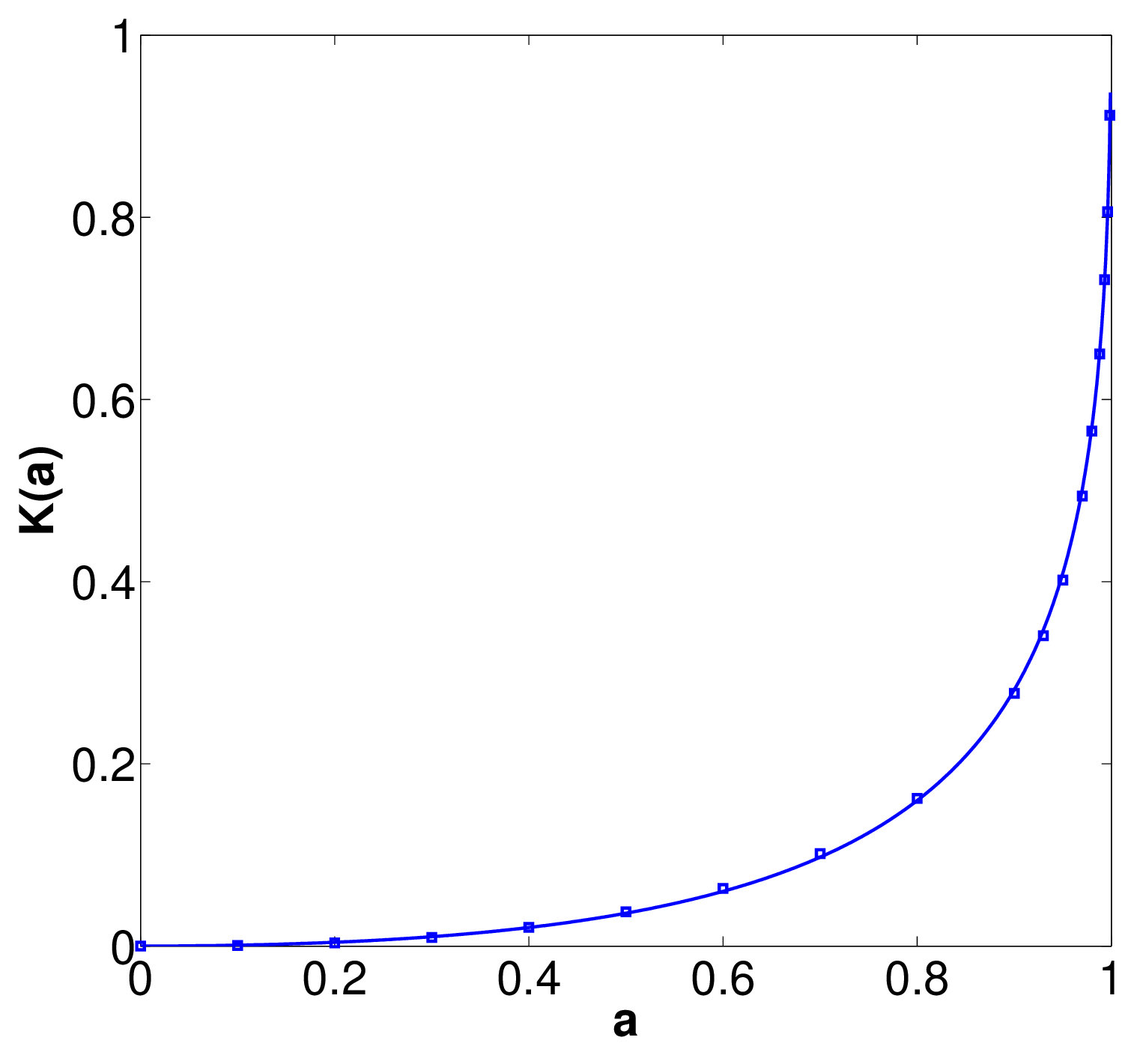}
 \end{center}
 \caption{Comparison between $K(a)$ computed using (\ref{Kalpha}) with an accuracy of at least $10^{-5}$ (squares) and $K(a)$ computed with the approximation (\ref{Kalpha2}) for $0\leq a\leq0.999$.}
 \label{Kfig}
\end{figure}

In this appendix we show how one can compute the average wind velocity profile and the apparent roughness from an exponentially decaying grain shear stress profile,
\begin{eqnarray}
 \tau_g(z)=\tau_{go}e^{-z/z_s}. \label{zsappendix}
\end{eqnarray}
The average wind velocity profile $u(z)$ can be obtained from Prandtl's mixing length approximation \cite{Raupach91,Prandtl25,DuranHerrmann06,Sauermannetal01,AnderssonHaff91} as
\begin{eqnarray}
 \frac{\d u(z)}{dz}=\frac{u_a(z)}{\kappa z}=\frac{u_*}{\kappa z}\sqrt{1-\tau_g(z)/\tau}, \label{defvelprofile} 
\end{eqnarray}
with
\begin{eqnarray}
 u(z_o)=0 \label{boundary},
\end{eqnarray}
where $\tau=\rho_w u_*^2$. The apparent roughness $z_o^*$, is the roughness of the asymptotic fluid velocity profile $\tilde v(z)=v(z)|_{z\rightarrow\infty}$, giving
\begin{eqnarray}
 \tilde u(z)=\frac{u_*}{\kappa}\ln\frac{z}{z_o^*}. \label{assympvelprofile}
\end{eqnarray}
In order to integrate (\ref{defvelprofile}) analytically, we Taylor-expand the square root in (\ref{defvelprofile}) in the argument $a\exp(-z/z_s)$, where $a=\tau_{go}/\tau$. It becomes
\begin{eqnarray}
 \sqrt{1-x}=1-\sum_{j=1}^\infty f_jx^j, \label{Taylor} 
\end{eqnarray}
where $f_j$ is given by
\begin{eqnarray}
 f_j=\frac{(2j-3)!!}{(2j)!!}.
\end{eqnarray}
Then (\ref{defvelprofile}) becomes
\begin{eqnarray}
 u(z)=\frac{u_*}{\kappa}\left(\ln\frac{z}{z_o}-\sum_{j=1}^\infty f_ja^j\int\limits_{z_o}^z\frac{\d z'}{z'}e^{-jz'/z_s}\right). \label{velprofile1}
\end{eqnarray}
Within the integral we transform the $z$-coordinate using
\begin{eqnarray}
 x=\frac{jz'}{z_s},
\end{eqnarray}
such that the integral transforms to 
\begin{eqnarray}
 \int\limits_{z_o}^z\frac{\d z'}{z'}e^{-jz'/z_s}=\int\limits_{jz_o/z_s}^{jz/z_s}\frac{\d x}{x}e^{-x}=\mathrm{E_1}(jz_o/z_s)-\mathrm{E_1}(jz/z_s), \label{E1} 
\end{eqnarray}
where $\mathrm{E_1}(x)$ is called the exponential integral. It can be expressed as
\begin{eqnarray}
 \mathrm{E_1}(x)=-0.577-\ln x+\mathrm{Ein}(x), \label{seriesE1} 
\end{eqnarray}
where $0.577$ is the Euler-Mascheroni constant,
\begin{eqnarray}
 0.57721...=\lim_{n\rightarrow\infty}\left(\sum_{k=1}^n\frac{1}{k}-\ln(n)\right),
\end{eqnarray}
and $\mathrm{Ein}(x)$ is an analytic function with the series-expansion
\begin{eqnarray}
 \mathrm{Ein}(x)=\sum_{l=1}^\infty\frac{(-1)^{l+1}x^l}{ll!}. \label{seriesEin}
\end{eqnarray}
$\mathrm{E_1}(x)$ vanishes for $x\rightarrow\infty$. Using (\ref{E1}) the velocity profile finally writes
\begin{eqnarray}
 u(z)=\frac{u_*}{\kappa}\left(\ln\frac{z}{z_o}-\sum_{j=1}^\infty f_ja^j\left(\mathrm{E_1}(jz_o/z_s)-\mathrm{E_1}(jz/z_s)\right)\right). \label{velprofileapp}
\end{eqnarray}
The comparison with the asymptotic profile (\ref{assympvelprofile}) then yields
\begin{eqnarray}
 \ln\frac{z_o^*}{z_o}=\sum_{j=1}^\infty f_ja^j\mathrm{E_1}(jz_o/z_s). \label{mainequation}
\end{eqnarray}
Since $z_o\ll z_s$, (\ref{mainequation}) can further be written as
\begin{eqnarray}
 \ln\frac{z_o^*}{z_o}=-K(a)+(1-\sqrt{1-a})\left(\ln\frac{z_s}{z_o}-0.577\right), \label{mainequation2} 
\end{eqnarray}
where $K(a)$ is defined by
\begin{eqnarray}
 K(a)=\sum_{j=2}^\infty f_j\ln(j)a^j. \label{Kalpha} 
\end{eqnarray}
To our knowledge, the sum on the right hand side is not analytically treatable with common methods. However, we found that it can be very well approximated by
\begin{eqnarray}
 K(a)&\approxeq&1.154(1-\sqrt{1-a})^{2.56}(1+\sqrt{1-a}\ln \sqrt{1-a}), \label{Kalpha2} 
\end{eqnarray}
where both constants, $2.56$ and $1.154$, are fit constants, ensuring very good agreement between the infinite sum and the approximation. This approximation performs very well over the whole range of $a$, the relative errors being typically below $1\%$, as is shown in Figure \ref{Kfig}.
Now we can finally write
\begin{eqnarray}
 \ln\frac{z_o^*}{z_o}=\left(1-\frac{u_b}{u_*}\right)\ln\frac{z_s}{1.78z_o}+G\left(\frac{u_b}{u_*}\right), \label{mainequation3} 
\end{eqnarray}
where $u_b=u_*\sqrt{1-a}$ was used (see (\ref{ub})), $1.78=\exp(0.577)$, and $G(x)$ is defined by
\begin{eqnarray}
 G(x)=1.154(1+x\ln x)(1-x)^{2.56}. \label{Gapp}
\end{eqnarray}
\subsection{Approximation for large $z$}
Now we further calculate an approximative expression for the wind profile at large heights $z$. The velocity profile in (\ref{velprofileapp}) can be rewritten as
\begin{eqnarray}
 u(z)=\frac{u_*}{\kappa}\left(\ln\frac{z}{z_o^*}+\sum_{j=1}^\infty f_ja^j\mathrm{E_1}(jz/z_s)\right). \label{velprofileapp2}
\end{eqnarray}
where we inserted (\ref{mainequation}). If $z/z_s$ is large enough, it is sufficient to only consider the first term of the Taylor-expansion, because $\mathrm{E_1}(x)$ decreases strongly with increasing argument $x$. This eventually yields
\begin{eqnarray}
 u(z)=\frac{u_*}{\kappa}\ln\frac{z}{z_o^*}+\frac{u_*^2-u_b^2}{2\kappa u_*}\mathrm{E_1}\left(\frac{z}{z_s}\right). \label{velprofileapp3}
\end{eqnarray}
where we also inserted (\ref{ub}) and (\ref{mainequation3}).
\subsection{Approximation for $z<z_s$}
\begin{figure}
 \begin{center}
  \includegraphics[scale=0.29]{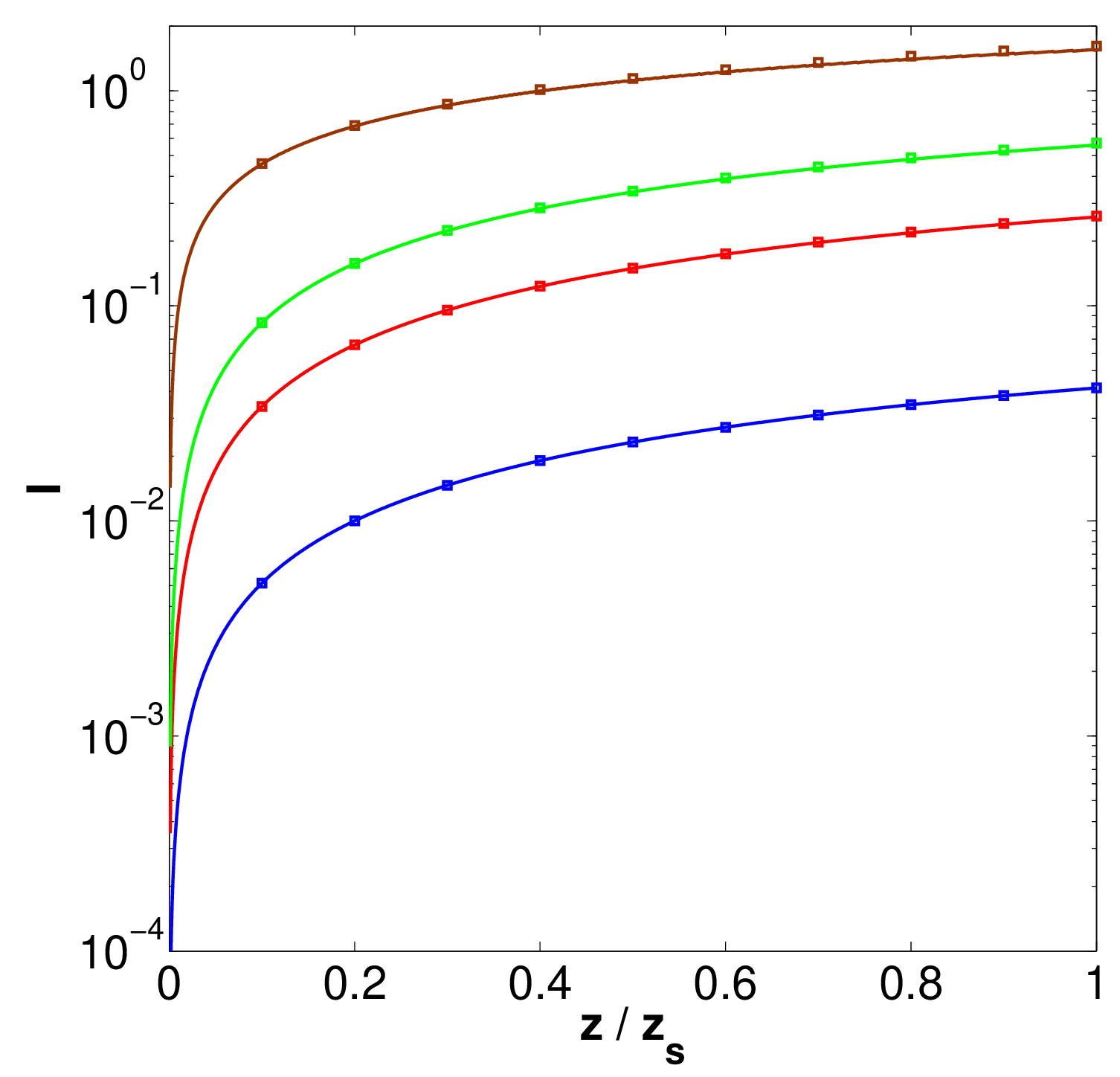}
 \end{center}
 \caption{Comparison between $I$ as function of $z/z_s$ computed using (\ref{infinitesum}) with an accuracy of at least $10^{-5}$ (squares) and $I$ computed using the approximation (\ref{infinitsumapprox}). The different lines correspond to $a=0.1$ (blue), $a=0.5$ (red), $a=0.8$ (green), and $a=0.999$ (brown).}
 \label{Ifig}
\end{figure}

\begin{figure}
 \begin{center}
  \includegraphics[scale=0.29]{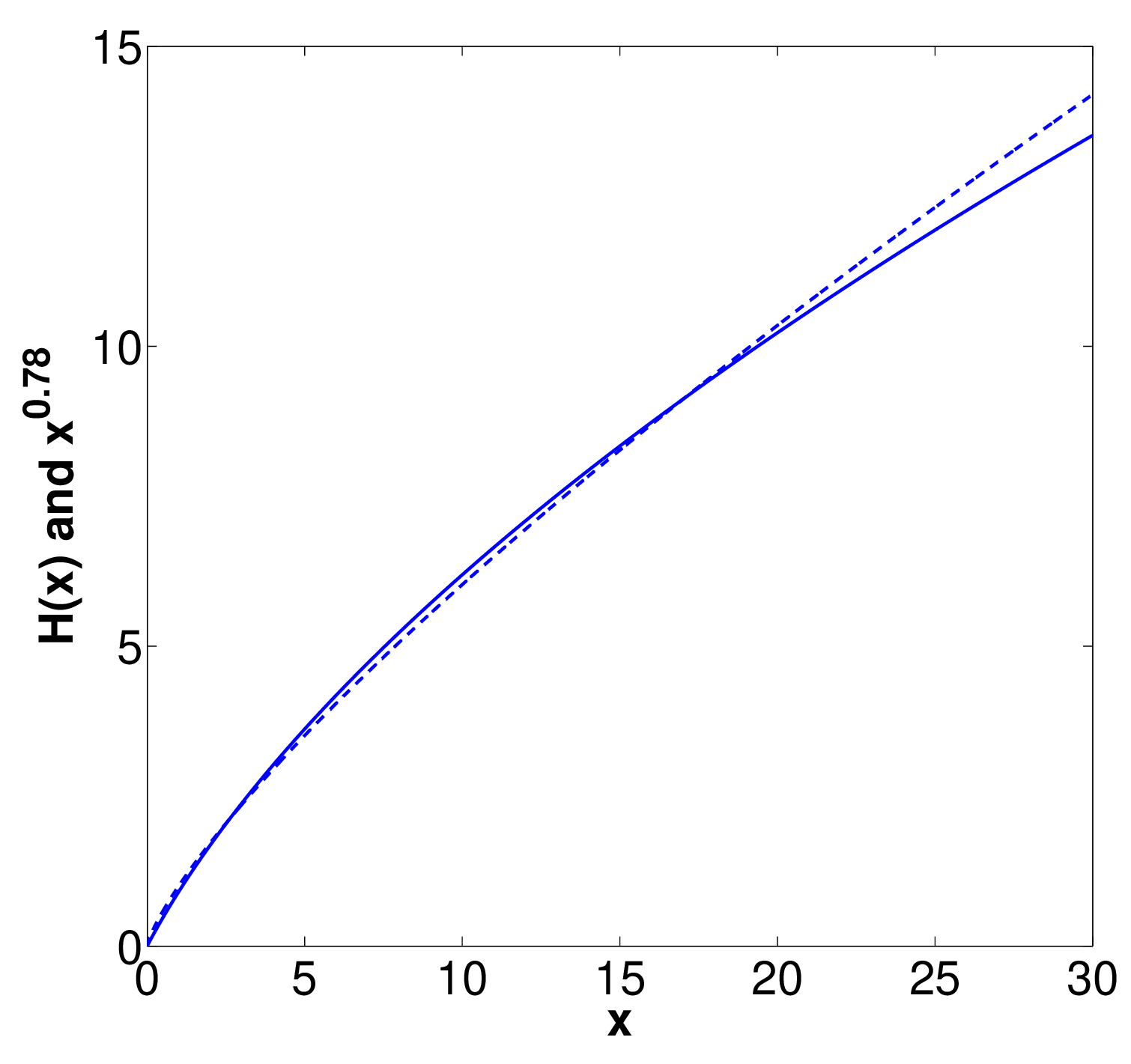}
 \end{center}
 \caption{The function $H(x)$ (solid line) compared with the function $x^{0.78}$ (dashed line).}
 \label{Hfig}
\end{figure}

\begin{figure}
 \begin{center}
  \includegraphics[scale=0.29]{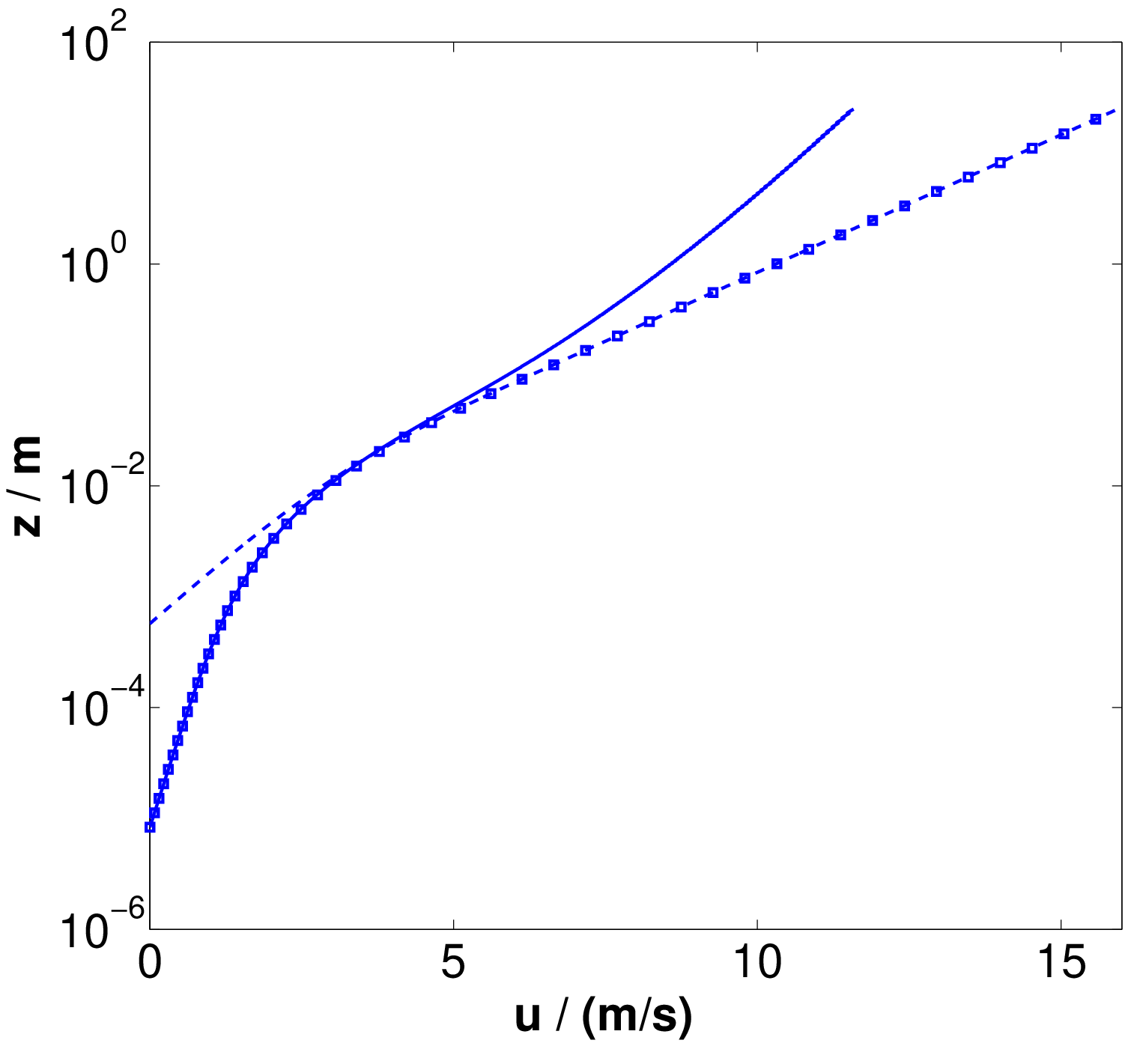}
 \end{center}
 \caption{The exact solution $u(z)$ of the boundary value problem (squares), calculated using (\ref{velprofileapp}) with an accuracy of at least $10^{-5}$, plotted versus the approximation (\ref{windapprox}) (solid line) and approximation (\ref{velprofileapp3}) (dashed line). Here $d=250\mu m$, $u_*=0.7m/s$, $u_b=0.1m/s$, and $z_s=100d$ are used.}
 \label{ufig}
\end{figure}

Here we motivate an approximation of the wind profile $u(z)$ for small heights $z<z_s$. For this purpose, we first make an approximation for the infinite sum
\begin{eqnarray}
 I\left(a,\frac{z}{z_s}\right):=\sum_{j=1}^\infty f_ja^j\mathrm{Ein}(jz/z_s), \label{infinitesum} 
\end{eqnarray}
and then insert it into the velocity profile, given by (\ref{velprofileapp2}). Inserting the series-expansion (\ref{seriesEin}) of $\mathrm{Ein}$ in (\ref{infinitesum}) and exchanging the order of the sums gives
\begin{eqnarray}
 I=\sum_{k=1}^\infty\sum_{j=1}^\infty f_ja^j\frac{(-1)^{k+1}j^k(z/z_s)^k}{kk!}.
\end{eqnarray}
The inner sum can be written as
\begin{eqnarray}\nonumber
 \sum_{j=1}^\infty f_ja^jj^k=(a\partial_a)^k[1-\sqrt{1-a}]. \label{innersum} 
\end{eqnarray}
It can be shown that
\begin{eqnarray}\nonumber
 (a\partial_a)^k=\sum_{l=1}^kS_k^{(l)}a^l\partial_a^l, \label{stirlingsum} 
\end{eqnarray}
where the Stirling numbers of the second kind $S_k^{(l)}$ are defined by
\begin{eqnarray}\nonumber
 S_k^{(l)}=\frac{1}{l!}\sum_{m=0}^l(-1)^m{l \choose m}(l-m)^k. \label{stirlingsumdef} 
\end{eqnarray}
Using
\begin{eqnarray}\nonumber
 \partial_a^l[1-\sqrt{1-a}]=\frac{(2l-3)!!}{2^l(1-a)^{l-1/2}}, \label{ableitungk} 
\end{eqnarray}
we can write
\begin{eqnarray}
 I=\sum_{k=1}^\infty\frac{(-1)^{k+1}(z/z_s)^k}{kk!}\sum_{l=1}^kS_k^{(l)}\frac{(2l-3)!!a^l}{2^l(1-a)^{l-1/2}}.
\end{eqnarray}
The above form of $I$ and the fact that $I$ can be approximated by the first-order Taylor-expansion $I=\frac{1}{2}a\mathrm{Ein}(z/z_s)$ for small $a$ allows the following approximation,
\begin{eqnarray}
 I&\approxeq&\sum_{k=1}^\infty\frac{(-1)^{k+1}(\mathrm{Ein}(z/z_s))^k}{kk!}\frac{(2k-3)!!a^k}{2^k(1-a)^{k-1/2}},
\end{eqnarray}
which is the Taylor-expansion of the function
\begin{eqnarray}
 I=\frac{a\mathrm{Ein}(z/z_s)~_3\mathrm{F}_2\left(\frac{1}{2},1,1;2,2,-\frac{a\mathrm{Ein}(z/z_s)}{1-a}\right)}{2\sqrt{1-a}}. \label{infinitsumapprox}
\end{eqnarray}
The function $~_3\mathrm{F}_2$ is a generalized hypergeometric series \cite{AskeyDaalhuis10}. This approximation performs very well for $z<z_s$ and the whole range of $a$, as can be seen in Figure \ref{Ifig}.
The relative errors of this approximation are typically below $1\%$. Even for larger values of $z$ up to $10z_s$ the approximation is still good, with relative errors below $10\%$. We can rewrite (\ref{infinitsumapprox}) in more compact form as
\begin{eqnarray}
 I=\frac{u_b}{2u_*}H\left(\frac{(u_*^2-u_b^2)\mathrm{Ein}(z/z_s)}{u_b^2}\right), \label{infinitsumapprox2}
\end{eqnarray}
where we used ({\ref{ub}}) and the function $H(x)$ is defined by
\begin{eqnarray}
 H(x)=x~_3\mathrm{F}_2(\frac{1}{2},2,2;2,2,-x). \label{H}
\end{eqnarray}
If accuracy is not crucial, the complicated function $H(x)$ can be replaced by $x^{0.78}$ for small arguments $x<20$. This is shown in Figure \ref{Hfig}.
As outlined before, we now insert the approximation (\ref{infinitsumapprox}) in (\ref{velprofileapp2}). By further using $z_o\ll z_s$, the approximated wind profile $u(z)$ can be written as
\begin{eqnarray}
 u(z)=\frac{u_b}{\kappa}\ln\frac{z}{z_o}+\frac{u_b}{2\kappa}H\left(\frac{(u_*^2-u_b^2)\mathrm{Ein}\left(\frac{z-z_o}{z_s}\right)}{u_b^2}\right). \label{windapprox}
\end{eqnarray}
The approximated wind profile calculated using (\ref{windapprox}) (solid line) and (\ref{velprofileapp3}) (dashed line) is exemplary plotted in Figure \ref{ufig} versus the exact solution of the initial boundary value problem (squares), calculated by (\ref{velprofileapp}) with an accuracy of at least $10^{-5}$.
It can be seen that (\ref{windapprox}) is an excellent approximation of the exact solution of the boundary value problem for small $z$ and (\ref{velprofileapp3}) an excellent approximation for large $z$. Since both approximations underestimate the analytic solution, the maximum value of $u(z)$ calculated using (\ref{velprofileapp3}) and (\ref{windapprox}), represents an excellent approximation for the whole range of $z$.
\section{Surface roughness of a quiescent sand bed} \label{appzo}
\begin{figure}
 \begin{center}
  \includegraphics[scale=0.29]{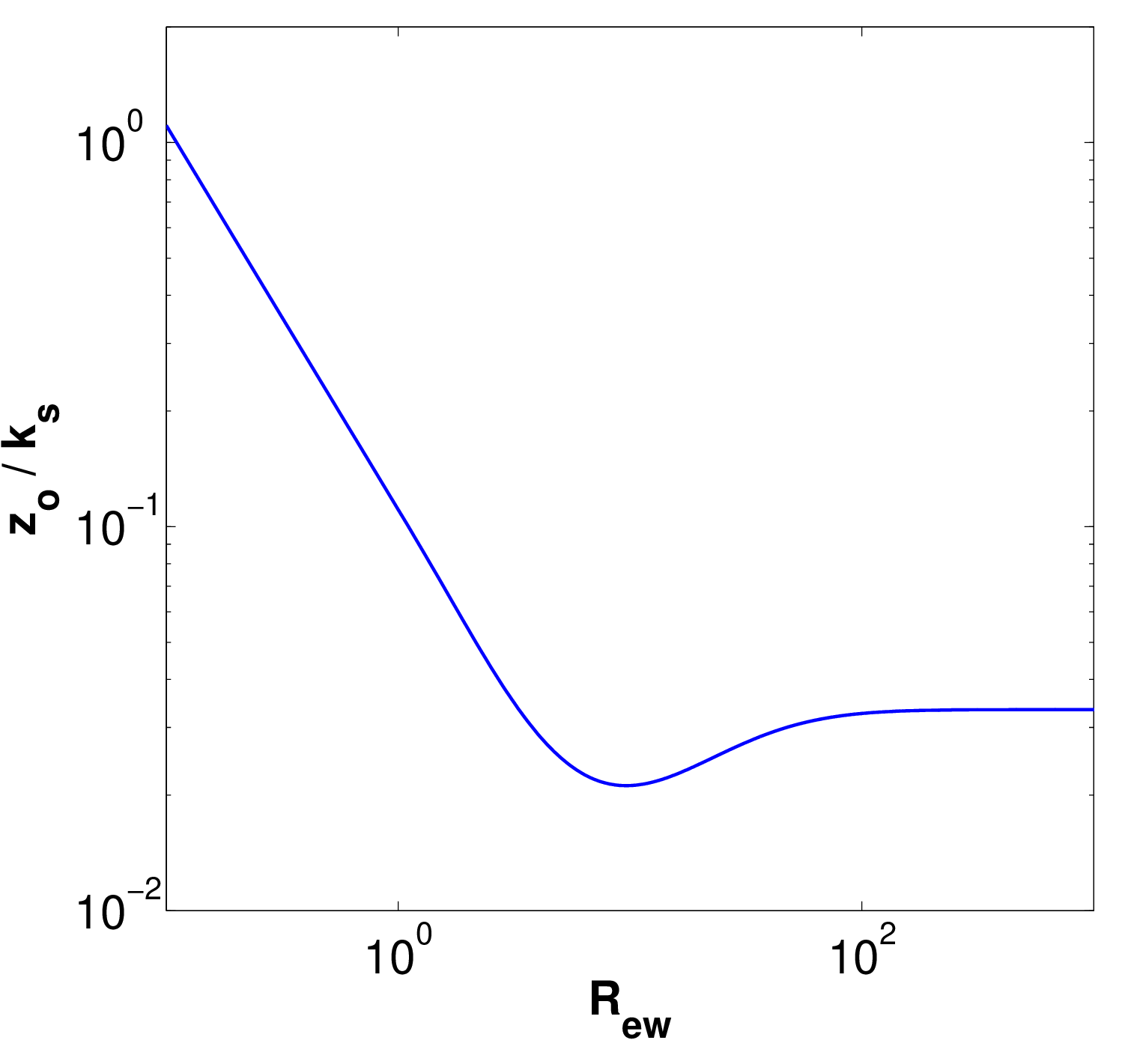}
 \end{center}
 \caption{Plot of $z_o/k_s$ over $R_{ew}$ according to the relation of Cheng and Chiew \cite{ChengChiew98}, (\ref{zo}) and (\ref{B}).}
 \label{zoRew}
\end{figure}
The surface roughness $z_o$ in the absence of saltation depends on the roughness Reynolds number
\begin{eqnarray}
 R_{ew}=\frac{u_*k_s\rho_w}{\mu}, \label{Rew}
\end{eqnarray}
where $k_s$ is the equivalent Nikuradse roughness \cite{Nikuradse33,Keulegan38}. $k_s$ equals the grain diameter $d$, if the grains of the sand bed are monodisperse, spherical, and very well arranged, meaning that the center point of each particle of the topmost layer is at the same height. However, under more natural conditions $k_s$ can be larger, depending on the grain size distribution and the arrangement of the sand bed, and the shape of the grains. A typical value, which is used by engineers, is $k_s=3d_{84}$ for water flows in pipes and flumes \cite{Keulegan38}, where $d_{84}$ denotes the diameter value which is larger than $84\%$ of the grains of the grain size distribution, or $k_s=2d$ for wind flows \cite{Sherman92}. However, since we validate our model with experiments, which used narrowly distributed sand \cite{Creysselsetal09,Rasmussenetal96}, we use $k_s=d$. Note that the value of $k_s$ does not much influence the final results in most cases. The well known and widely used roughness law
\begin{eqnarray}
 z_o=k_s/30 \label{zorough}
\end{eqnarray}
is obtained for large roughness Reynolds numbers $R_{ew}>100$, which is called the aerodynamic rough regime. On the other hand, in the limit of low roughness Reynolds numbers $R_{ew}<3$, the roughness is proportional to the size of the viscous sublayer
\begin{eqnarray}
 z_o=\mu/(9\rho_wu_*), \label{zosmooth}
\end{eqnarray}
which is called the aerodynamic smooth regime. Most natural conditions for aeolian saltation on Earth fall between those regimes, what is called the aerodynamic transitional regime. For instance we obtain $R_{ew}\approx6$ for wind with a shear velocity of $u_*=0.4m/s$ over a typical sand surface with a mean diameter of $d=250\mu m$ and $k_s=d$. The behavior of $z_o$ as function of $R_{ew}$ was measured by Nikuradse \cite{Nikuradse33} for pipe flows and described by Cheng and Chiew \cite{ChengChiew98} by the following empirical relation
\begin{eqnarray}
 z_o=k_s\exp(-\kappa B), \label{zo}
\end{eqnarray}
where
\begin{eqnarray}
 B=8.5+(2.5\ln R_{ew}-3)\exp\left(-0.11(\ln R_{ew})^{2.5}\right). \label{B}
\end{eqnarray}
$z_o/k_s$ is plotted in Figure \ref{zoRew} as function of $R_{ew}$. It describes a function, which has a minimum in the transitional regime at $R_{ew}=9.6$ and converges against (\ref{zorough}) for large $R_{ew}$ and against (\ref{zosmooth}) for low $R_{ew}$. The same behavior was measured by Dong et al. \cite{Dongetal01} in wind tunnels. Since the variance of $z_o/k_s$ between the transitional and rough regime is not very large and the measurement errors of $z_o$ are large, it is appropriate for saltation models to use a constant value for $z_o/k_s$ like $z_o=k_s/30$ in these regimes. However one cannot neglect the very strong increase of $z_o$ for Reynolds numbers below $R_{ew}=3$ in the smooth regime. Since we consider very small particle diameters in our model, for which $R_{ew}<3$, we use (\ref{zo}) and (\ref{B}) to compute $z_o$.
\glossary{name={$z$}, description={height above the sand bed = vertical coordinate}}
\glossary{name={$x$}, description={horizontal coordinate in direction of the wind flow}}
\glossary{name={$u_*$}, description={fluid shear velocity}}
\glossary{name={$\kappa$}, description={von K\'arm\'an constant}}
\glossary{name={$\mathbf{u}(z)$}, description={average vectorial fluid velocity at height $z$}}
\glossary{name={$z_o^*$}, description={apparent roughness of the transported saltation layer}}
\glossary{name={$z_o$}, description={surface roughness of a quiescent sand bed}}
\glossary{name={$g$}, description={gravity constant}}
\glossary{name={$\tilde g$}, description={buoyancy-reduced gravity constant}}
\glossary{name={$v_l$}, description={average lift-off velocity}}
\glossary{name={$\theta_l$}, description={average lift-off angle}}
\glossary{name={$v_i$}, description={average impact velocity}}
\glossary{name={$\theta_i$}, description={average impact angle}}
\glossary{name={$h$}, description={average saltation height}}
\glossary{name={$z_r$}, description={height of the low-energy layer}}
\glossary{name={$u_t$}, description={impact threshold = shear velocity above which saltation can be sustained}}
\glossary{name={$z_s$}, description={decay height of an exponentially decreasing grain shear stress profile}}
\glossary{name={$z_q$}, description={decay height of an exponentially decreasing mass flux profile}}
\glossary{name={$\tau_g(z)$}, description={grain shear stress profile}}
\glossary{name={$\tau_{go}$}, description={$\tau_{go}=\tau_g(0)$}}
\glossary{name={$d$}, description={mean particle diameter}}
\glossary{name={$\rho_s$}, description={particle density}}
\glossary{name={$\rho_w$}, description={air density}}
\glossary{name={$s$}, description={$s=\rho_s/\rho_w$}}
\glossary{name={$\mu$}, description={kinematic viscosity of the air}}
\glossary{name={$Q$}, description={mass flux}}
\glossary{name={$\alpha$}, description={first model parameter}}
\glossary{name={$\alpha'$}, description={a parameter proportional to $\alpha$}}
\glossary{name={$\beta$}, description={second model parameter}}
\glossary{name={$\beta'$}, description={parameter proportional to $\beta$}}
\glossary{name={$\gamma$}, description={third model parameter, $\gamma=z_m/z_s$}}
\glossary{name={$z_m$}, description={effective height, where the mean motion takes place}}
\glossary{name={$\eta$}, description={fourth model parameter, defined by (\ref{gammadef})}}
\glossary{name={$\rho(z)$}, description={transported mass per volume at height $z$}}
\glossary{name={$\rho_o$}, description={$\rho_o=\rho(0)$}}
\glossary{name={$M$}, description={transported mass per unit soil area}}
\glossary{name={$\mathbf{v}(z)$}, description={average vectorial particle velocity at height $z$}}
\glossary{name={$v_{xo\uparrow(\downarrow)}$}, description={$v_{xo\uparrow(\downarrow)}=v_{x\uparrow(\downarrow)}(0)$}}
\glossary{name={$v_{zo\uparrow(\downarrow)}$}, description={$v_{zo\uparrow(\downarrow)}=v_{z\uparrow(\downarrow)}(0)$}}
\glossary{name={$\mathbf{v_r}(z)$}, description={$\mathbf{u}(z)-\mathbf{v}(z)$}}
\glossary{name={$\Delta v_x(z)$,$\Delta v_z(z)$,$\Delta v_x^2(z)$,$\Delta v_z^2(z)$}, description={velocity differences defined by (\ref{deltavx}-\ref{deltavz2})}}
\glossary{name={$\Delta v_{xo}$,$\Delta v_{zo}$,$\Delta v_{xo}^2$,$\Delta v_{zo}^2$}, description={$\Delta v_x(0)$,$\Delta v_z(0)$,$\Delta v_x^2(0)$,$\Delta v_z^2(0)$}}
\glossary{name={$\phi(z)$}, description={local vertical mass flux at height $z$}}
\glossary{name={$p_g(z)$}, description={grain normal stress profile}}
\glossary{name={$p_{go}$}, description={$p_{go}=p_g(0)$}}
\glossary{name={$\mathbf{f}$}, description={sum of average vectorial drag force and gravity}}
\glossary{name={$C_d(V)$}, description={drag coefficient as function of a velocity difference $V$}}
\glossary{name={$C_o$}, description={parameter in the drag law}}
\glossary{name={$C_\infty$}, description={$C_\infty=C_d(\infty)$}}
\glossary{name={$\mathbf{\mathrm{e_z}}$}, description={unit vector in $z$-direction}}
\glossary{name={$\overline V_r$}, description={$\overline V_r=\overline U-\overline V$}}
\glossary{name={$\overline V$}, description={average particle velocity}}
\glossary{name={$\overline{V_z^2}$}, description={$\overline{V_z^2}=\overline{\langle v_z^2\rangle}$}}
\glossary{name={$\overline U$}, description={$\overline U=u(z_m)$}}
\glossary{name={$G$}, description={function defined by (\ref{G})}}
\glossary{name={$H$}, description={function defined by (\ref{H})}}
\glossary{name={$\mathrm{E_1}$}, description={function defined by (\ref{expint})}}
\glossary{name={$\mathrm{Ein}$}, description={function defined by (\ref{seriesEin})}}
\glossary{name={$~_3F_2$}, description={generalized hypergeometric series}}
\glossary{name={$u_a(z)$}, description={reduced shear velocity profile, defined by (\ref{ua})}}
\glossary{name={$u_b$}, description={$u_b=u_a(0)$}}
\glossary{name={$V_o$}, description={average particle slip velocity, $V_o=(v_{xo\uparrow}+v_{xo\downarrow})/2$}}
\glossary{name={$\overline V_t$}, description={$\overline V_t=\overline V(u_t)$}}
\glossary{name={$\overline U_t$}, description={$\overline U_t=\overline U(u_t)$}}
\glossary{name={$z_{mt}$}, description={$z_{mt}=z_m(u_t)$}}
\glossary{name={$\tau$}, description={wind shear stress, $\tau=\rho_wu_*^2$}}
\glossary{name={$a$}, description={$a=\tau_{go}/\tau$}}
\glossary{name={$f_j$}, description={Taylor-coefficients of $1-\sqrt{1-x}$}}
\glossary{name={$K(a)$}, description={function defined by (\ref{Kalpha})}}
\glossary{name={$I$}, description={infinite sum defined by (\ref{infinitesum})}}
\glossary{name={$S_k^{(l)}$}, description={Stirling numbers of the second kind}}
\glossary{name={$g_f$}, description={effective gravity incorporating the effect of cohesion}}
\glossary{name={$\zeta$}, description={dimensional cohesion parameter}}
\printglossary
\section*{References}
\bibliographystyle{unsrt}
\bibliography{model}

\begin{thebibliography}{10}

\bibitem{Bagnold41}
R.~A. Bagnold.
\newblock {\em The physics of blown sand and desert dunes}.
\newblock Methuen, New York, 1941.

\bibitem{Kroyetal02a}
K.~Kroy, G.~Sauermann, and H.~J. Herrmann.
\newblock Minimal model for aeolian sand dunes.
\newblock {\em Physical Review E}, 66:031302, 2002.

\bibitem{Kroyetal02b}
K.~Kroy, G.~Sauermann, and H.~J. Herrmann.
\newblock A minimal model for sand dunes.
\newblock {\em Physical Review Letters}, 64:054301, 2002.

\bibitem{ParteliHerrmann07}
E.~J.~R. Parteli and H.~J. Herrmann.
\newblock Dune formation on the present mars.
\newblock {\em Physical Review E}, 76:041307, 2007.

\bibitem{Partelietal07}
E.~J.~R. Parteli, O.~Duran, and H.~J. Herrmann.
\newblock Minimal size of a barchan dune.
\newblock {\em Physical Review E}, 75:011301, 2007.

\bibitem{Owen64}
P.~R. Owen.
\newblock Saltation of uniform grains in air.
\newblock {\em Journal of Fluid Mechanics}, 20(2):225--242, 1964.

\bibitem{Charnock55}
H.~Charnock.
\newblock Wind stress on a water surface.
\newblock {\em The Quarterly Journal Of The Royal Meteorological Society},
  81:639--640, 1955.

\bibitem{Namikas03}
S.~L. Namikas.
\newblock Field measurement and numerical modelling of aeolian mass flux
  distributions on a sandy beach.
\newblock {\em Sedimentology}, 50(2):303–326, 2003.

\bibitem{RasmussenSorensen08}
K.~R. Rasmussen and M.~Sorensen.
\newblock Vertical variation of particle speed and flux density in aeolian
  saltation: Measurement and modeling.
\newblock {\em Journal of Geophysical Research}, 113:F02S12, 2008.

\bibitem{Creysselsetal09}
M.~Creysells, P.~Dupont, A.~Ould el~Moctar, A.~Valance, I.~Cantat, J.~T.
  Jenkins, J.~M. Pasini, and K.~R. Rasmussen.
\newblock Saltating particles in a turbulent boundary layer: experiment and
  theory.
\newblock {\em Journal of Fluid Mechanics}, 625:47--74, 2009.

\bibitem{Hoetal11}
T.~D. Ho, A.~Valance, P.~Dupont, and A.~Ould~El Moctar.
\newblock Scaling laws in aeolian sand transport.
\newblock {\em Physical Review Letters}, 106:094501, 2011.

\bibitem{Rasmussenetal96}
K.~R. Rasmussen, J.~D. Iversen, and P.~Rautahemio.
\newblock Saltation and wind-flow interaction in a variable slope wind tunnel.
\newblock {\em Geomorphology}, 17:19--28, 1996.

\bibitem{Dongetal03}
Z.~Dong, X.~Liu, and H.~Wang.
\newblock The aerodynamic roughness with a blowing sand boundary layer (bsbl):
  A redefinition of the owen effect.
\newblock {\em Geophysical Research Letters}, 30(2):1047, 2003.

\bibitem{ShermanFarrell08}
D.~J. Sherman and E.~J. Farrell.
\newblock Aerodynamic roughness lengths over movable beds: Comparison of wind
  tunnel and field data.
\newblock {\em Journal of Geophysical Research}, 113:F02S08, 2008.

\bibitem{Sherman92}
D.~J. Sherman.
\newblock An equilibrium relationship for shear velocity and roughness length
  in aeolian saltation.
\newblock {\em Geomorphology}, 5:419--431, 1992.

\bibitem{Kok10a}
J.~F. Kok.
\newblock Difference in the wind speeds required for initiation versus
  continuation of sand transport on mars: Implications for dunes and dust
  storms.
\newblock {\em Physical Review Letters}, 104:074502, 2010.

\bibitem{Raupach91}
M.~R. Raupach.
\newblock Saltation layers, vegetation canopies and roughness lengths.
\newblock {\em Acta Mechanica Supplementum}, 1:83--96, 1991.

\bibitem{Prandtl25}
L.~Prandtl.
\newblock \"uber die ausgebildete turbulenz.
\newblock {\em Zeitschrift f\"ur Angewandte Mathematik und Mechanik},
  5:136--139, 1925.

\bibitem{Andreotti04}
B.~Andreotti.
\newblock A two-species model of aeolian sand transport.
\newblock {\em Journal of Fluid Mechanics}, 510:47--70, 2004.

\bibitem{DuranHerrmann06}
O.~Dur\'an and H.~J. Herrmann.
\newblock Modelling of saturated sand flux.
\newblock {\em Journal of Statistical Mechanics}, page P07011, 2006.

\bibitem{Bagnold73}
R.~A. Bagnold.
\newblock The nature of saltation and "bed-load" transport in water.
\newblock {\em Proceedings of the Royal Society London Series A}, 332:473–504,
  1973.

\bibitem{Sorensen91}
M.~Sorensen.
\newblock An analytic model of wind-blown sand transport.
\newblock {\em Acta Mechanica Supplement}, 1:67--81, 1996.

\bibitem{Sauermannetal01}
G.~Sauermann, K.~Kroy, and H.~J. Herrmann.
\newblock A continuum saltation model for sand dunes.
\newblock {\em Physical Review E}, 64:31305, 2001.

\bibitem{Bagnold37}
R.~A. Bagnold.
\newblock The transport of sand by wind.
\newblock {\em The Geographical Journal}, 89(5):409--438, 1937.

\bibitem{Chepil45}
W.~S. Chepil.
\newblock Dynamics of wind erosion: Ii. initiation of soil movement.
\newblock {\em Soil Science}, 60:397--411, 1945.

\bibitem{Rasmussenetal94}
K.~R. Rasmussen, J.~D. Iversen, and P.~Rautahemio.
\newblock The effect of surface slope on saltation threshold.
\newblock {\em Sedimentology}, 41:721--728, 1994.

\bibitem{KokRenno09}
J.~F. Kok and N.~O. Renno.
\newblock A comprehensive numerical model of steady state saltation (comsalt).
\newblock {\em Journal of Geophysical Research}, 114:D17204, 2009.

\bibitem{SorensenMcEwan96}
M.~Sorensen and I.~McEwan.
\newblock On the effect of mid-air collisions on aeolian saltation.
\newblock {\em Sedimentology}, 43:65--76, 1996.

\bibitem{Dongetal05}
Z.~Dong, N.~Huang, and X.~Liu.
\newblock Simulation of the probability of midair interparticle collisions in
  an aeolian saltating cloud.
\newblock {\em Journal of Geophysical Research}, 110:D24113, 2005.

\bibitem{Huangetal07}
N.~Huang, Y.~Zhang, and R.~D'Adomo.
\newblock A model of the trajectories and midair collision probabilities of
  sand particles in a steady state saltation cloud.
\newblock {\em Journal of Geophysical Research}, 112:D08206, 2007.

\bibitem{RenHuang10}
S.~Ren and N.~Huang.
\newblock A numerical model of the evolution of sand saltation with
  consideration of two feedback mechanisms.
\newblock {\em European Physical Journal E}, 33:351--358, 2010.

\bibitem{Jenkinsetal10}
J.~T. Jenkins, I.~Cantat, and A.~Valance.
\newblock Continuum model for steady, fully developed saltation above a
  horizontal particle bed.
\newblock {\em Physical Review E}, 82:020301(R), 2010.

\bibitem{AnderssonHaff91}
R.~S. Andersson and P.~K. Haff.
\newblock Wind modification and bed response during saltation of sand in air.
\newblock {\em Acta Mechanica Supplementum}, 1:21--51, 1991.

\bibitem{UngarHaff87}
J.~E. Ungar and P.~K. Haff.
\newblock Steady state saltation in air.
\newblock {\em Sedimentology}, 34:289--299, 1987.

\bibitem{Rubey33}
W.~W. Rubey.
\newblock Settling velocity of gravel, sand and silt particles.
\newblock {\em American Journal of Science}, 5(25):325--338, 1933.

\bibitem{Julien95}
P.~Y. Julien.
\newblock {\em Erosion and Sedimentation}.
\newblock Press Syndicate of the University of Cambridge, 1995.

\bibitem{Cheng97}
N.~S. Cheng.
\newblock Simplified settling velocity formula for sediment particle.
\newblock {\em Journal of Hydraulic Engineering}, 123(2):149--152, 1997.

\bibitem{Riceetal95}
M.~A. Rice, B.~B. Willetts, and I.~K. McEwan.
\newblock An experimental study of multiple grain-size ejecta produced by
  collisions of saltating grains with a flat bed.
\newblock {\em Sedimentology}, 42(4):695--706, 1995.

\bibitem{Ogeretal08}
L.~Oger, M.~Ammi, A.~Valance, and D.~Beladjine.
\newblock Study of the collision of one rapid sphere on 3d packings:
  Experimental and numerical results.
\newblock {\em Computers and Mathematics with Applications}, 55:132--148, 2008.

\bibitem{Kok10b}
J.~F. Kok.
\newblock An improved parameterization of wind blown sand flux on mars that
  includes the effect of hysteresis.
\newblock {\em Geophysical Research Letters}, 37:L12202, 2010.

\bibitem{Greeleyetal96}
R.~Greeley, D.~G. Blumberg, and S.~H. Williams.
\newblock Field measurements of the flux and speed of wind-blown sand.
\newblock {\em Sedimentology}, 43:41--52, 1996.

\bibitem{Dongetal04}
Z.~Dong, H.~Wang, X.~Liu, and X.~Wang.
\newblock The blown sand flux over a sandy surface: a wind tunnel investigation
  on the fetch effect.
\newblock {\em Geomorphology}, 57:117--127, 2004.

\bibitem{DongQian07}
Z.~Dong and G.~Qian.
\newblock Characterizing the height profile of the flux of wind-eroded
  sediment.
\newblock {\em Environmental Geology}, 54(5):835--845, 2007.

\bibitem{Bagnold38}
R.~A. Bagnold.
\newblock The measurement of sand storms.
\newblock {\em Proceedings of the Royal Society London Series A}, 167:282--291,
  1938.

\bibitem{RasmussenMikkelsen98}
K.~R. Rasmussen and H.~E. Mikkelsen.
\newblock On the efficiency of vertical array aeolian field traps.
\newblock {\em Sedimentology}, 45(4):789--800, 1998.

\bibitem{IversenRasmussen99}
J.~D. Iversen and K.~R. Rasmussen.
\newblock The effect of wind speed and bed slope on sand transport.
\newblock {\em Sedimentology}, 46(4):723--731, 1999.

\bibitem{Almeidaetal07}
M.~P. Almeida, J.~S. Andrade, and H.~J. Herrmann.
\newblock Aeolian transport of sand.
\newblock {\em The European Physical Journal E}, 22:195--200, 2007.

\bibitem{Almeidaetal08}
M.~P. Almeida, E.~J.~R. Parteli, J.~S. Andrade, and H.~J. Herrmann.
\newblock Giant saltation on mars.
\newblock {\em Proceedings of the National Academy of Science},
  105(17):6222--6226, 2008.

\bibitem{ShaoLu00}
Y.~Shao and H.~Lu.
\newblock A simple expression for wind erosion threshold friction velocity.
\newblock {\em Journal of Geophysical Research}, 105(D17):22437--22443, 2000.

\bibitem{AskeyDaalhuis10}
R.~A. Askey and A.~B.~Olde Daalhuis.
\newblock {\em "Generalized hypergeometric function" in NIST Handbook of
  Mathematical Functions}.
\newblock Cambridge University Press, http://dlmf.nist.gov/16, 2010.

\bibitem{ChengChiew98}
N.~S. Cheng and Y.~M. Chiew.
\newblock Modified logarithmic law for velocity distribution subjected to
  upward seepage.
\newblock {\em Journal of Hydraulic Engineering}, 124(12):1235--1241, 1998.

\bibitem{Nikuradse33}
J.~Nikuradse.
\newblock {\em "Str\"omungsgesetze in rauhen R\"ohren" in Forschungsheft 361,
  Teil B}.
\newblock VDI Verlag, Berlin, 1933.

\bibitem{Keulegan38}
G.~H. Keulegan.
\newblock Laws of turbulent flow in open channels.
\newblock {\em Journal of the National Bureau of Standards}, 21(Research Paper
  1151):707--741, 1938.

\bibitem{Dongetal01}
Z.~Dong, X.~Wang, A.~Zhao, L.~Liu, and X.~Liu.
\newblock Aerodynamic roughness of fixed sand beds.
\newblock {\em Journal of Geophysical Research}, 106:11001--11011, 2001.

\end{thebibliography}
\end{document}